\documentclass[useAMS,usenatbib]{mn2e}

\usepackage{morefloats}
\usepackage{array}

\usepackage{epsfig}

\title[AKARI/IRC source catalogues and source counts for three extragalactic fields]{AKARI/IRC source catalogues and source counts for the IRAC Dark Field, ELAIS North and the AKARI Deep Field South}
\author[H. Davidge et al.]{H. Davidge$^{1}$\thanks{E-mail:
helen.davidge@open.ac.uk}, S. Serjeant$^{1}$, C. Pearson$^{1,2,3}$, H. Matsuhara$^{4,5}$, T. Wada$^{4}$, B. Dryer$^{1}$, \vspace*{0.25cm}\\ \LARGE \rm \hspace*{-0.1cm}L. Barrufet$^{1,2}$\\
$^{1}$School of Physical Sciences, The Open University, Milton Keynes, MK7 6AA, UK\\
$^{2}$RAL Space, Rutherford Appleton Laboratory, Chilton, Didcot, Oxfordshire OX11 0QX, UK\\
$^{3}$Oxford Astrophysics, Denys Wilkinson Building, University of Oxford, Keble Rd, Oxford OX1 3RH, UK\\
$^{4}$Institute of Space and Astronautical Science, Japan Aerospace Exploration Agency, Sagamihara, Kanagawa 229-8510, Japan\\
$^{5}$Department of Space and Astronautical Science, The Graduate University for Advanced Studies, Hayama, Kanagawa 240-0193, Japan}
\begin{document}

\date{Accepted YYYY month DD. Received YYYY month DD; in original form YYYY month DD}

\pagerange{\pageref{firstpage}--\pageref{lastpage}} \pubyear{2014}

\maketitle

\label{firstpage}

\begin{abstract}
We present the first detailed analysis of three extragalactic fields (IRAC Dark Field, ELAIS-N1, ADF-S) observed by the infrared satellite, {\it AKARI}, using an optimised data analysis toolkit specifically for the processing of extragalactic point sources. The InfaRed Camera (IRC) on {\it AKARI} complements the {\it Spitzer} space telescope via its comprehensive coverage between  $8-24\,\mu$m filling the gap between the {\it Spitzer} IRAC and MIPS instruments. Source counts in the {\it AKARI} bands at 3.2, 4.1, 7, 11, 15 and 18\,$\mu$m are presented. At near-infrared wavelengths, our source counts are consistent with counts made in other {\it AKARI} fields and in general with {\it Spitzer}/IRAC (except at 3.2\,$\mu$m where our counts lie above). In the mid-infrared (11\,$-$\,18\,$\mu$m) we find our counts are consistent with both previous surveys by {\it AKARI} and the {\it Spitzer} peak-up imaging survey with the InfraRed Spectrograph (IRS). Using our counts to constrain contemporary evolutionary models we find that although the models and counts are in agreement at mid-infrared wavelengths there are inconsistencies at wavelengths shortward of 7\,$\mu$m, suggesting either a problem with stellar subtraction or indicating the need for refinement of the stellar population models. We have also investigated the {\it AKARI}/IRC filters, and find an AGN selection criteria out to $z<2$ on the basis of {\it AKARI} 4.1, 11, 15 and 18\,$\mu$m colours.

\end{abstract}


\section{Introduction}
One of the most basic statistical properties in the analysis of galaxy populations is galaxy source counts. Originally suggested as a method of determining the geometry of the Universe, early discoveries using the results from counts at radio wavelengths, showed that the Universe is inconsistent with a steady-state model \citep{RowanRobinson1967}. Subsequently galaxy source counts have been used to study galaxy evolution, star formation history and the epoch of galaxy formation (e.g., \citet{Ellis1987}).

Given that around half of the energy in the Universe is emitted at infrared wavelengths, models of infrared galaxy counts have become important in analysing the dusty star-formation history of the Universe (\citealt{Franceschini91}, \citealt{blain93}, \citealt{Pearson1996}). One key infrared wavelength regime to study galaxy evolution is the mid-infrared, as this probes the Polycyclic Aromatic Hydrocarbon (PAH), silicate feature and AGN emission as a function of redshift. Large area multi-wavelength surveys have been carried out by {\it Spitzer} \citep{Werner2004}, {\it AKARI} \citep{Murakami2007} and {\it WISE} \citep{Wright2010}.

{\it AKARI} was Japan's first satellite dedicated to infrared astronomy; launched on 21 February 2006, it was operational until 24 November 2011 \citep{Murakami2007}. {\it AKARI} carried a Ritchey-Chr\'etien telescope, with effective diameter of 68.5\,cm, and two instruments: Far-Infrared Surveyor (FIS, \citealt{Kawada2007}) observing from 50\,-\,180\,$\umu$m and the InfraRed Camera (IRC, \citealt{Onaka2007}) observing at 1.5\,-\,26.5\,$\umu$m.

The {\it AKARI} archive has over 5000 individual observations, and contains multiple pointings of many extragalactic deep fields. The {\it AKARI}/IRC imaging covers the 8\,-\,24\,$\umu$m band gap between {\it Spitzer}/MIPS and {\it Spitzer}/IRAC, and is able to observe deeper over this wavelength range than any other telescope \citep{Werner2004}.

In this paper, we present an optimised data processing chain for {\it AKARI}/IRC data, specifically tailored for the production of high quality extragalactic images and the extraction of galaxy point sources. In Section~\ref{instrumentpipeline} we review the original and new archival pipelines in the context of the IRC instrument and the {\it AKARI} mission. In Section \ref{optimisedpipeline}, the new optimised toolkit is described, step-by-step. In Section~\ref{results} the optimised toolkit is applied to a deep early to mid-Phase 2 data (Section~\ref{irac_dark_field}), deep late-Phase 2 data (Section~\ref{ELAIS_N1}) and a shallow (Section~\ref{ADFS}) field respectively. Our results, including the galaxy source counts are presented in Section~\ref{results2} and discussed in the wider context of observational galaxy surveys and phenomenological source count models in Section~\ref{discussion}. The {\it AKARI}/IRC colour-colour space is explore, with the view to find an AGN selection criteria. Finally our conclusions are presented in Section~\ref{conclusions}. Throughout this work we assume a concordance cosmology of a Hubble constant of  $H_0=67.8$\,km\,s$^{-1}$\,Mpc$^{-1}$ and density parameters of $\Omega_{\rm M}=0.3$ and $\Omega_\Lambda=0.7$.

\section{The IRC instrument and pipeline}
\label{instrumentpipeline}
\subsection{Instrumentation}
\label{instrumentation}
The IRC contained three detectors, the NIR (near-infrared), the MIR-S (mid-infrared short) and the MIR-L (mid-infrared long), each of which had three filters and one prism or grism. The imaging area is smaller than the array, because the area of the array around the slit was masked. The NIR and MIR-S detectors share the same field of view (FoV); the FoV of the MIR-L detector is offset by $\sim$\,20 arcmin. The imaging specifications of the nine {\it AKARI}/IRC filters are presented in Table \ref{table:IRC_detectors}. These specifications did not alter during the mission. 

\begin{table*}
  \caption{Specifications of the 9 filters in the IRC.}
  \label{table:IRC_detectors}
  \begin{center}
    \begin{tabular}{llllllllll}
      \hline \hline
 Channel&Name&Filter&Wavelength&Centre&Effective width&Detector&Array size&Imaging FoV&Pixel scale\\
 &&&$\umu$m&$\umu$m&$\umu$m&&pixels&arcmin&arcsec\\
 \hline
 NIR&N2&filter&1.9-2.8&2.34&0.71&InSb&$512 \times 412$&$9.3 \times 10.0$&$1.46 \times 1.46$\\
 NIR&N3&filter&2.7-3.8&3.19&0.87&InSb&$512 \times 412$&$9.3 \times 10.0$&$1.46 \times 1.46$\\
 NIR&N4&filter&3.6-5.3&4.33&1.53&InSb&$512 \times 412$&$9.3 \times 10.0$&$1.46 \times 1.46$\\
 MIR-S&S7&filter&5.9-8.4&7.12&1.75&Si:As&$256 \times 256$&$9.1 \times 10.0$&$2.34 \times 2.34$\\
 MIR-S&S9W&filter&6.7-11.6&8.61&4.10&Si:As&$256 \times 256$&$9.1 \times 10.0$&$2.34 \times 2.34$\\
 MIR-S&S11&filter&8.5-13.1&10.45&4.12&Si:As&$256 \times 256$&$9.1 \times 10.0$&$2.34 \times 2.34$\\
 MIR-L&L15&filter&12.6-19.4&15.58&5.98&Si:As&$256 \times 256$&$10.3 \times 10.2$&$2.51 \times 2.39$\\
 MIR-L&L18W&filter&13.9-25.6&18.39&9.97&Si:As&$256 \times 256$&$10.3 \times 10.2$&$2.51 \times 2.39$\\
 MIR-L&L24&filter&20.3-26.5&22.89&5.34&Si:As&$256 \times 256$&$10.3 \times 10.2$&$2.51 \times 2.39$\\
    \hline
    \end{tabular}
  \end{center}
\end{table*}

The {\it AKARI} mission was divided into three phases. In Phase 1, from May 2006 to November 2006, {\it AKARI} performed an All-Sky Survey at far infrared \citep{Yamamura2010} and mid infrared \citep{Ishihara2010} wavelengths. Phase 2 spanned November 2006 to August 2007, and was predominantly populated by guaranteed time observations (referred to as Mission Programmes) and Open Time observations. During Phases 1 and 2, the telescope and instruments, including the IRC, were cryogenically cooled to $\sim$\,6\,K. The supply of liquid helium coolant was exhausted in August 2007. A warm Phase 3 operating only the NIR detector at 40\,K, consisted mainly of Open Time observations \citep{Murakami2007}.

\subsection{The Original Archival Pipeline}
\label{the_standard_pipeline}
The original archival IRC imaging pipeline (version 20110304) used to populate the {\it AKARI} data archive, runs in the IRAF \footnote{IRAF is distributed by the National Optical Astronomy Observatory, which is operated by the Association of Universities for Research in Astronomy, Inc., under cooperative agreement with the National Science Foundation.} environment. The original archival pipeline was written to process all of the {\it AKARI}/IRC pointings, both Galactic and extragalactic. The original archival pipeline is sub-divided into three parts: the pre-pipeline, the pipeline and the post-pipeline. The pre-pipeline slices the raw 3D data cubes into 2D image frames and creates an observation log file for use in subsequent steps. The main pipeline is comprised of eleven processing steps, which correct for numerous instrumental artefacts, perform dark subtraction and flat fielding. The post-pipeline has four steps to co-add the individual frames from each pointing together and a step to correct for any offset in the applied World Coordinate System (WCS) \citep{Lorente2008}.

There were several outstanding issues with the original archival IRC pipeline, that prompted the development of the reanalysis in this paper, and in parallel drove some of the changes to the instrument team's IRC pipeline (itself informed in part by the work in this paper), discussed in the next section. The raw frames suffer from an astrometry error; to be aligned with the correct WCS, each frame needs to be astrometry corrected. The new and updated archival pipeline (discussed below) makes a partial correction for this error. The raw IRC frames are also warped; this warping is due to both an optical distortion and because the detectors are not completely square on the sky.  Frames from several of the IRC filters can also remain badly affected by reflected Earthshine light, that creates a flux gradient across the frame. The original archival pipeline also only makes partial corrections for the image warping and the Earthshine gradient artefact.

\subsection{Updated Archival Pipeline}
\label{updatedpipeline}
There exists a major improvement to the original archival pipeline. This new archival pipeline is discussed in detail in \cite{Egusa2016}. This updated pipeline was written to process all IRC pointings taking during Phases 1 and 2, totalling $\sim\,4000$ pointings. All the pointings have been processed through this new archival pipeline and are available for public download from the {\it AKARI} archive. The new archival pipeline has done much work on the dark and flat field corrections, but has not corrected for several artefacts, including: {\it column pull-down}, {\it muxbleed}, (both discussed below), `ghosts'. memory effects and Earthshine light. For fields with enough sources, the new archival pipeline has been able to perform an astrometry correction. It should be noted that the new archival pipeline is written to process generic IRC observations, both Galactic and extragalactic, and hence is not optimised for the specific processing of extragalactic images for point source extraction, which is the specific objective of this work. 

\section{The Optimised Toolkit}
\label{optimisedpipeline}
\subsection{Overview}
\label{The_optimised_Pipeline}
This paper presents a new toolkit (hereafter referred to as the optimised toolkit) for the specific use of processing extragalactic {\it AKARI} deep field images which the archival pipelines were unable to do to a high enough scientific level. The optimised toolkit was written in the Interactive Data Language (IDL) \footnote{Interactive  data Language: http://www.exelisvis.com/Products Services/IDL.aspx}, in order to take advantage of IDL's array based processing, which is well suited to the IRC data sets. This new toolkit has been created with the assistance of the {\it AKARI}/IRC archival team.

\begin{figure}
\centering
\includegraphics{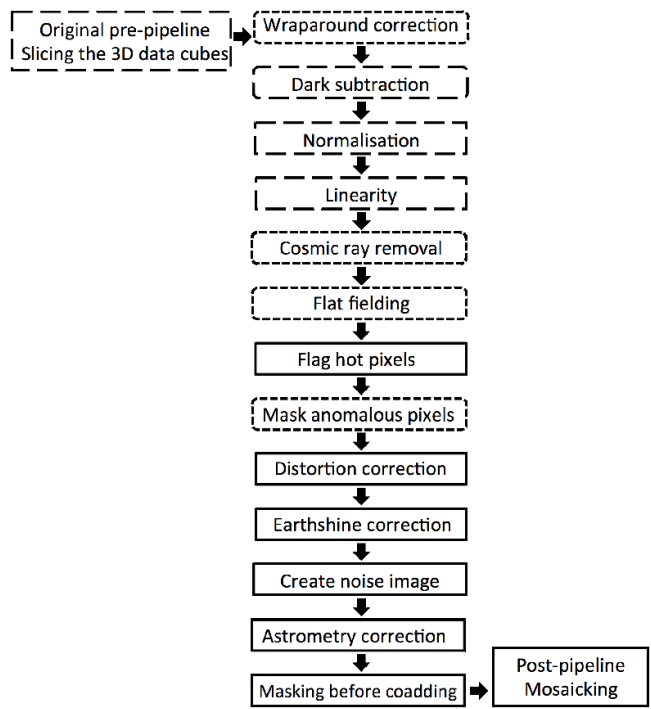}
\caption{Outline of the steps in the optimised toolkit. The dashed line boxes are steps copied directly from the original archival pipeline. The dotted line boxes are based on steps from the original archival pipeline, with some changes, e.g. a more comprehensive masking in the wraparound correction, cosmic ray and mask anomalous pixels steps, and incorporating a time dependent flat field. The solid line boxes are steps created for the optimised toolkit.}
\label{fig:Outline_of_IDL_pipeline_steps}
\end{figure}

The raw {\it AKARI} data is accessed as individual pointings from the archive, consisting of 3D data cubes. The raw 3D data cubes are sliced into 2D frames using the original archival pre-pipeline. The first step of the optimised toolkit creates a structure to hold the: frame, associated header, noise array, noise header, mask array and  mask header. Neither archival pipelines create a mask or noise image from each frame. In order to ensure that the optimised toolkit is efficient and to avoid excessive reading and writing to/from storage, all frames are passed as IDL structures between processing steps. The subsequent steps of the optimised toolkit are shown in Figure \ref{fig:Outline_of_IDL_pipeline_steps}.

It should be noted that the dark current subtraction, the normalisation, linearity correction, anomalous pixels and flat fielding steps of the optimised toolkit, see dashed boxes in Figure \ref{fig:Outline_of_IDL_pipeline_steps}, replicate the same steps in the original archival pipeline. The new archival pipeline uses the dark subtraction method from \cite{Tsumura2011}, using time dependent dark frames for each pointing. This optimised toolkit also uses time dependant dark frames, selected from suitable pointings over the entire Phase 2 period. The normalisation steps correct for data compression and Fowler sampling. The linearity step corrects for a non-linear relationship between the number of electrons and Analogue to Digital Unit (ADU) \citep{Lorente2008}.

The optimised toolkit modules are described in turn below.

\subsection{Wraparound Correction}
\label{wraparound_correction}
\begin{figure}
\begin{tabular}{cc}
  \includegraphics[width=65mm]{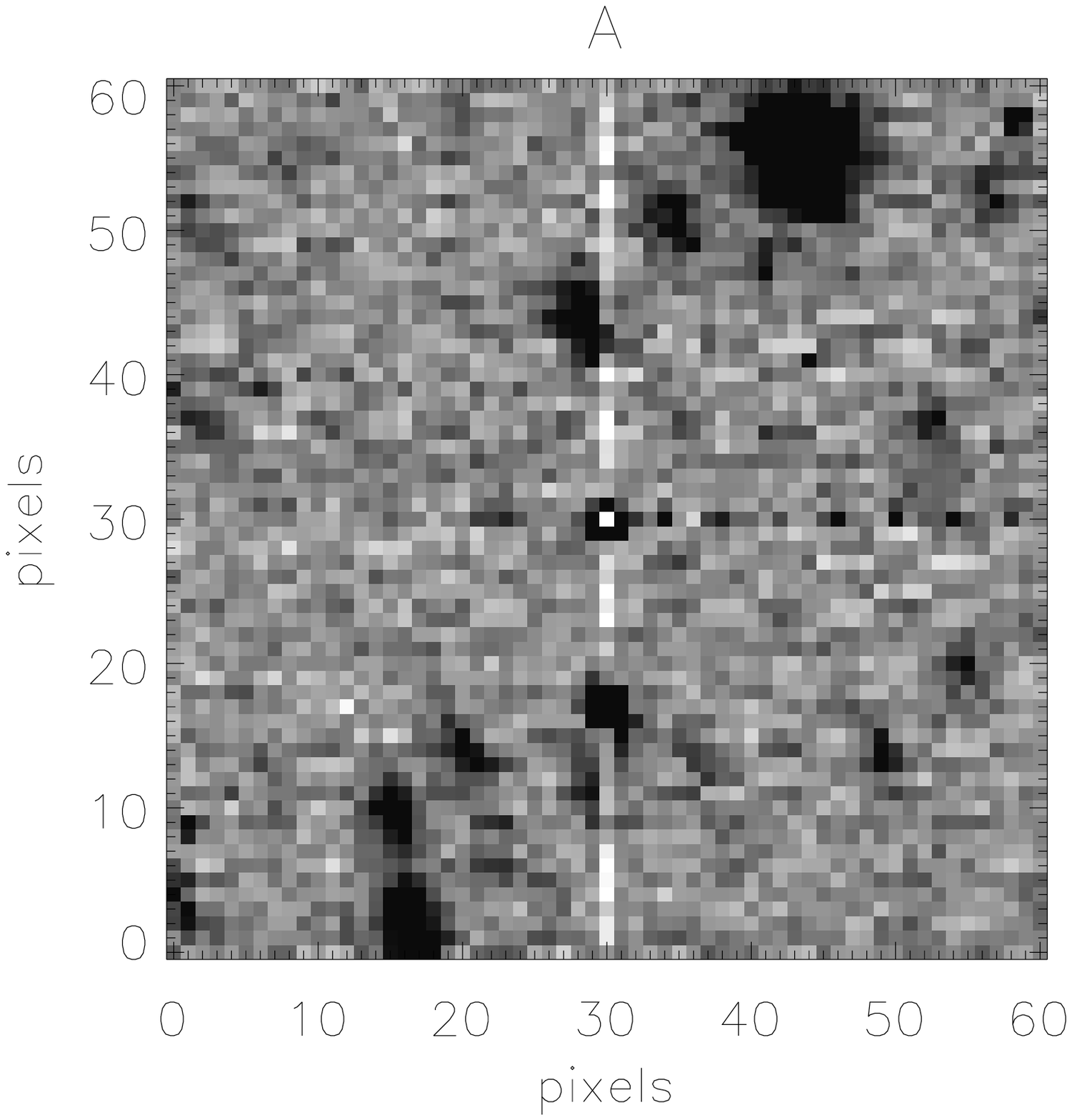}\\
  \includegraphics[width=65mm]{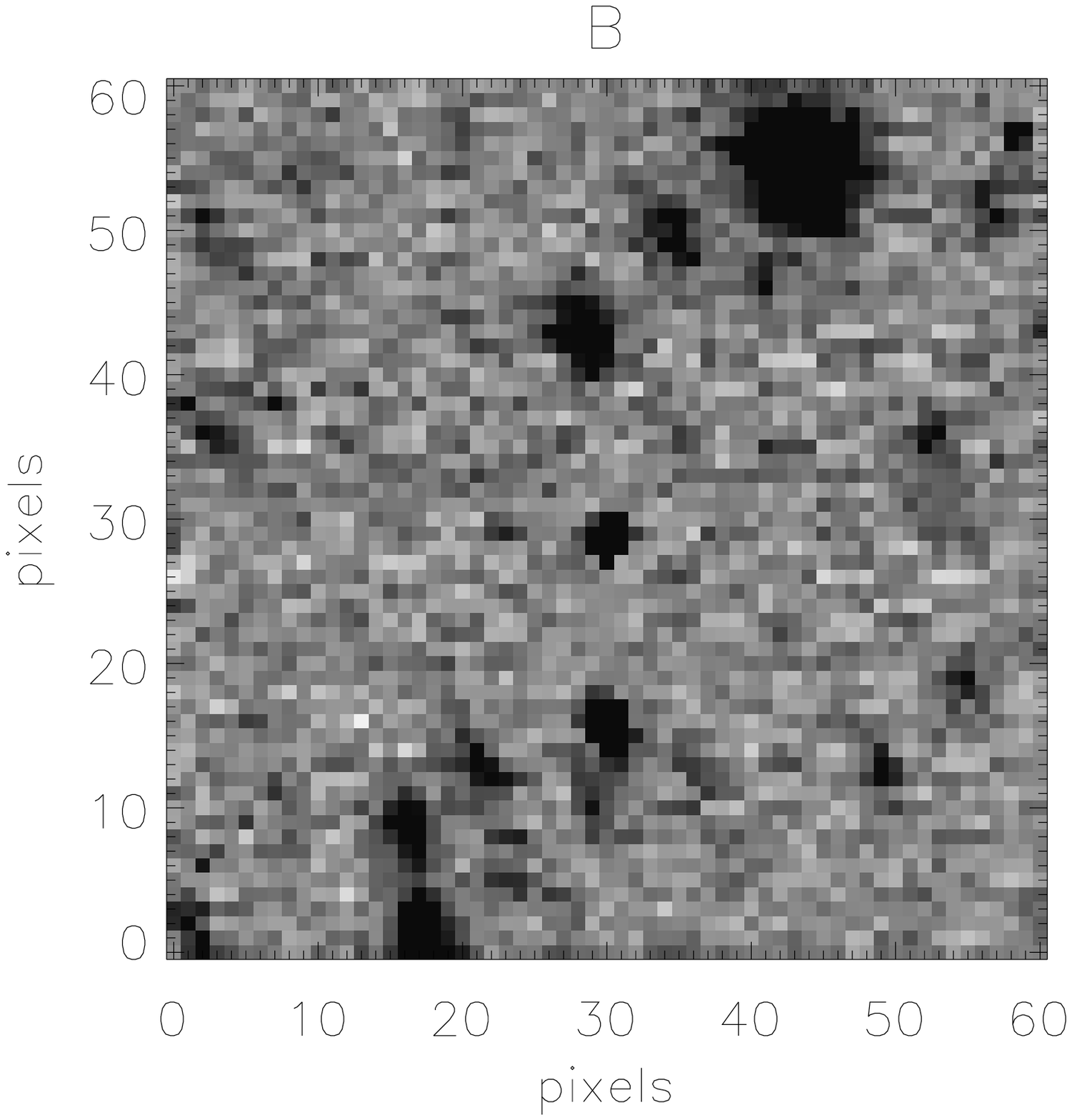}\\
\end{tabular}
\caption{A close up of a single N3 frame; (A) shows the raw frame, (B) shows the processed frame with muxbleed removed. The saturated pixel causing the artefact is at pixel number (30, 30). The masked pixels have been assigned an interpolated value.}
\label{fig:pixels_affected_by_wraparound}
\end{figure}

Due to telemetry constraints on the data size for downlink, data was compressed on board; all pixels with a flux greater than $2^{16}$ ADUs are `wrapped around' to a pixel value less than $-$\,11953.8\,ADU \citep{Lorente2008}. 
Figure \ref{fig:pixels_affected_by_wraparound} shows examples of artefacts created by a bright (many times brighter than the detector full well) object (star, cosmic ray etc.) viewed by the detector. There are two clear effects: every fourth pixel after the pixel which the bright object is incident on, shows a brighter value; and the column in which the bright object pixel is situated also shows an increase in signal. The first effect, termed {\it muxbleed}, is understood to be due to the extreme signal charge not being completely reset from the output chain, resulting in an offset applied to subsequent reads from that output (of which there are 4 between those pixels that are multiplexed) for some time until the offending signal has decayed in the electronics. The second effect, often referred to as {\it column pull-down}, is likely to be a blooming effect during the integration time. Blooming is when a large number of photons (or signal from a very high energy photon) fills the full well capacity of the pixel it is incident on. The electron signal then blooms to the next pixel and so on, until all the electrons have been captured in potential wells.

{\it Column pull-down} and {\it muxbleed} are not corrected for in either the original or new archival pipelines. This optimised toolkit masks the affected pixels, similar to \cite{Murata2013}. In the paper of \cite{Murata2013}, all columns with {\it Column pull-down} and rows with {\it muxbleed} are masked. All columns affected by {\it Column pull-down} are masked, but only every fourth pixel in rows affected by {\it muxbleed} are masked. This is performed by the optimised toolkit by first checking each NIR frame for pixels with a value less than $-$\,11953.8\,ADU. The location of the pixel is flagged in the mask image, as are all the pixels in the same column (affected by the {\it column pull-down} effect) and every fourth pixel for the following two rows (due to the {\it muxbleed} effect).

\begin{figure}
 \centering
  \includegraphics [angle=90] {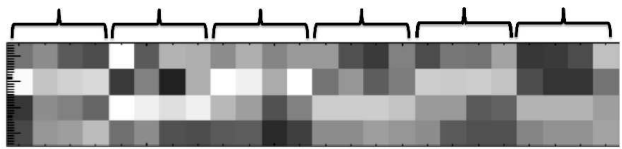}
 \caption{Close up of a NIR frame, showing the sets of four pixels and the increasing and decreasing flux.}
 \label{fig:four_pixels_thing}
\end{figure}

During the work of creating the optimised toolkit, a new artefact in many of the NIR images was discovered. The artefact is a pattern, with sets of four pixels with increased flux and decreased flux, as shown in Figure \ref{fig:four_pixels_thing}. This is not mentioned in the archival pipelines or previous toolkits. As the pattern is in sets of four, it is presumed to be linked to reset changing of the four read-out nodes. A limitation of the optimised toolkit is that it does not remove this artefact. Further work is required to research how to remove this new artefact.

\subsection{Cosmic Ray Removal}
\label{cosmic_ray_removal}
Frames from all of the nine filters contain cosmic rays. The cosmic ray detection was found not to work well in the original archival pipeline, and is not mentioned in the new archival pipeline. The optimised toolkit detects cosmic rays in two places, in the cosmic ray detection step (i.e. in individual frames) and additionally during the later co-adding stage. 
Quite often, the first NIR frame in a pointing contains many more cosmic rays than subsequent frames. This is due to the detector not being read out between pointings causing build up of cosmic ray signals. The IDL procedure LA\_COSMIC \citep{Dokkum2001}, an algorithm to detect cosmic rays using Laplacian edge detection, is used to remove the cosmic rays in the individual NIR frames.

In the MIR-S and -L bands the cosmic rays appear very similar to faint point sources and LA\_COSMIC was found to remove some of the point sources as well. In order to preserve the flux from the point sources, in the cosmic ray removal step for the MIR-S and L bands, a simple source extraction algorithm was run on the individual frames to detect connected pixels. The threshold was set low, so as to mask as many point sources as possible. Each frame with the connected pixels masked, was then put through a sigma clipping algorithm to detect spuriously bright single pixels. The algorithm was written for the optimised toolkit. A different sigma level was used for each detector. Before the implementation of the masking of connected pixels, flux was lost from point sources. The masking of the connected pixels minimises the loss of flux. The locations of detected cosmic rays are flagged in the array mask. A second cosmic ray detection is performed during the frame co-adding stage. This is mainly performed to detect any remaining cosmic rays that were incorrectly masked as point sources. The same sigma clipping algorithm is performed in the coadding stage.

\subsection{Flat fielding}
\label{flat_fielding}
\begin{figure}
\begin{tabular}{c}
  \includegraphics[width=65mm]{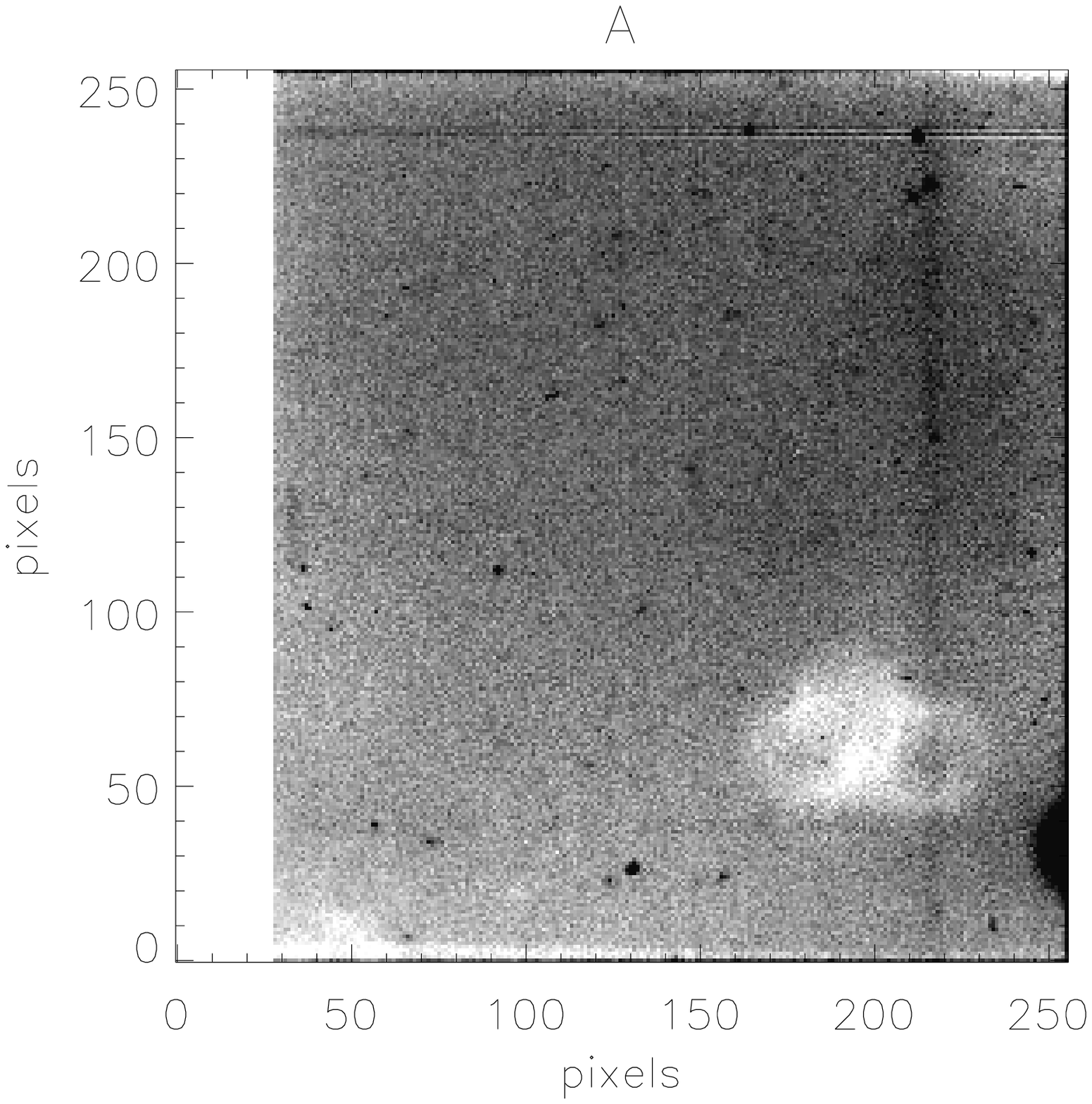}\\
  \includegraphics[width=65mm]{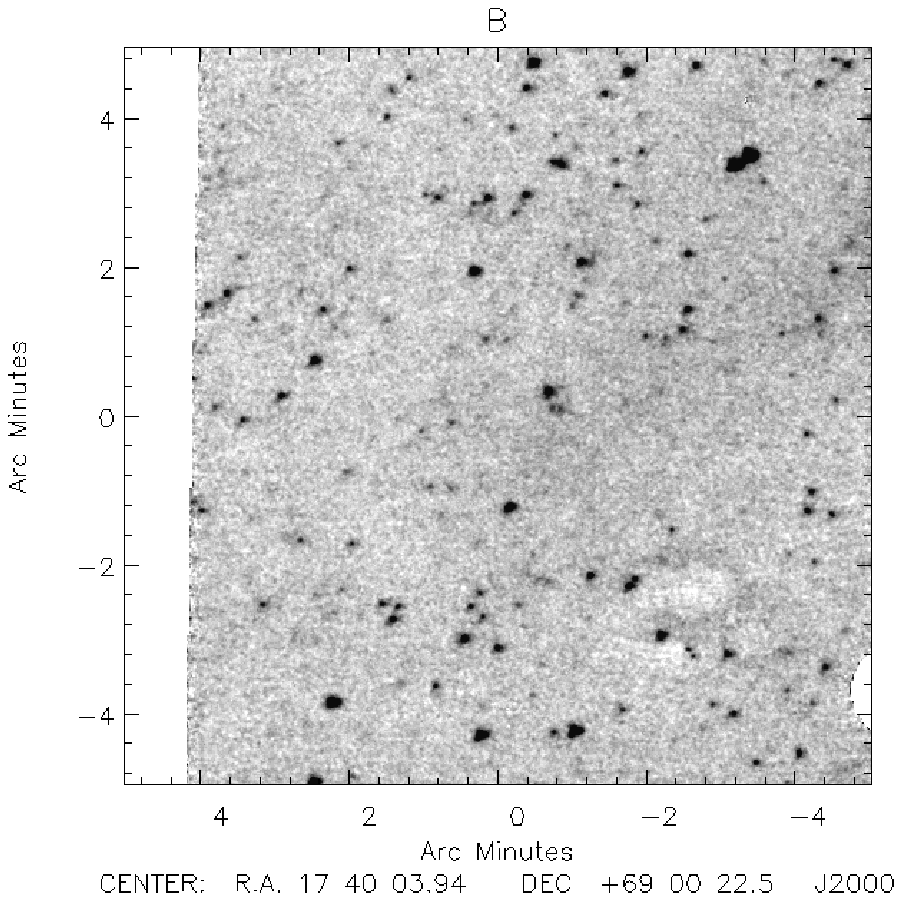}\\
  \includegraphics[width=65mm]{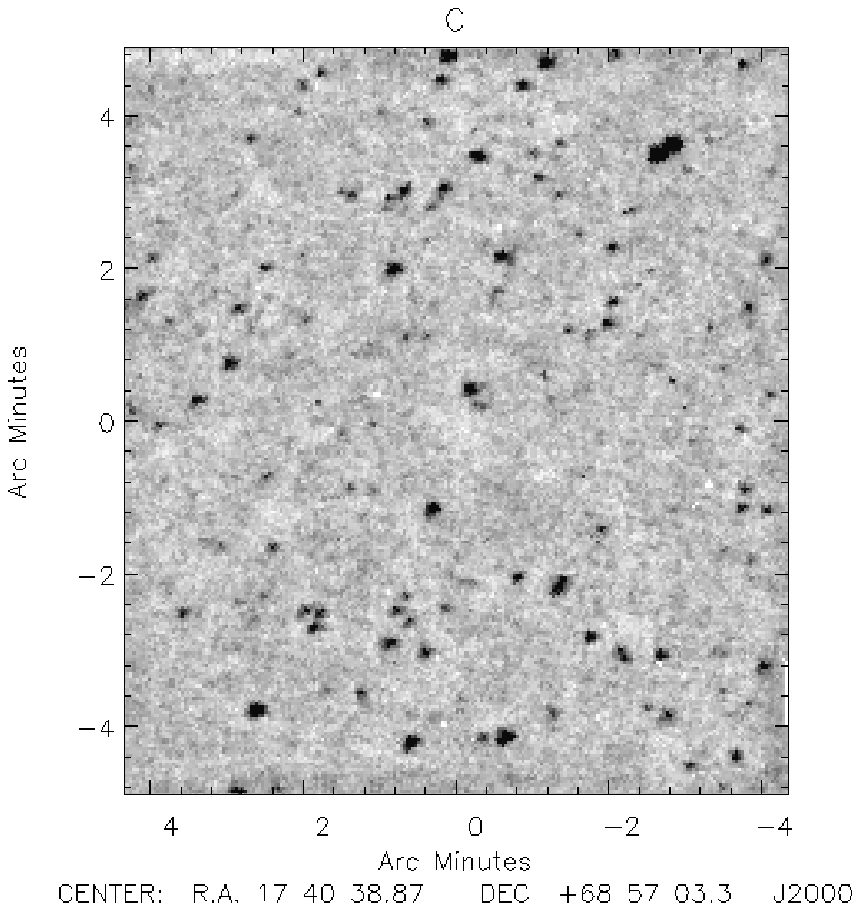}\\
\end{tabular}
\caption{An example of the improvement provided by the optimised toolkit for the bean removal, following the method used in \citealt{Murata2013}. For reference (A) shows a raw single frame note the sky bean is the white patch in the bottom right. (B) shows a single pointing processed by the new archival pipeline \citep{Egusa2016}, note that the bean has not been fully removed. (C) shows the same pointing processed by the optimised toolkit, the sky bean has been fully removed. This image has been cropped, so as not to contain the slit area.}
\label{fig:bean} 
\end{figure}

The optimised toolkit creates a bespoke flat, using frames time-stamped with a similar date to the observation. To remove as many instrumental artefacts as possible, we have found it advisable to create time dependent flats for each date range of a set of pointings. One prominent artefact removed by such a bespoke flat field is the so-called {\it sora-mame} (Japanese for sky-bean due to the artefact's shape, hereafter called `the bean'). The bean appears in MIR-S detector images from the beginning of the mission to part way through Phase 2, where it disappeared on 07/01/2007. The bean is a pattern of three almost circular shapes (see Figure \ref{fig:bean}.a). The shape of the bean is time dependant, and there is currently no generic model to remove it. The time dependent flats created for use with the new archival pipeline were found not to remove the bean in the frames processed in Section \ref{results}. Following the method outlined in \cite{Murata2013}, a very time dependent bespoke flat (of order of a couple of days) was found to remove the bean. Comparison between the use of the generic flat field and a bespoke time dependant flat field can be seen in Figure \ref{fig:bean}.

\subsection{Flag Hot Pixels}
\label{amonalous_pixel_detection}
Analysis of images from the IRC shows that there is an increase in the number of hot/bright pixels over the duration of the mission. These hot pixels are visible in frames as having a high dark current. This is likely to have been caused by displacement damage from impacting protons. This damage is ever-more evident in later-Phase 2 MIR-S and MIR-L images. 
In addition, in the MIR-S detector images there is a spread of hotter pixels in the lower right hand side of the frame. This is the part of the image nearest to the amplifier, which increases dark current in surrounding pixels. This is worsened by the warming up of the telescope, which gets worse over the mission. This area needs to be masked in later-Phase 2 MIR-S images. Both of these artefacts are evident in the images and a bespoke time dependent flat is unable to correct for them.  

\begin{figure}
 \centering
   \includegraphics[width=40mm]{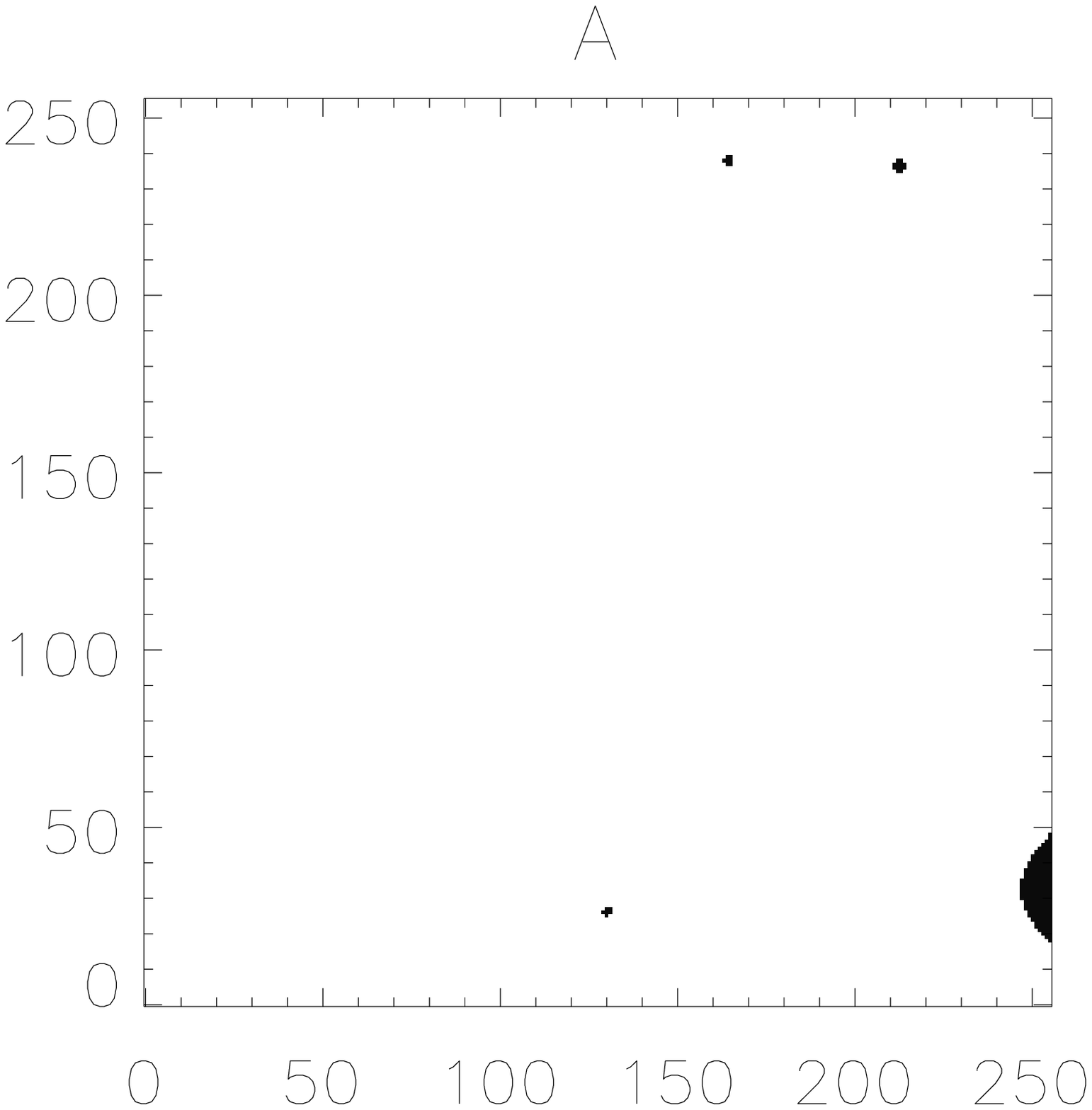}
  \includegraphics[width=40mm]{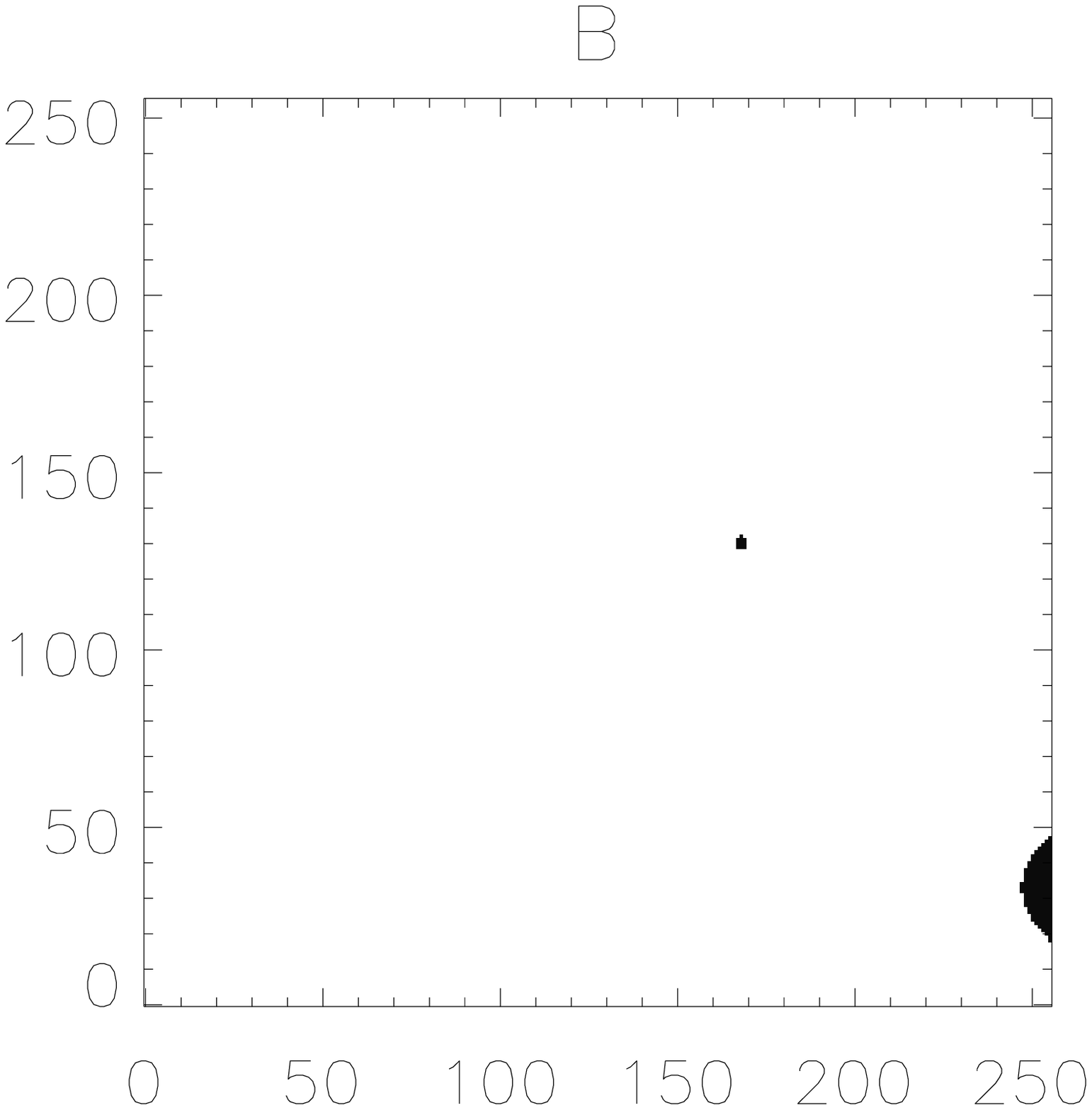} \\
  \includegraphics[width=40mm]{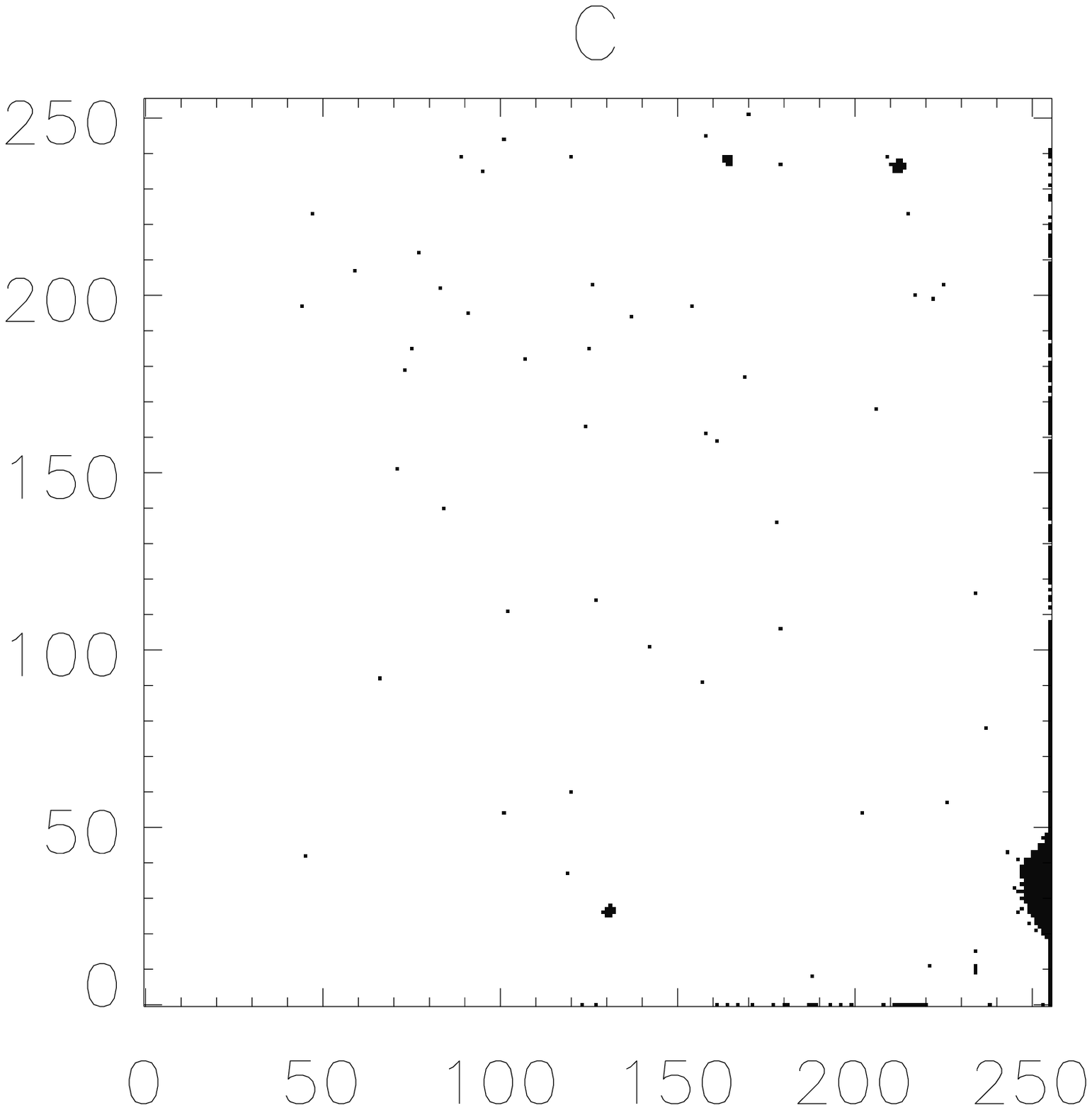}
  \includegraphics[width=40mm]{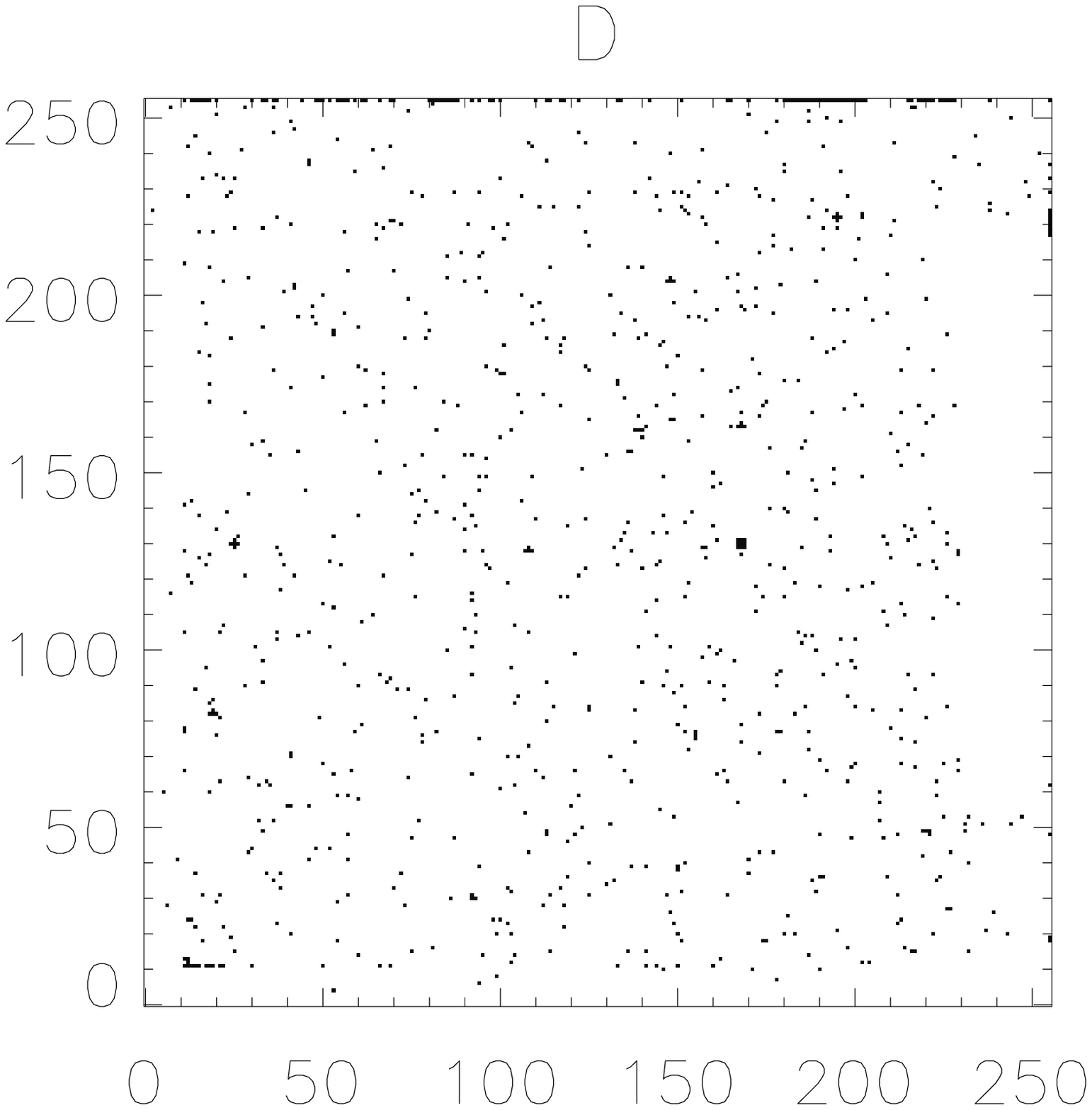} \\
 \caption{Hot pixel masks, (A) shows the original MIR-S hot pixel mask, (B) shows the original MIR-L hot pixel mask. (C) shows the optimised toolkit hot pixel mask for MIR-S and (D) shows the optimised toolkit hot pixel mask for MIR-L.}
 \label{fig:hot_pixel_mask}
\end{figure}

\begin{figure}
\begin{tabular}{cc}
  \includegraphics[width=65mm]{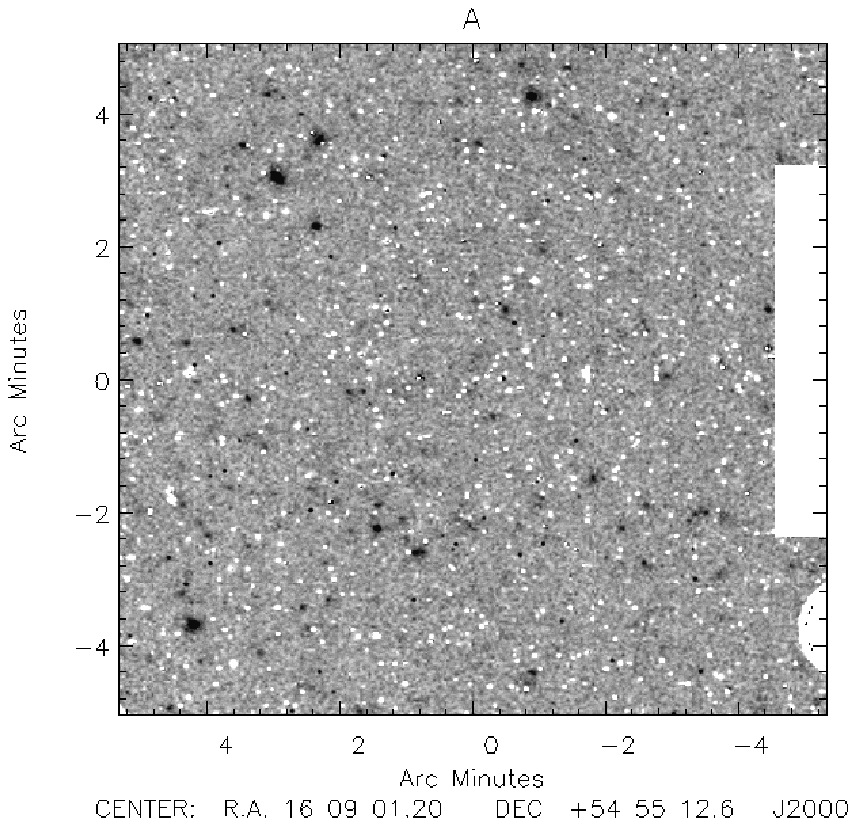}\\
  \includegraphics[width=65mm]{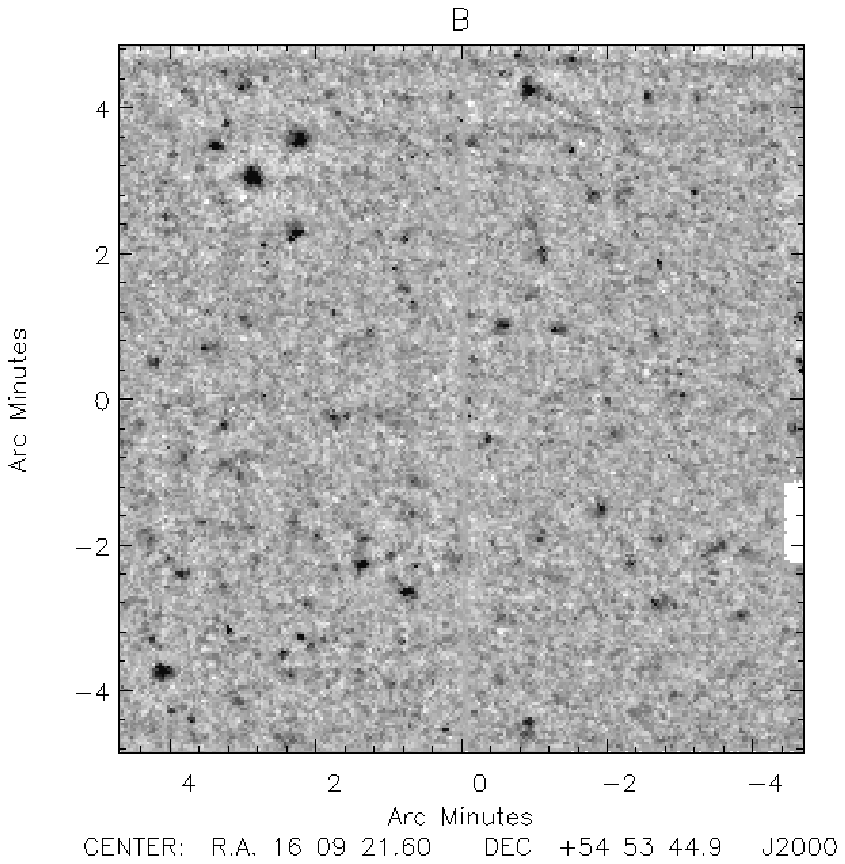}\\
\end{tabular}
\caption{An example of the improvement provided by the optimised toolkit for masking hot pixels. Hot pixels are most evident in late-Phase 2 frames, due to accumulation of radiation damage. (A) shows a single pointing processed by the new archival pipeline \citep{Egusa2016}, note the white pixels, which are the hot pixels. (B) shows the same pointing processed by the optimised toolkit. This image has been cropped, so as not to contain the slit area.}
\label{fig:hot_pixel_comparison} 
\end{figure}

The original hot pixel masks provided by the {\it AKARI}/IRC instrument team (shown in Figure \ref{fig:hot_pixel_mask}.a and Figure \ref{fig:hot_pixel_mask}.b) for use with the original archival pipeline were found not to mask all the hot pixels. Also the method of hot pixel detection used by the new archival pipeline was found not to work on the pointings processed in Section \ref{results}. For this toolkit a different method was used. The optimised toolkit uses a new template to flag the location of hot pixels in the mask array associated with each image. These new templates to mask the locations of hot pixels in in MIR-S and MIR-L frames are shown in Figure \ref{fig:hot_pixel_mask}.c and Figure \ref{fig:hot_pixel_mask}.d. 
The hot pixels cannot be removed by a time dependent flat, due to the fact that the background flux of the image varies considerably during a single pointing (see Section \ref{change_of_background_flux}). 
The new hot pixel mask was created from late-Phase 2 {\it AKARI} observations. Only the first two thirds of the frames from each pointing were used. The NIR detector does not appear to have the same later-Phase 2 artefacts. Figure \ref{fig:hot_pixel_comparison} shows the improvement the optimised toolkit does in removing the hot pixels. The figure gives a comparison between a co-added pointing processed by the new archival pipeline suffering from hot pixels, and the same pointing co-added using the optimised pipeline.

\subsection{Mask Anomalous Pixels}
\label{mask_image}
In the original archival pipeline, image pixels identified as `bad' are set to $-$\,9999.90\,ADU, whereas the optimised toolkit flags the locations of bad pixels in the associated mask array. The steps in the optimised toolkit, which flag bad pixels are: wraparound correction, cosmic ray removal and mask anomalous pixels. Both archival pipelines also masks the slit area, whereas the optimised toolkit does not, as it contains information about the dark current. The slit area is masked during the co-adding stage.

\subsection{Distortion Correction}
\label{Distortion_correction}
The raw AKARI data suffer from an image distortion. This distortion depends on the detector (NIR, MIR-S or MIR-L, and to a lesser degree on the filter, see Table \ref{table:IRC_detectors}). Neither archival pipelines discuss the image distortion, and do not correct for it. This distortion is corrected for in the optimised toolkit. To correct for this distortion, true sky positions are obtained using the {\it 2MASS} catalogue \citep{Skrutskie2006} for NIR detector frames and the {\it WISE} catalogue \citep{Wright2010} for the MIR-S and MIR-L frames. Using these {\it AKARI}/ancillary point source pairs, a $\chi^2$ multi-parameter fitting program \citep{Markwardt2009} was used to simultaneously fit the distortion polynomial using the astrometry. The distortion correction and the astrometry were then iterated until the $\chi^2$ of the fit of the difference between the {\it AKARI} positions and {\it 2MASS}/{\it WISE} positions were at a minimum. Using this method, a second order polynomial distortion correction was created for each filter;

 \begin{equation}
X'=\sum_{i=0}^{N}\sum_{j=0}^{n}P_{ij}x^{j}y^{i}
\label{x_polynomial}
\end{equation}
 \begin{equation}
Y'=\sum_{i=0}^{N}\sum_{j=0}^{n}Q_{ij}x^{j}y^{i}
\label{y_polynomial}
\end{equation}
 
 The second order polynomial distortion correction for the $x$-axis and $y$-axis are given respectively by Equation \ref{x_polynomial} and Equation \ref{y_polynomial}, where $x$ is the original distorted $x$ pixel position, $y$ is the original distorted $y$ pixel position, $X'$ is the undistorted $x$ pixel position, $Y'$ is the undistorted $y$ pixel position, $P_{ij}$ is the $x$ matrix transformation, $Q_{ij}$ is the $y$ matrix transformation, $N$ is the order of polynomial and $n$ is the square root of the number of elements of the matrix.

\begin{figure*}
\begin{tabular}{ccc}
  \includegraphics[width=50mm]{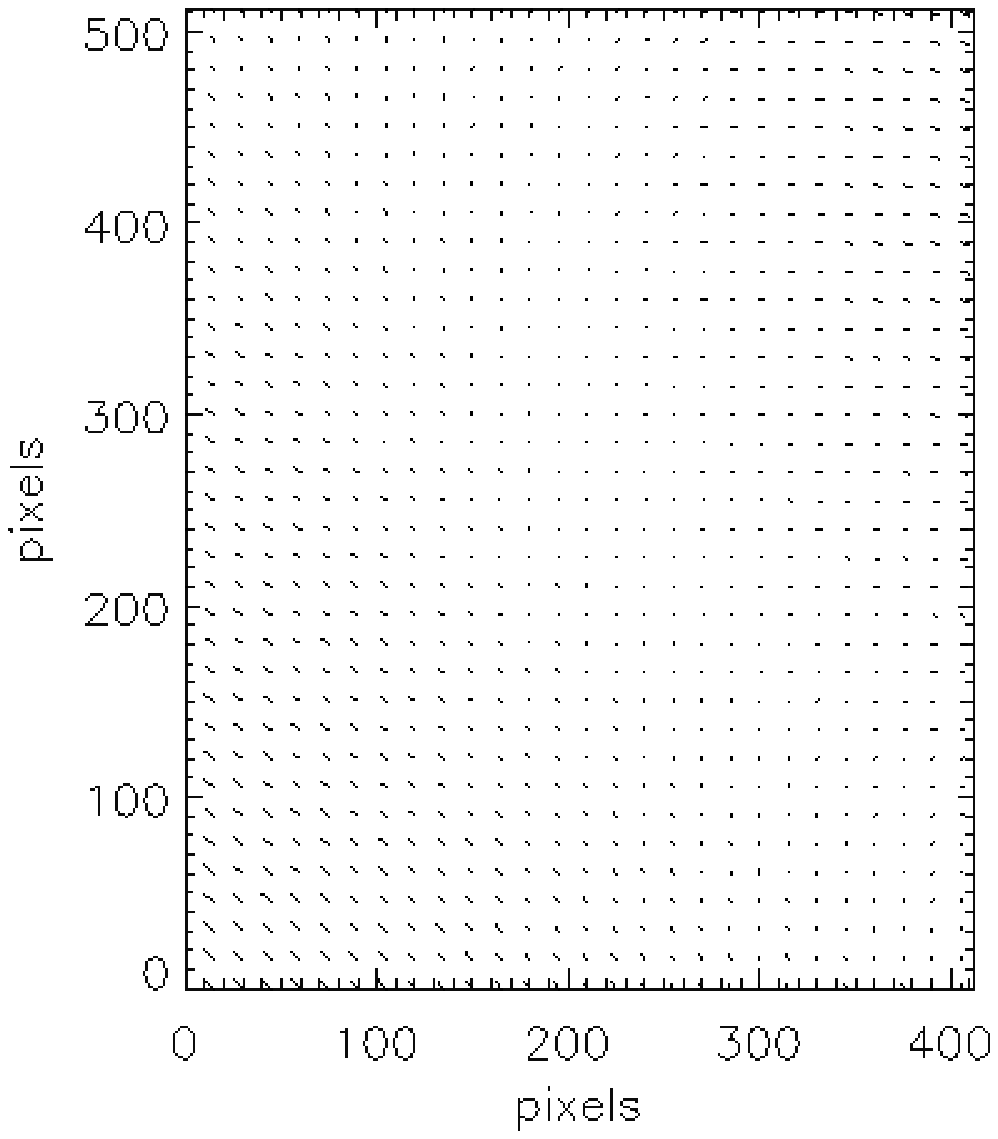} & 
   \includegraphics[width=50mm]{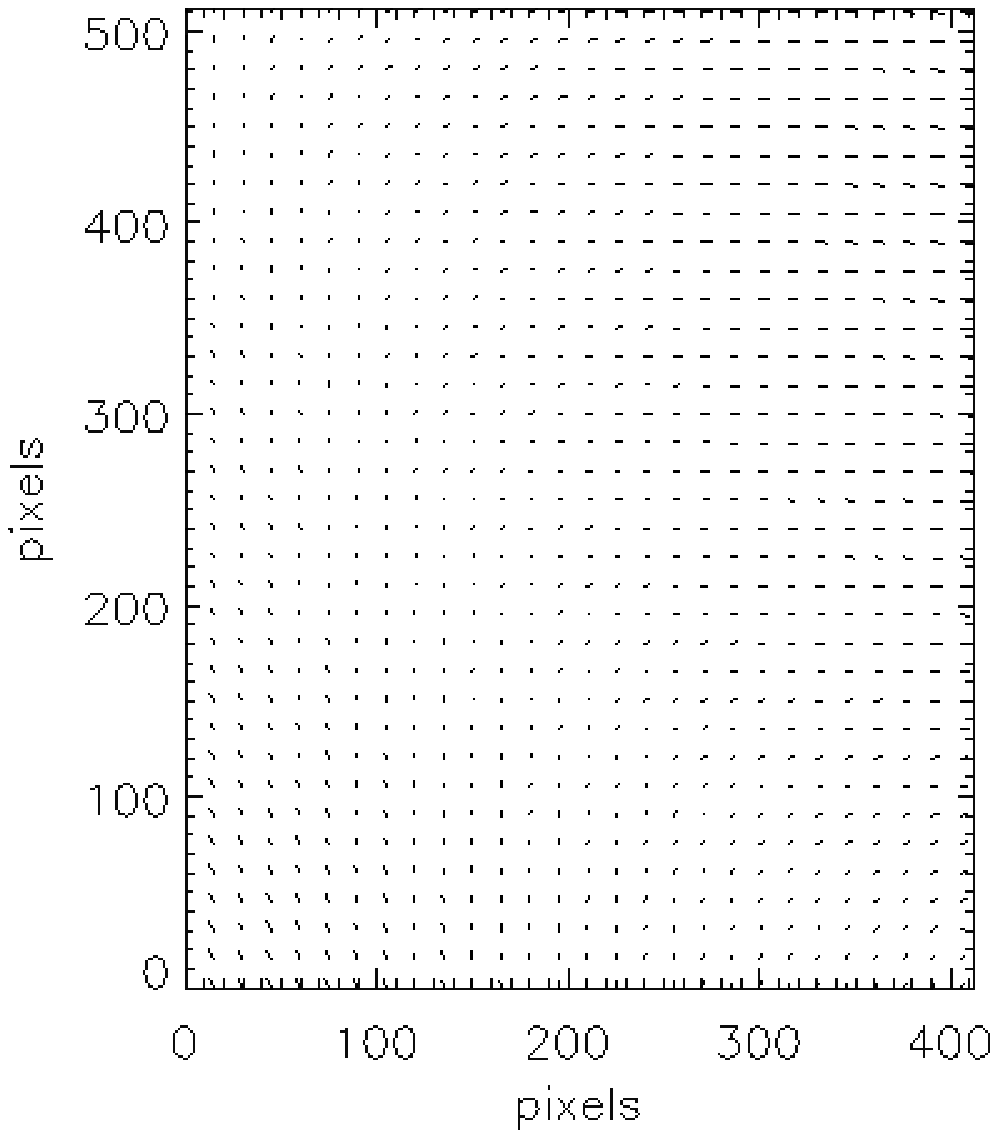} & 
  \includegraphics[width=50mm]{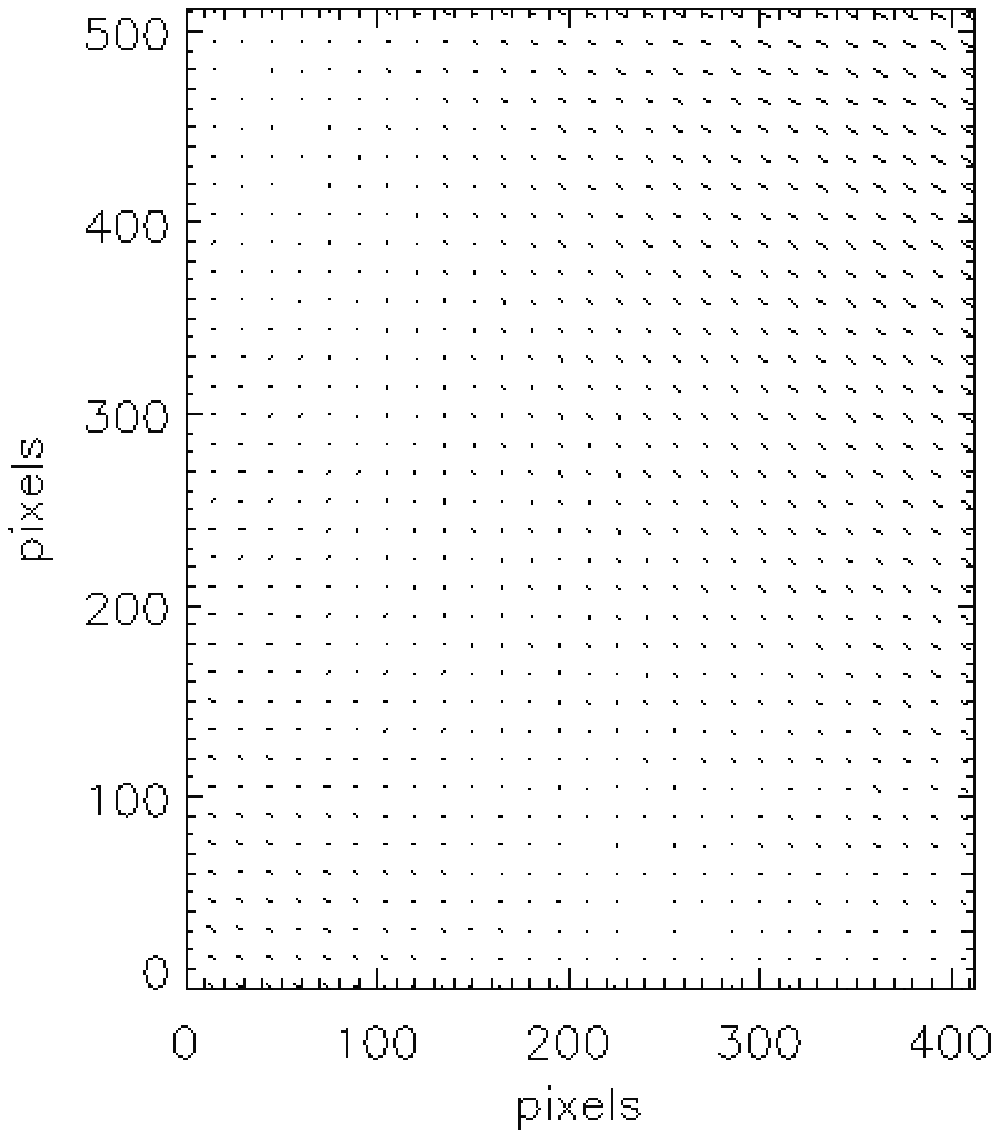} \\
  (a) N2 & (b) N3 & (c) N4 \\[6pt]
  \includegraphics[width=50mm]{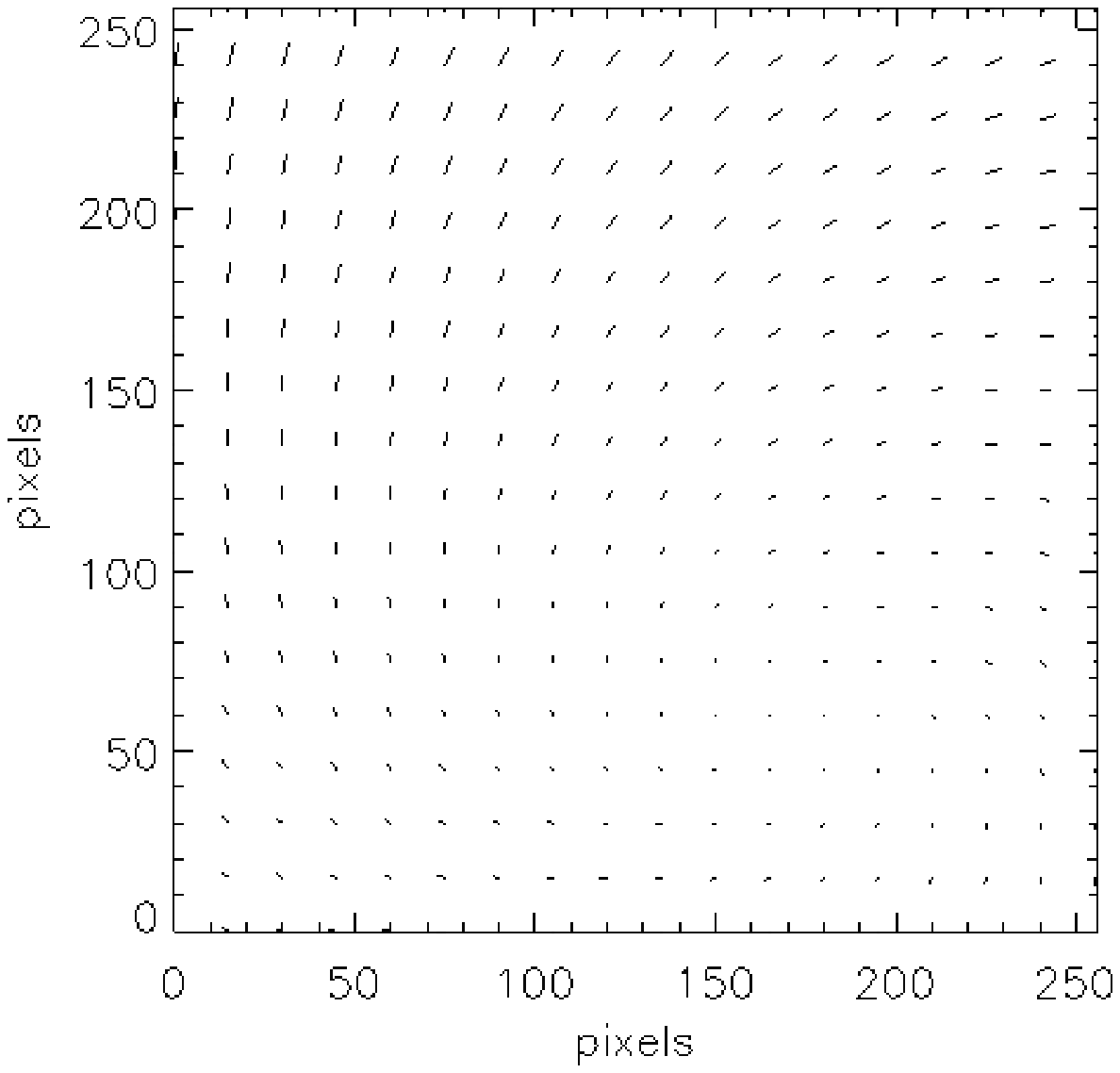} & 
  \includegraphics[width=50mm]{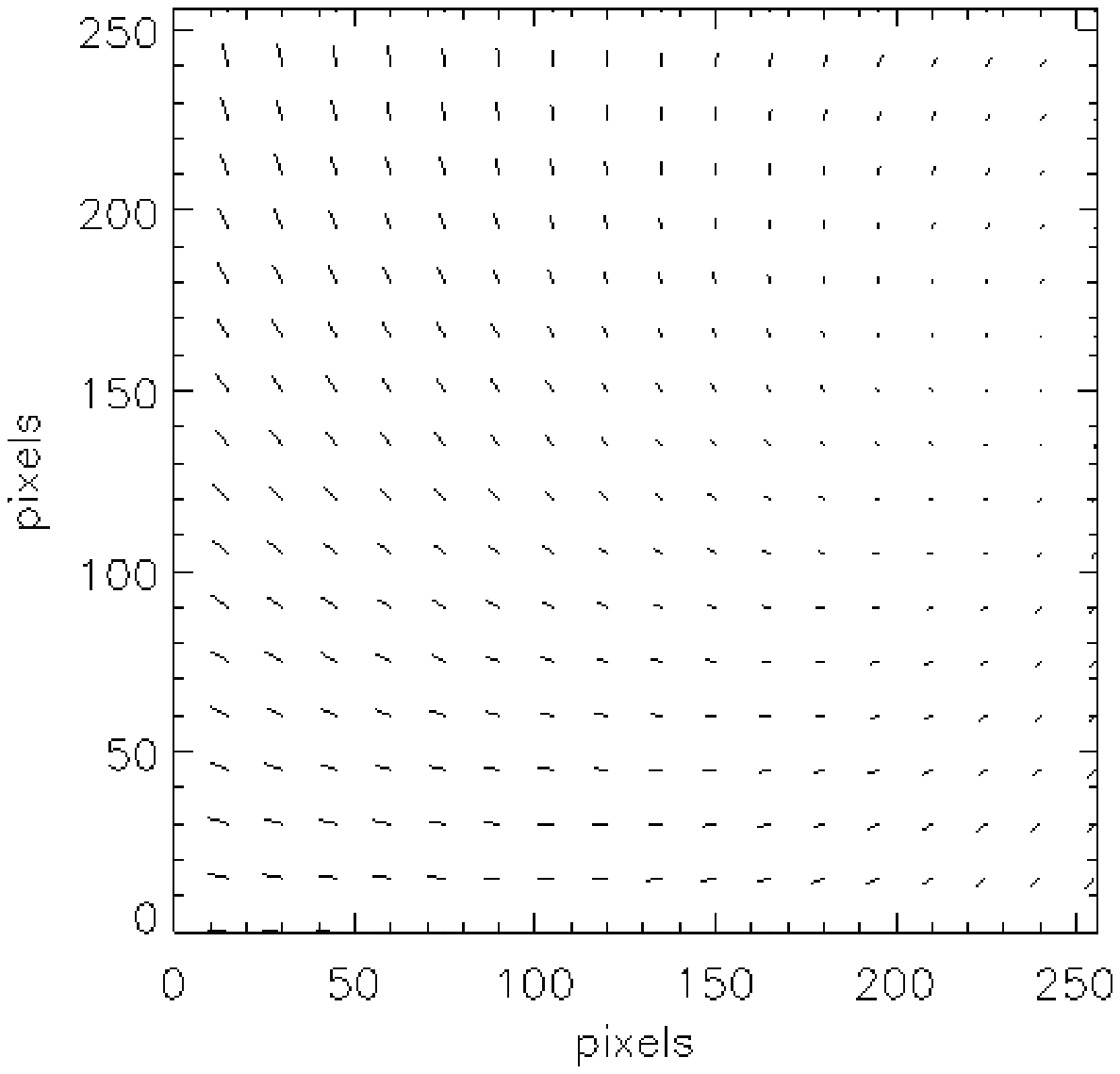} & 
  \includegraphics[width=50mm]{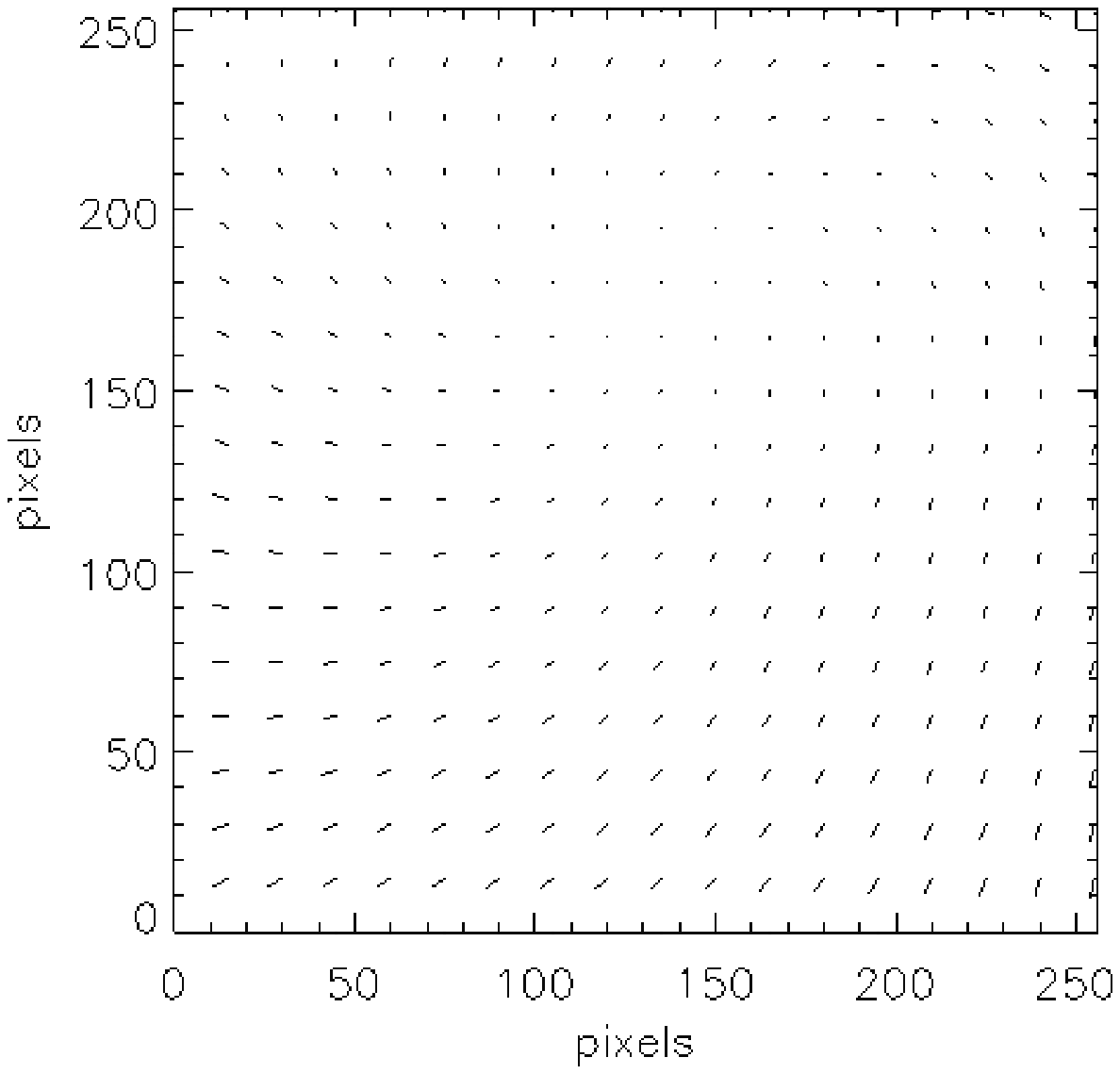} \\
  (d) S7 & (e) S9W & (f) S11 \\[6pt]
    \includegraphics[width=50mm]{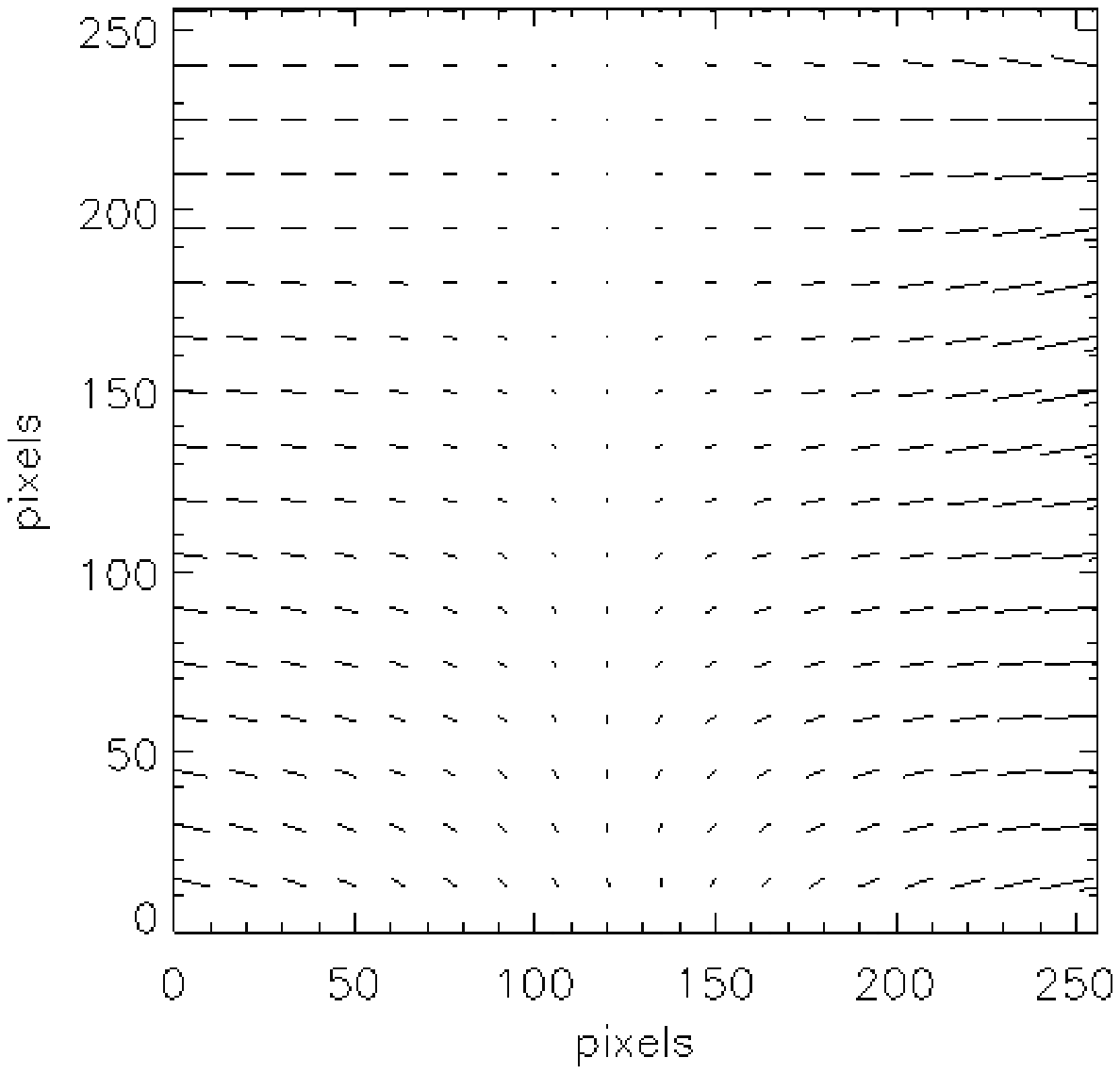} & 
  \includegraphics[width=50mm]{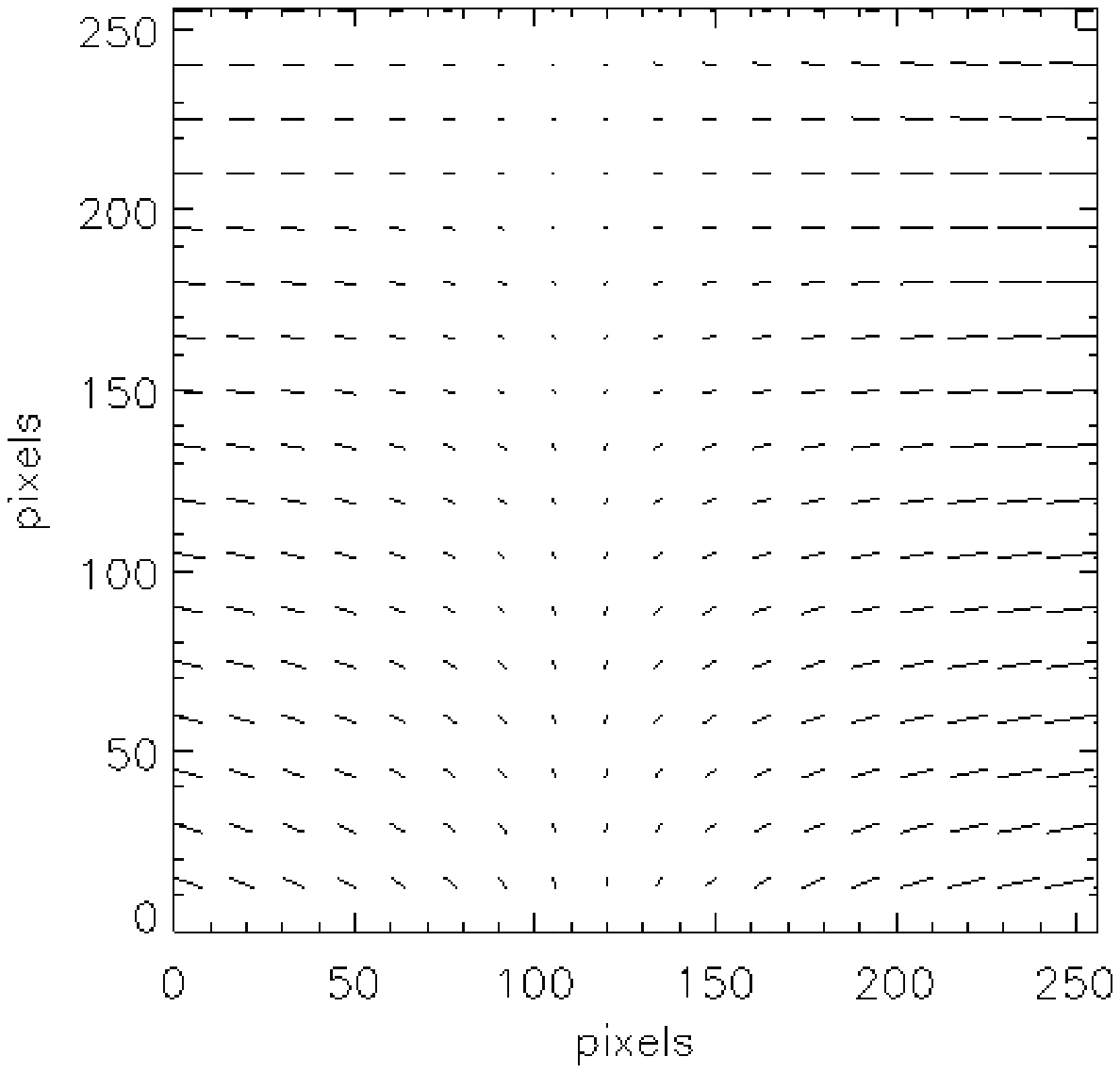} & 
  \includegraphics[width=50mm]{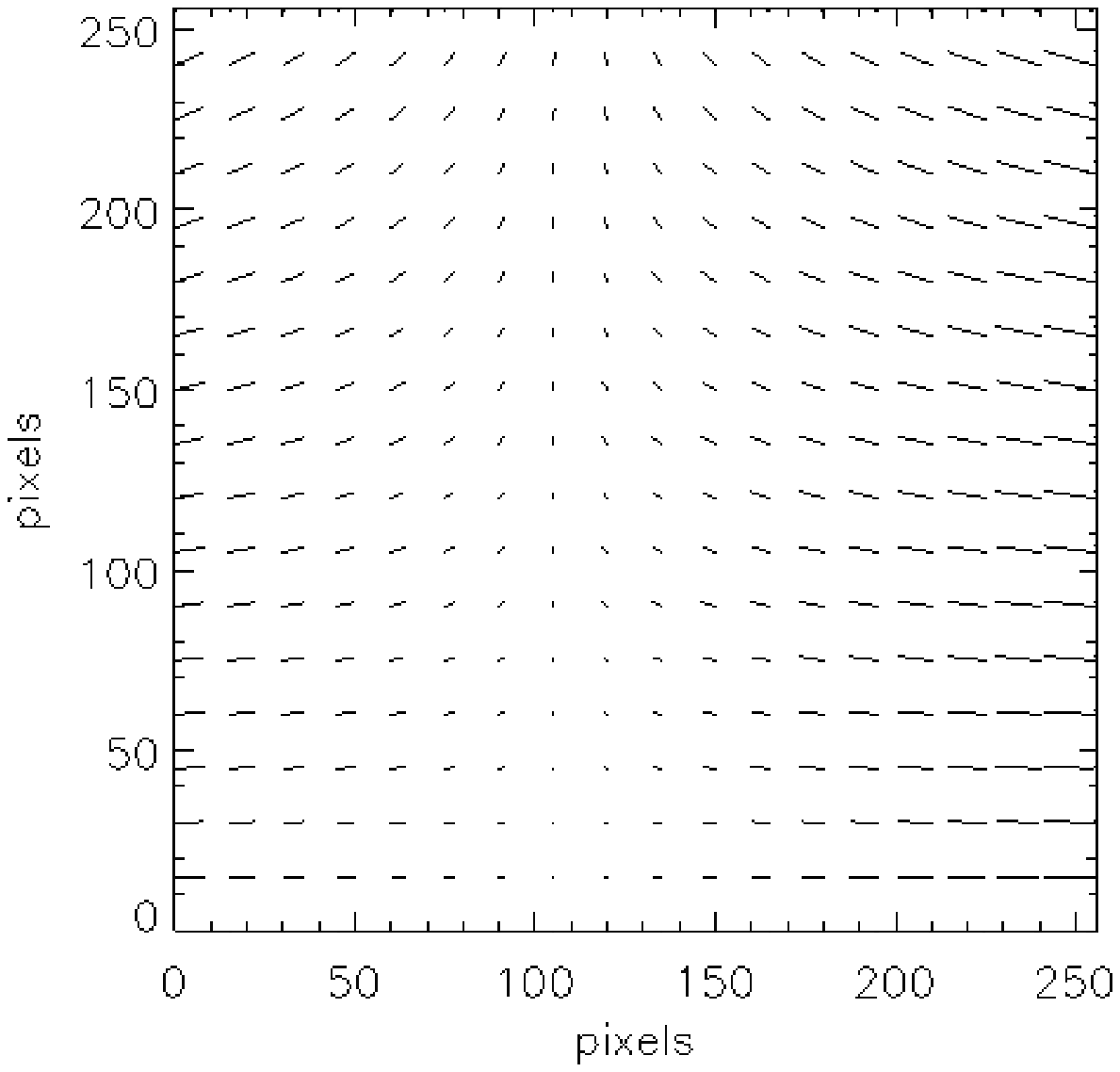} \\
  (g) L15 & (h) L18W & (i) L24 \\[6pt]
\end{tabular}
\caption{Distortion correction polynomials for the nine filters calculated using Equation \ref{x_polynomial} and Equation \ref{y_polynomial} and the coefficients in Appendix~\ref{appendixA}.}
\label{fig:distortion_correction_polynomials} 
\end{figure*}

The vector plot for the distortion correction for each of the nine filters are shown in Figure \ref{fig:distortion_correction_polynomials} and the distortion polynomial coefficients themselves are listed in the Appendix~\ref{appendixA}. The distortion correction polynomials created for the optimised toolkit, used in the processing of the three extragalactic fields discussed in Section \ref{results} have all been found to be accurate to within the Nyquist scale i.e. less than half of the Full Width Half Maximum (FWHM). Table \ref{tab:pixel_error} shows the pixel error for the IRC filters in Section \ref{results}.

The above method assumes that for each filter the distortion is not time dependant during Phase 2 of the {\it AKARI} mission. The distortion correction was tested on frames from November 2006 (early-Phase 2), February 2007 (mid-Phase 2) and July 2007 (late-Phase 2), no time dependancy was found. Approximately 20 frames of different extragalactic deep fields were used to create the NIR distortion polynomials. Individual frames of extragalactic deep fields could not be used to create the MIR-S and MIR-L distortion correction polynomials. There are two main reasons for this: firstly, both raw and processed individual frames do not show many point sources(see Figure \ref{fig:bean}.a for an example of an individual MIR frame); secondly, in individual frames, point source extraction often incorrectly labels hot pixels as point sources, which would be removed in the coadding stage. For the $\chi^2$ multi-parameter fitting to work correctly, each individual frame was required to have $20+$ galaxies. Therefore frames of Galactic targets were used instead for MIR-S and MIR-L images. The distortion correction also automatically corrects the aspect ratio.
 
\subsection{Earthshine Correction}
\label{scattered_light}
\subsubsection{Earthshine Artefact}
\label{earthshine_light}
\begin{figure}
 \centering
 \includegraphics{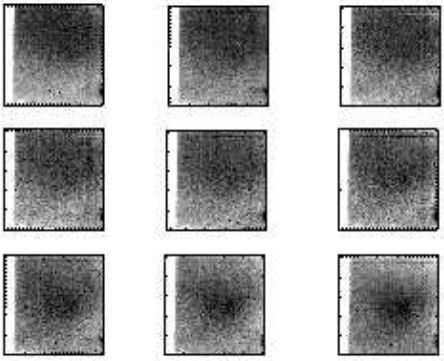}
 \caption{A sample of 9 raw frames from one S11 30 frame pointing from later-Phase 2 data, Note how the Earthshine artefact `moves' over the image from frame to frame.}
\label{fig:earthshine_30_frames}
\end{figure}

\begin{figure}
\begin{tabular}{cc}
  \includegraphics[width=65mm]{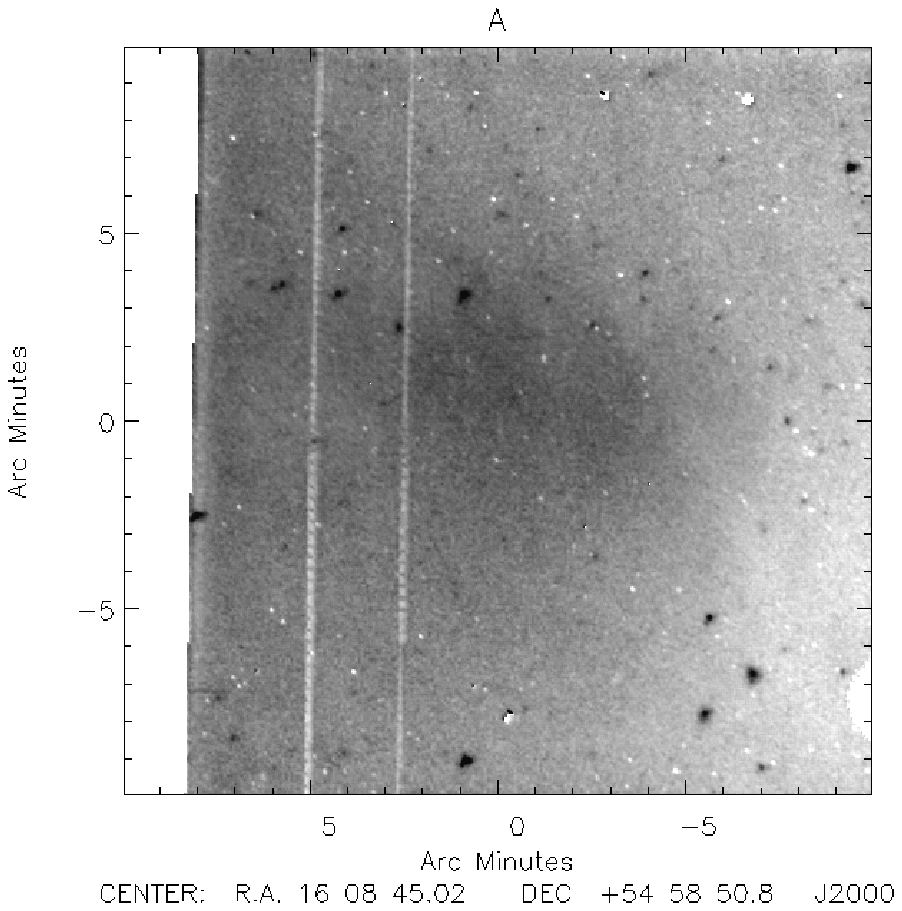}\\
  \includegraphics[width=65mm]{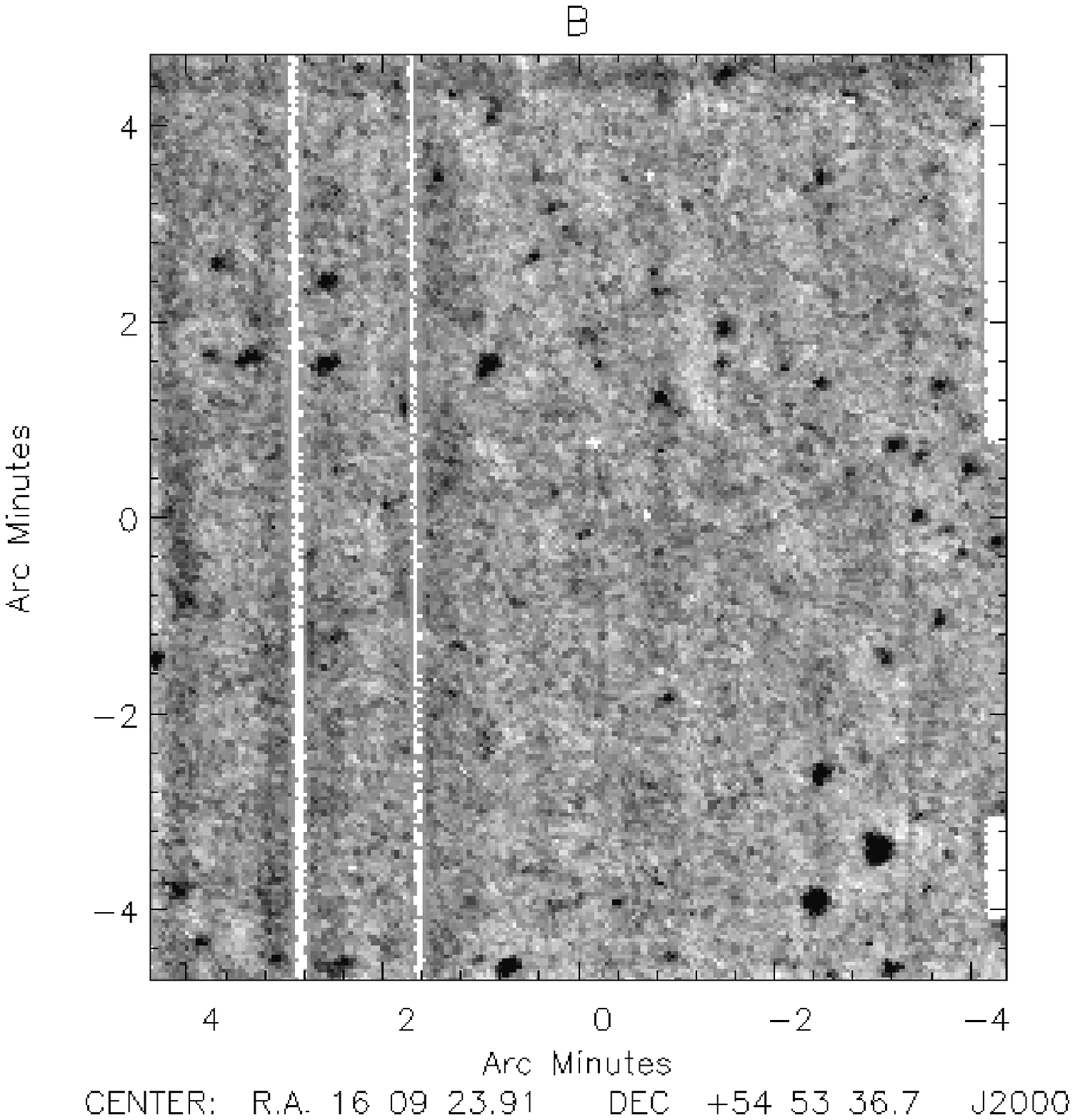}\\
\end{tabular}
\caption{Example of the improvement provided by the optimised toolkit for removal of the Earthshine light. (A) shows a single pointing processed by the new archival pipeline \citep{Egusa2016}, with the Earthshine evident as a strong flux effect; (B) shows the same pointing processed by the optimised toolkit, with Earthshine light successfully removed. The image has been cropped, so as not to contain the slit area, also the white horizontal lines in image (B) are masked. These lines have not been masked in image (A).}
\label{fig:earthshine_s11} 
\end{figure}

IRC frames suffer from Earthshine, caused by sunlight reflected by the Earth onto the telescope. This artefact appears as an area of increased incident flux, which can move around the image from frame to frame. The Earthshine artefact can be seen in images from all of the nine IRC filters. The Earthshine effect is worse in the MIR detector images and also during later-Phase 2 observations. Figure \ref{fig:earthshine_30_frames} shows 9 individual MIR-S raw frames from a 30 frame single S11 pointing observed during late-Phase 2. The figure clearly shows that the Earthshine artefact moves around the image, from frame to frame, during a single pointing. A template or time dependant flat is not able to remove this artefact. Neither of the archival pipelines were able to fully remove the Earthshine artefact. \cite{Egusa2016} state that they have a template which is able to remove this artefact when it is fairly low level. The optimised toolkit removes this artefact by creating a boxcar median filtered image for each frame and subtracts this from the original image. Figure \ref{fig:earthshine_s11} shows the improvement to a co-added pointing, after the removal of the Earthshine artefact from each individual frame.

\begin{table}
  \caption{The positional error of the six IRC filters used in the work of section \ref{results}.}
  \label{tab:pixel_error}
  \begin{center}
    \begin{tabular}{lllllllll}
      \hline \hline
Filter&positional error/pixels\\
 \hline
N3&0.42\\
N4&0.63\\
S7&0.57\\
S11&0.69\\
L15&0.56\\
L18W&0.59\\
    \hline
    \end{tabular}
  \end{center}
\end{table}

\subsubsection{Temperature Change of the Detectors}
\label{change_of_background_flux}
\begin{figure}
 \centering
 \includegraphics[width=85mm]{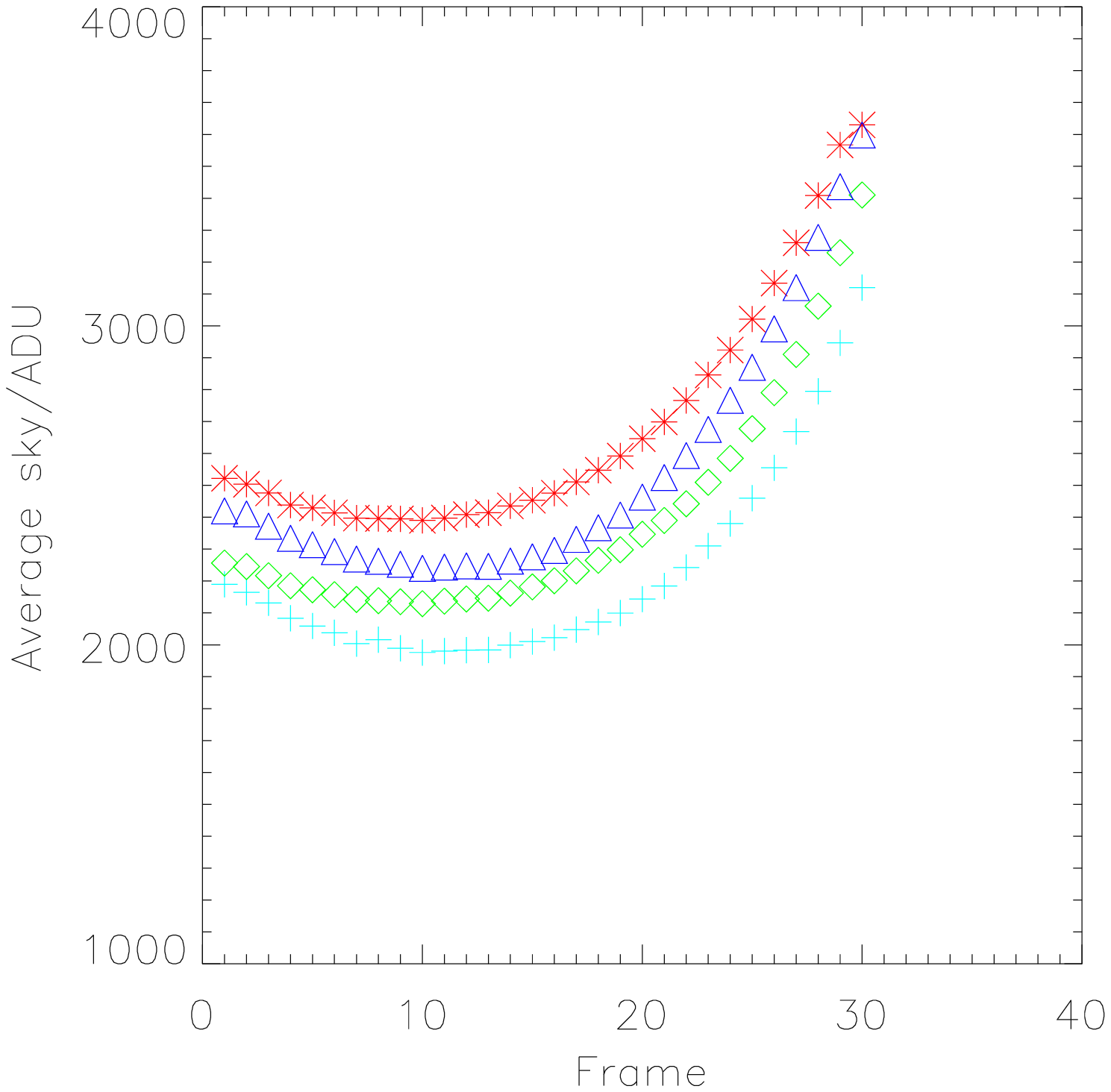}
 \caption{Image showing the change in average background flux of the frames over a pointing. Each different shape/coloured represents a different observation.}
\label{fig:background_flux}
\end{figure}

Present in all observations, over a single pointing the background flux of the detector decreases and then increases. This is worse for pointings taken at later stages of the mission and for MIR-S and MIR-L images. This is due to the Sun's light warming the telescope, causing the entire detector to change temperature. Figure \ref{fig:background_flux} shows how much the average sky background flux alters over a single pointing. The flux variation follows the same pattern for different pointings but the magnitude of the effect varies. Neither of the archival pipelines discusses this artefact. The optimised toolkit corrects for this, by masking the point sources and calculating the average background flux for each frame. During the coadding stage (Sections \ref{irac_dark_field}, \ref{ADFS} and \ref{ELAIS_N1}) the corresponding average value is subtracted from each frame.

\subsubsection{Extended Ghosting}
\begin{figure}
\begin{tabular}{cc}
  \includegraphics[width=65mm]{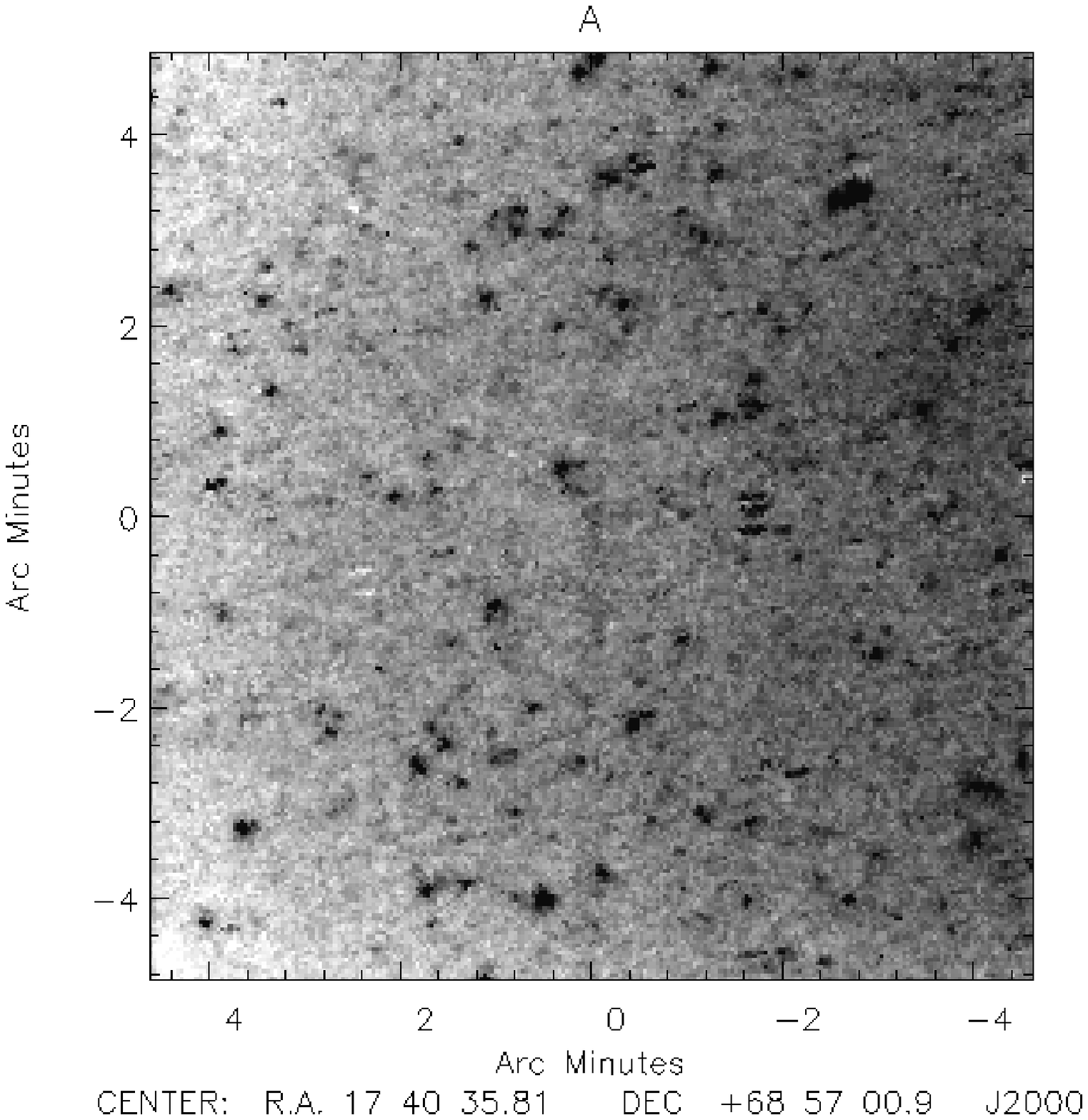}\\
  \includegraphics[width=65mm]{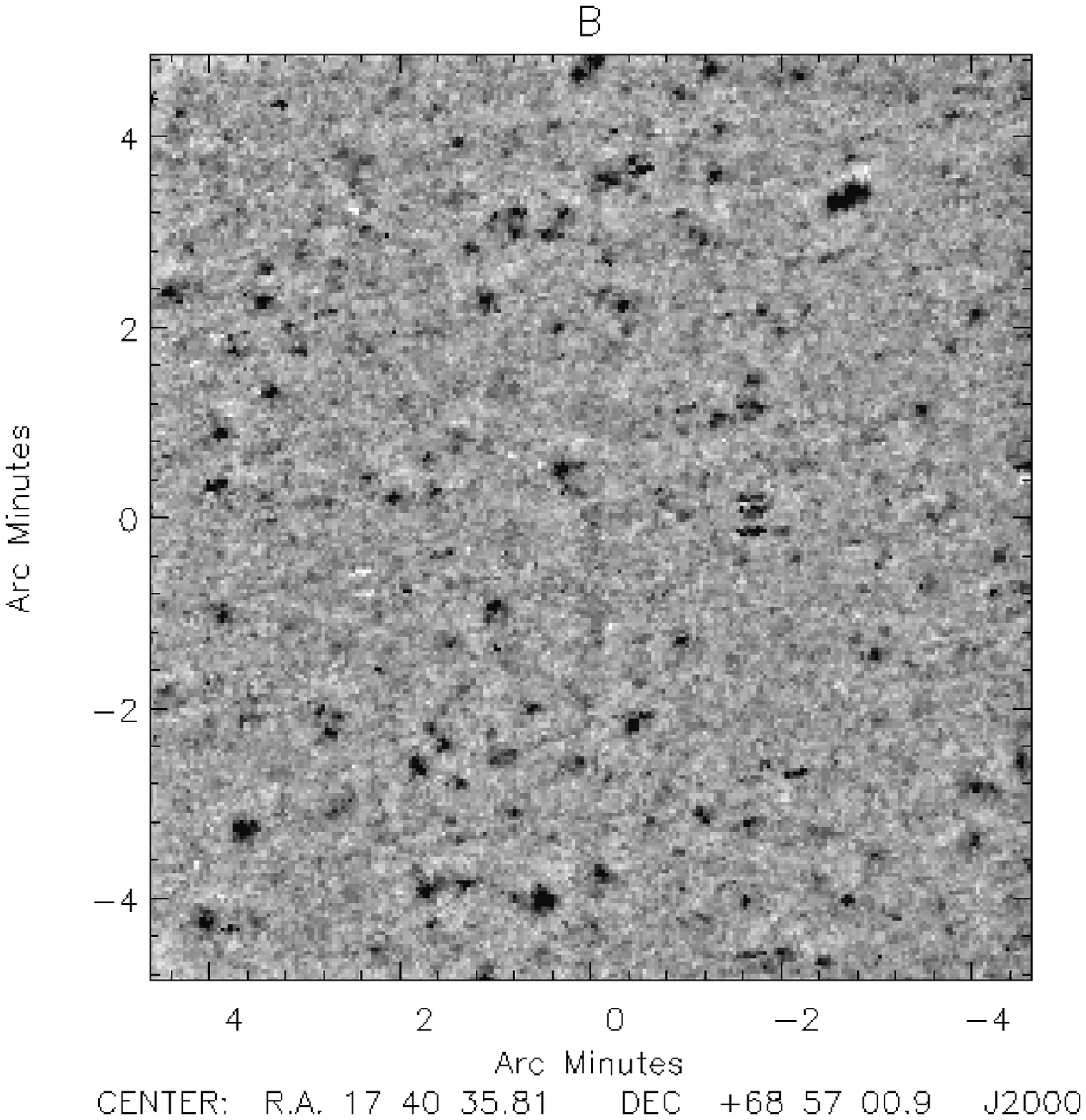}\\
\end{tabular}
\caption{(A) shows the MIR-L detector affected by extended ghosting (light scattered reflected off the array or filter, onto the detector). (B) shows the removal of the scattered light artefact by subtracting a boxcar median filtered image from the original frames (the co-added image is shown here for clarity).}
\label{fig:background_gradient} 
\end{figure}

The MIR-L detector images suffer an additional artefact, referred to as extended ghosting, see Figure \ref{fig:background_gradient}.a, caused by light reflecting off the array or filter, onto the detector. This artefact is removed in the same toolkit step as the Earthshine light (see section \ref {earthshine_light}), by subtracting a boxcar median filtered image from the original frame. The removal of this artefact is demonstrated in Figure \ref{fig:background_gradient}.

\subsection{Create Noise Image}
The optimised toolkit creates a separate noise image for each frame. Neither archival pipelines creates an individual noise image for each frame, but only creates a noise image from each co-added pointing. Individual noise images for each frame are required when performing noise weighted coadding. 
The noise value for each pixel is calculated using Equation \ref{noise_equation} \citep{Mortara1981};
\begin{equation}
N_T=\sqrt{N_* + n_{pix} (N_S + N_D + N_R^2)}
\label{noise_equation}
\end{equation}
 where $N_T$ is the total noise of the pixel. {\it $N_{*}$} is derived from the total number of photons, {\it $n_{pix}$} is the number of pixels, in this case $n_{pix}=1$, {\it $N_{S}$} is derived from the background number of photons (the background flux of the image),  {\it $N_{D}$} is the total number of electrons caused by the dark current found from the dark image and {\it $N_{R}$} is the read out/shot noise of the pixel, calculated from the number of electrons.
Note that each noise image has the same distortion correction and astrometry correction as the associated image frame (see Sections \ref{Distortion_correction} and \ref{astrometry_correction}). Such analytic noise images are created for each associated image and used at the co-adding stage.

\subsection{Astrometry Correction}
\label{astrometry_correction}
The original archival pipeline performs an astrometry correction on a pointing, once it had been co-added, using the World Coordinate System (WCS). Due to the fact that most of the IRC frames have a positional offset, in turn due to incorrect telescope astrometry, which varies from frame to frame and can be as bad as 5 arcsec in the long wavelength channels. The optimised toolkit corrects for astronomical offsets on individual frames, before the co-adding of them, since the offset can vary from frame to frame even within the same pointing. The jitter is believed to have been caused by changes in the temperature of the star trackers. The new archival pipeline makes an astrometry correction, if greater than five sources have been detected in a single frame. Sources can be difficult to detect in deep extragalactic images, and many are only detectable after coadding. In the archive processed data, these astrometry offsets have prevented successful processing of the faintest extragalactic deep fields (notably the ELAIS-N1 field discussed in Section~\ref{ELAIS_N1}). The archival pipeline co-adds frames observed by the same filter in a single pointing and then corrects for any astrometric offset on the final co-added image. The optimised toolkit corrects for the astrometry offset on a frame-by-frame, filter-by-filter basis individually and independently of the other filters. The astrometry of the first frame in each pointing is corrected by aligning extracted point sources with the identical sources in the {\it 2MASS} or {\it WISE}  galaxy catalogues, for NIR  and MIR-S/MIR-L channels respectively. Each subsequent frame of the same filter in the pointing is aligned to the first frame by matching the point sources. After the frames have been through the astrometry correction stage, they are ready to be co-added with frames from the same or other pointings. 

\subsection{Masking before coadding}
\label{masking_before_coadding}
\begin{figure}
\begin{tabular}{cc}
  \includegraphics[width=65mm]{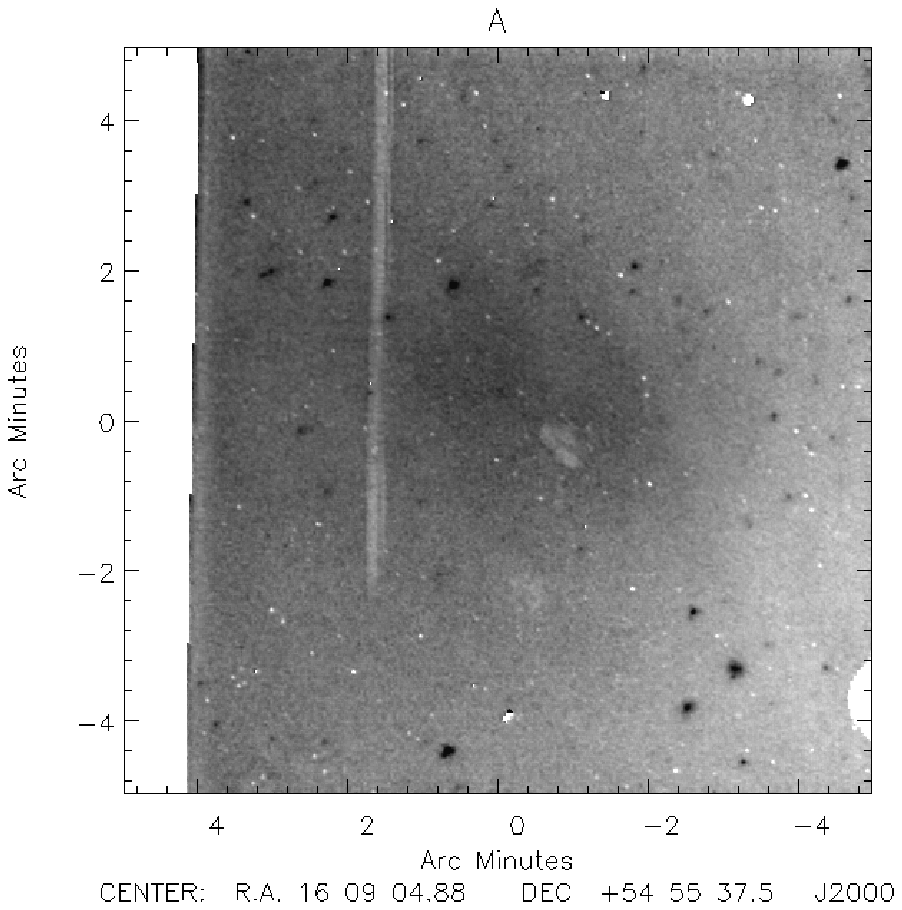}\\
  \includegraphics[width=65mm]{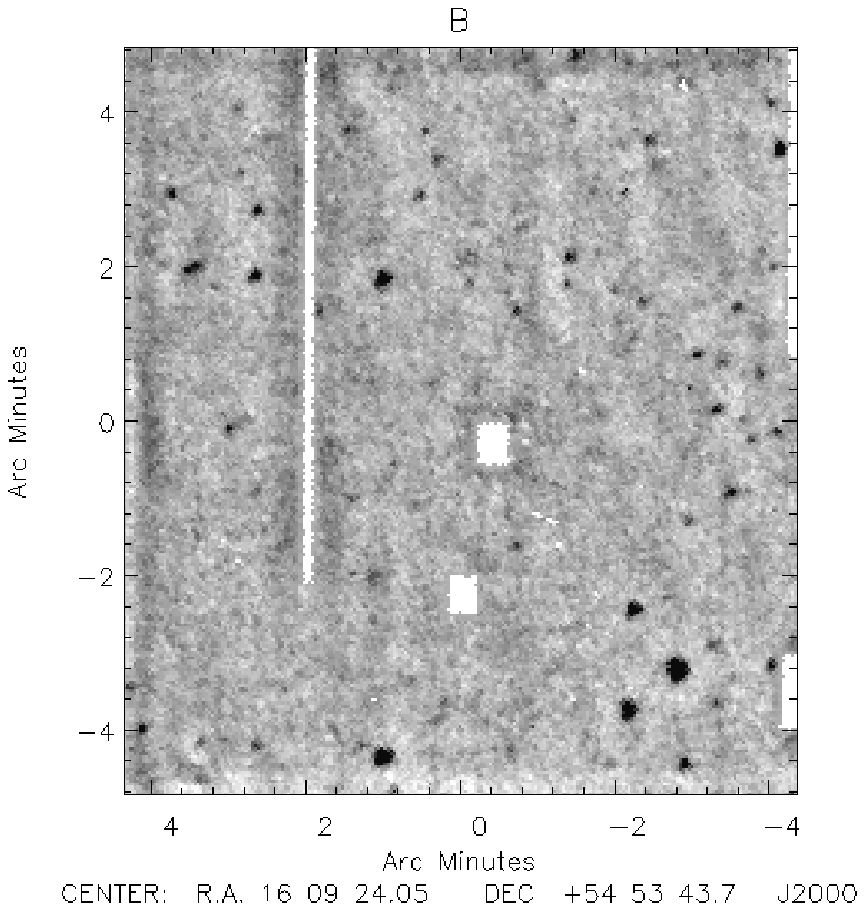}\\
\end{tabular}
\caption{Example of masking frames before coadding. (A) shows a single pointing processed by the new archival pipeline \citep{Egusa2016}. Note the `clover leaf' artefact in the centre of the image and just a little lower than centre, and the vertical line. (B) shows the same pointing processed by the optimised toolkit. The vertical line and two `clover leaf' memory effects have been masked in the individual frames before coadding. Also note the removal of the Earthshine light and masking of hot pixels. This image has been cropped, so as not to contain the slit area.}
\label{fig:mask_and_not_mask_comparison} 
\end{figure}

Due to the increasing temperature of the telescope, as discussed in Section \ref{amonalous_pixel_detection}, and memory effects, images from the later stages of Phase 2 have artefacts which neither a dedicated flat field nor hot pixel mask are able to remove. Two such artefacts are the: `clover leaf' pattern memory effect and vertical lines. These two artefacts can be seen in the coadded pointing in Figure \ref{fig:mask_and_not_mask_comparison}.a. The `clover leaf' memory effect is caused by imaging a very bright source in a previous pointing up to a few hours prior to the pointing in question. The artefact has the distinctive clover leaf pattern because the previous pointing observing the very bright source was dithered. The vertical lines are caused by a bright source observed on the IRC array during an {\it AKARI}/FIS scanning observation, hence the artefact appears as vertical lines. The same area is masked for all frames in a given pointing. Figure \ref{fig:mask_and_not_mask_comparison} demonstrates the improvement to an extragalactic deep field by performing masking on individual frames.

\begin{figure}
 \centering
 \includegraphics{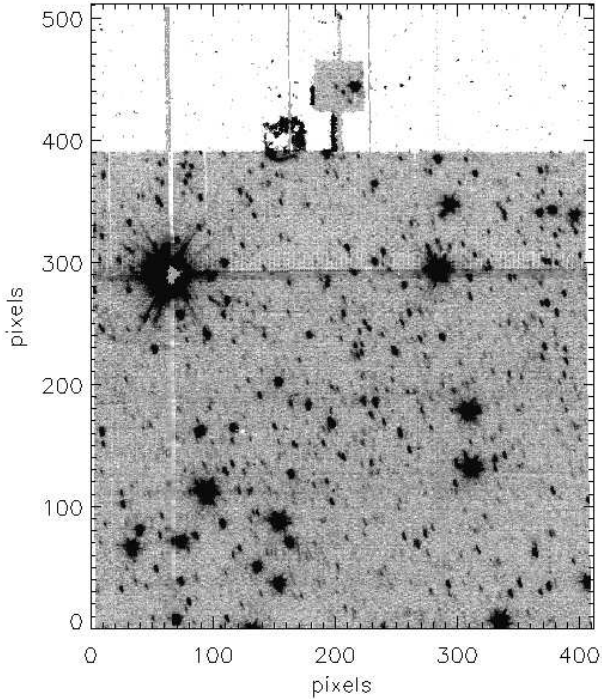}
 \caption{The figure shows one of the ghosts discussed in \citealt{Murata2013}. The ghost source can be seen at the bottom of the slit area, just left of centre. The source creating this ghost is the saturated source in the upper left of the image.}
 \label{fig:murata_ghost}
\end{figure}

Several ghost artefacts, found in {\it AKARI} images, have been discussed in pervious work, e.g., \cite{Murata2013} discuss a ghost artefact created by a bright source reflected about the image pixel position coordinates $x=115$, $y=350$. An example of this artefact is shown in Figure \ref{fig:murata_ghost}. This artefact was only present in the NIR images. The artefact was treated in a similar way to the memory effects discussed above, and masked where appropriate.

\cite{Arimatsu2011} discuss several ghost artefacts in the MIR bands, caused by bright sources. In the MIR-S channel, bright sources cause a ghost artefact, repeating at a period of about 24 pixels in the $y$-direction from the source. In both the MIR-S and MIR-L channels, bright sources produce a ghost artefact a little offset from the true source, and a set of two concentric artefacts, bigger than the original source. As the extragalactic fields discussed in this paper do not contain any overly bright sources in the MIR-S and MIR-L bands, it is observed that these artefacts are below the instrument noise level.

\section{Application of Optimised toolkit}
\label{results}
In order to quantify the improvement provided by the optimised toolkit compared to the archival data, the optimised toolkit was applied to three {\it AKARI} extragalactic survey fields.
A example early to mid-Phase 2 deep field was chosen, the IRAC Dark Field (Section \ref{irac_dark_field}) and an example late-Phase 2 deep field was also chosen, ELAIS North1 (hereafter ELAIS-N1, Section~\ref{ELAIS_N1}). The IRAC Dark Field, centred on 17h40m00s, +69$^\circ$00m00s (J2000), was observed as an AKARI Open Time programme (PI E. Egami) and is one of the deepest fields observed by {\it AKARI}. ELAIS-N1 was observed as part of the  FU-HYU (Follow-Up Hayai-Yasui-Umai, PI C.Pearson) {\it AKARI} Mission Programme  (RA = 16h09m20s, Dec = +54$^\circ$57'00" J2000) \citep{Pearson2010}. 
An example shallow field, was also selected, the {\it AKARI} Deep Field South (PI C. Pearson, proposal ID Open Time IRSEP, see Section \ref{ADFS}) near the south ecliptic pole (hereafter SEP).

\subsection{Case Study Deep Field Early-Phase 2: IRAC Dark Field}
\label{irac_dark_field}
The IRAC Dark Field is a $20 \times 20$\,arcminute area, in the north, centred on 17h40m00s, +69$^\circ$00m00s (J2000), close to the NEP. This survey area was used in calibrating the instrumental background for {\it Spitzer}/IRAC, observed every two to three weeks over a 5+ week period\citep{Krick2009}. The IRAC Dark Field was also a calibration field for the SPIRE instrument on the {\it Herschel} Space Observatory and hence is one of the deepest extragalactic fields observed by {\it Herschel}. There is also a large amount of multi-wavelength data for the IRAC Dark Field. With such a large amount of deep infrared data from {\it Spitzer}, {\it AKARI} and {\it Herschel}/SPIRE, the IRAC Dark Field is a survey area of great importance. Data from all three telescopes has not as yet been fully utilised.

The IRAC Dark Field was chosen as a test of the optimised toolkit, as the observations are of a deep field observed during early to mid-Phase 2. Though evident, the frames observed during this time period were not too badly affected by Earthshine light and hot pixels. One of the major issues for early-Phase 2 MIR-S frames is that they suffered from the so called bean artefact, discussed in more detail in Section \ref{flat_fielding}. Unlike the original and updated archival pipeline, the optimised toolkit is able to remove the bean artefact (see Figure \ref{fig:bean}).

Table \ref{tab:iracpointings} shows the {\it AKARI}/IRC pointings of the IRAC Dark Field. There were 34 successful pointings of the IRAC Dark Field. It was observed 19 times by the N4 and S11 filters, 10 times by the L15 filter and 5 times by the L18W filter. Note half of the observations were observed October 2006 (late-Phase 1) and the other half were observed April/May 2007 (mid-Phase 2). The N4 and S11 frames were obtained during the same pointings. All pointings were observed using IRC05, the Astronomically Observed Template (AOT) for deep pointings. IRC05 was used, which does not have an option for dithered pointings. Due to the fact that each of these pointings covered roughly the same area of sky, this is one of the deepest fields observed by {\it AKARI}/IRC.

\begin{figure*}
\begin{tabular}{cc}
  \includegraphics[width=70mm]{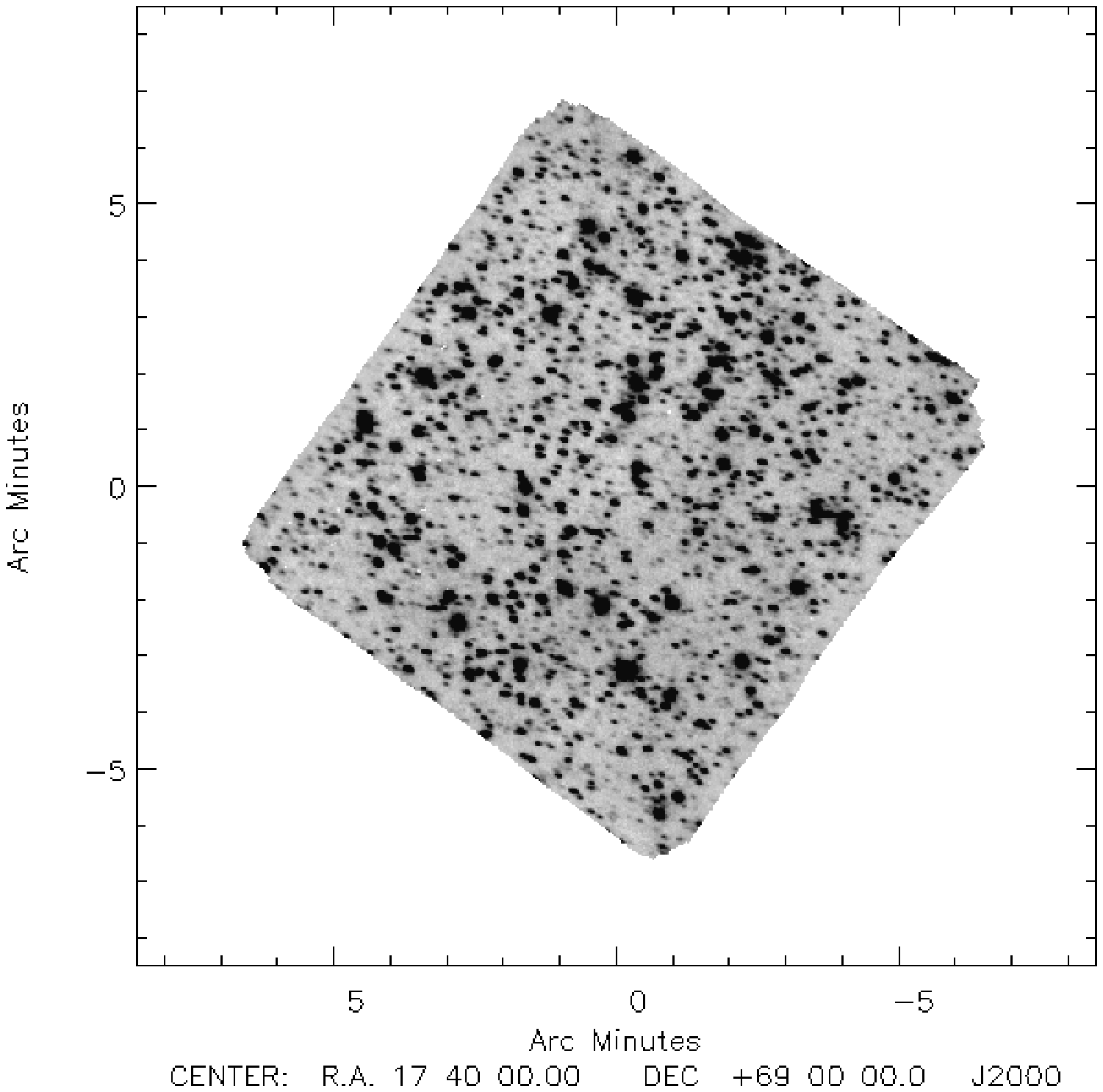} & 
  \includegraphics[width=70mm]{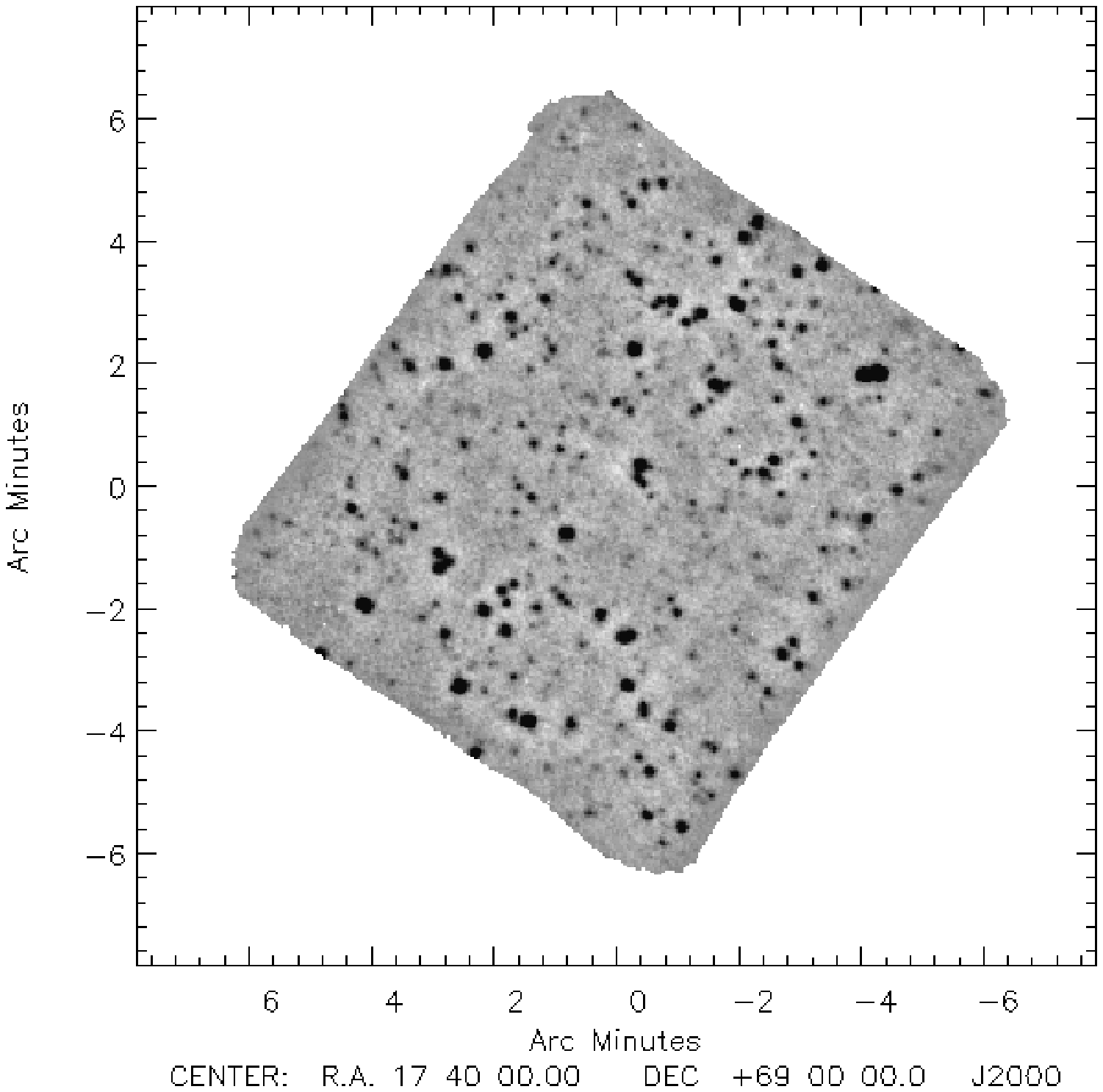} \\
  (a) N4 Filter & (b) S11 Filter \\[6pt]
  \includegraphics[width=70mm]{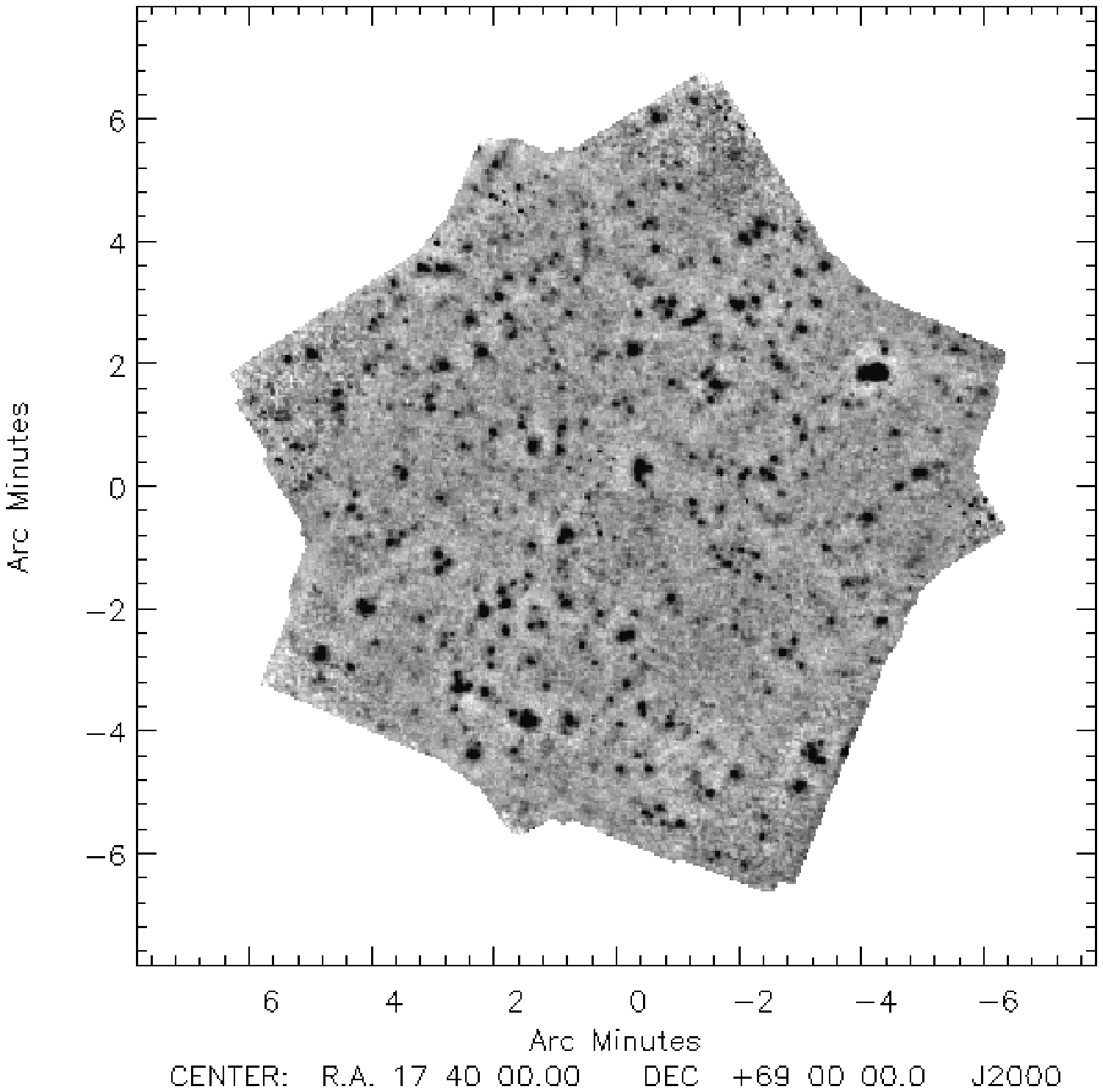} & 
  \includegraphics[width=70mm]{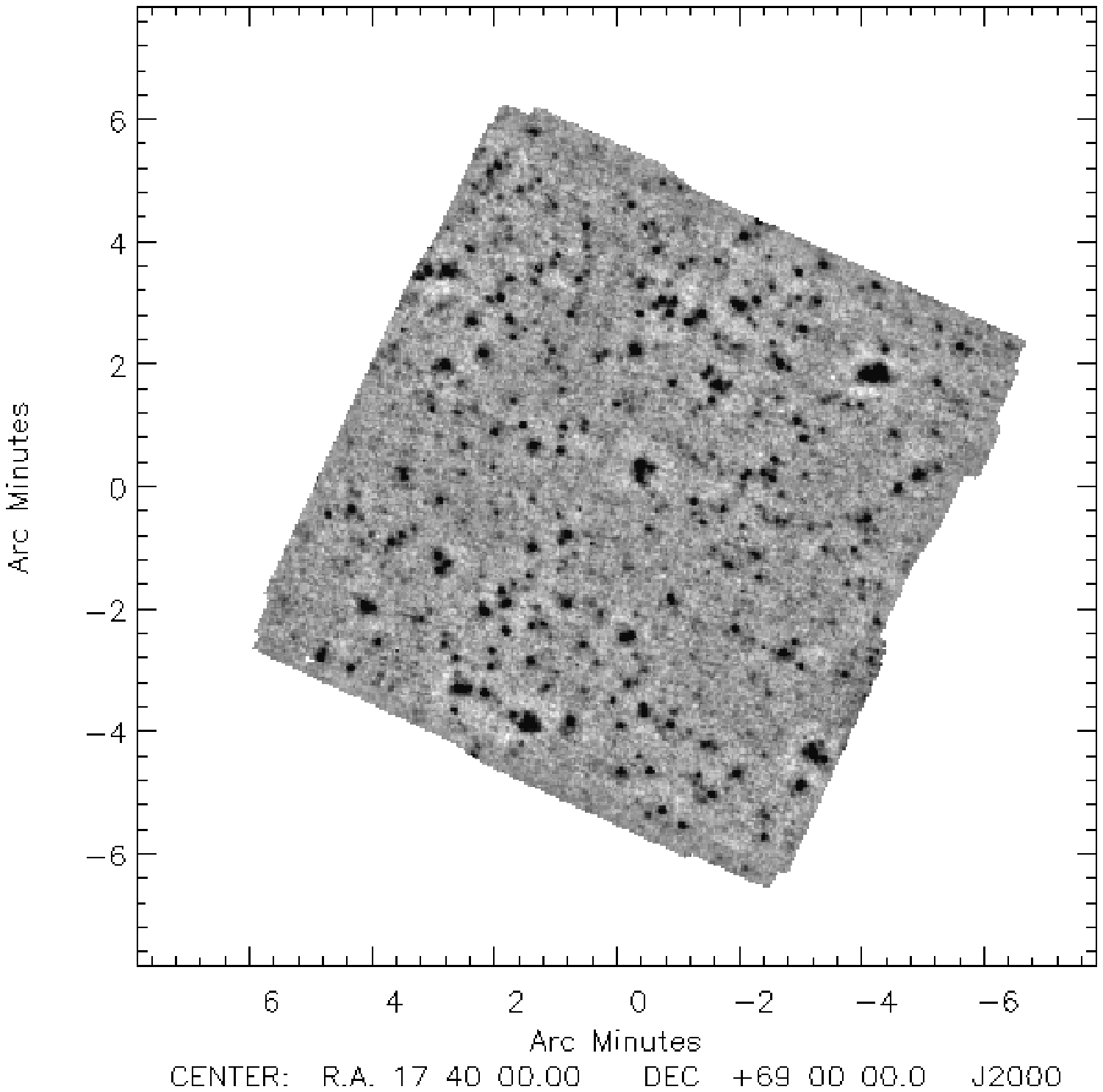} \\
  (c) L15 Filter & (d) L18 Filter\\[6pt]
\end{tabular}
\caption{IRAC Dark field mosaicked images}
\label{iracdarkfieldmosaickedimages} 
\end{figure*}

The individual frames were also checked for possible image ghosts discussed in Section \ref{masking_before_coadding}, but none were found. Table \ref{tab:discarrdedframes} gives a list of damaged frames which were not used to create the IRAC Dark Field mosaicked images. In the table, the stripe artefact was caused by a cosmic ray, hitting the detector and saturating the entire node. This effect lasts for a few minutes, and can be seen in subsequent frames. Figures \ref{iracdarkfieldmosaickedimages} show IRAC Dark Field final deep field images from the optimised toolkit at 4.1, 11, 15 and 18\,$\mu$m. The survey area of the IRAC Dark Field is 0.027 square degrees.

\begin{table}
  \caption{Observation log for the IRAC Dark Field pointings.}
  \label{tab:iracpointings}
  \begin{center}
    \begin{tabular}{lllllllll}
      \hline \hline
 Pointing ID&Date&Filter&AOT\\
 \hline
 3030001-001&18/10/2006&N4 \& S11&IRC05\\
 3030001-003&18/10/2006&N4 \& S11&IRC05\\
 3030001-004&16/04/2007&N4 \& S11&IRC05\\
 3030001-005&19/10/2006&N4 \& S11&IRC05\\
 3030001-006&19/04/2007&N4 \& S11&IRC05\\
 3030001-007&27/10/2006&N4 \& S11&IRC05\\
 3030001-008&27/10/2006&N4 \& S11&IRC05\\
 3030001-009&27/10/2006&N4 \& S11&IRC05\\
 3030001-010&27/10/2006&N4 \& S11&IRC05\\
 3030001-011&28/10/2006&N4 \& S11&IRC05\\
 3030001-012&19/04/2007&N4 \& S11&IRC05\\
 3030001-013&20/04/2007&N4 \& S11&IRC05\\
 3030001-014&31/10/2006&N4 \& S11&IRC05\\
 3030001-015&21/04/2007&N4 \& S11&IRC05\\
 3030001-016&21/04/2007&N4 \& S11&IRC05\\
 3030001-017&21/04/2007&N4 \& S11&IRC05\\
 3030001-018&22/04/2007&N4 \& S11&IRC05\\
 3030001-019&22/04/2007&N4 \& S11&IRC05\\
 3030001-020&22/04/2007&N4 \& S11&IRC05\\
 3030002-001&10/10/2006&L15&IRC05\\
 3030002-002&10/10/2006&L15&IRC05\\
 3030002-003&12/10/2006&L15&IRC05\\
 3030002-004&12/10/2006&L15&IRC05\\
 3030002-005&23/04/2007&L15&IRC05\\
 3030003-001&13/10/2006&L18W&IRC05\\
 3030003-002&13/10/2006&L18W&IRC05\\
 3030003-003&13/10/2006&L18W&IRC05\\
 3030003-004&13/10/2006&L18W&IRC05\\
 3030003-005&23/04/2007&L18W&IRC05\\
 3031001-001&14/05/2007&L15&IRC05\\
 3031001-002&14/05/2007&L15&IRC05\\
 3031001-003&14/05/2007&L15&IRC05\\
 3031001-004&15/05/2007&L15&IRC05\\
 3031001-005&15/05/2007&L15&IRC05\\
    \hline
    \end{tabular}
  \end{center}
\end{table}

\begin{table}
\scriptsize
  \caption{IRAC Dark Field discarded frames.}
  \label{tab:discarrdedframes}
  \begin{center}
    \begin{tabular}{lllllllll}
      \hline \hline
 Filter&Pointing ID&Frame Number&Reason\\
 \hline
 S11&3030001\_014&F004059008\_S004&Artificial stripe pattern\\
 L15&3031001\_001&F008043439\_L004&Cosmic ray\\
 L15&3031001\_004&F008048716\_L002&Artificial stripe pattern\\
 L15&3031001\_004&F008048716\_L003&Artificial stripe pattern\\
 L15&3031001\_004&F008048716\_L004&Artificial stripe pattern\\
 L15&3031001\_004&F008048718\_L002&Artificial stripe pattern\\
 L15&3031001\_004&F008048718\_L003&Artificial stripe pattern\\
 L15&3031001\_004&F008048718\_L004&Artificial stripe pattern\\
    \hline
    \end{tabular}
  \end{center}
\end{table}

\subsection{Case Study Deep Field Late-Phase 2: ELAIS-N1}
\label{ELAIS_N1}
The ELAIS-N1 was one of 11 fields making up the European Large Area ISO Survey (ELAIS  \citealt{Oliver2000}, the largest open time survey performed by the Infrared Space Observatory, ISO \citealt{kessler96}). The ELAIS fields were selected at high Galactic latitudes (for ELAIS-N1 $\beta>40^\circ$) for low cirrus emission (I$_{100\mu m}<1.5$\,MJy/sr cirrus level).  ELAIS-N1 is one of the deepest surveys performed by {\it AKARI}. Out of all of the FU-HYU fields, it had the best visibility during the {\it AKARI} mission. ELAIS-N1 has a large amount of multi-wavelength ancillary data (\citealt{Ciliegi1999}, \citealt{Mcmahon2001}, \citealt{Vaisanen2002}, \citealt{Basilakos2002}, \citealt{Manners2003}, \citealt{Chary2004} \& \citealt{Oliver2012}).

ELAIS-N1 was chosen as a test of the optimised toolkit of a deep field observed late-Phase 2; as apart from three of the pointings, which were observed mid-Phase 2, most of the observations (17 pointings) were made in late-Phase 2. As discussed earlier in the paper, late-Phase 2 observations were plagued by many artefacts, e.g. Earthshine light, hot pixels, memory effects. Many of which neither the original nor updated archival pipelines were able to fully remove.

Table \ref{tab:ELAISN1} shows the observation log for the ELAIS-N1 pointings. All the pointings use astronomical template IRC05, which was not dithered. As the majority  of the pointings are late-Phase 2, they suffer from more Earthshine light than earlier observations. Subsection \ref{scattered_light} shows how the most significant components of the Earthshine light were removed. The late-Phase 2 images also suffered from more detector deterioration. 
 The time dependent flat field images and hot-pixel masks were created from independent observations taken over the same period as the ELAIS-N1 observations from 12/01/2007 to 21/01/2007 and 19/07/2007 to 22/07/2007. Due to the large areas of some pointings which required masking, the same area of each frame in a pointing was masked before coadding the pointing. Even though the ELAIS-N1 pointings were not dithered, since the frames are masked before co-adding the intrinsic jitter of the telescope can be used effectively instead, to dither the observations to increase the total useable area of the frames. 

\begin{figure}
\begin{tabular}{cc}
  \includegraphics[width=65mm]{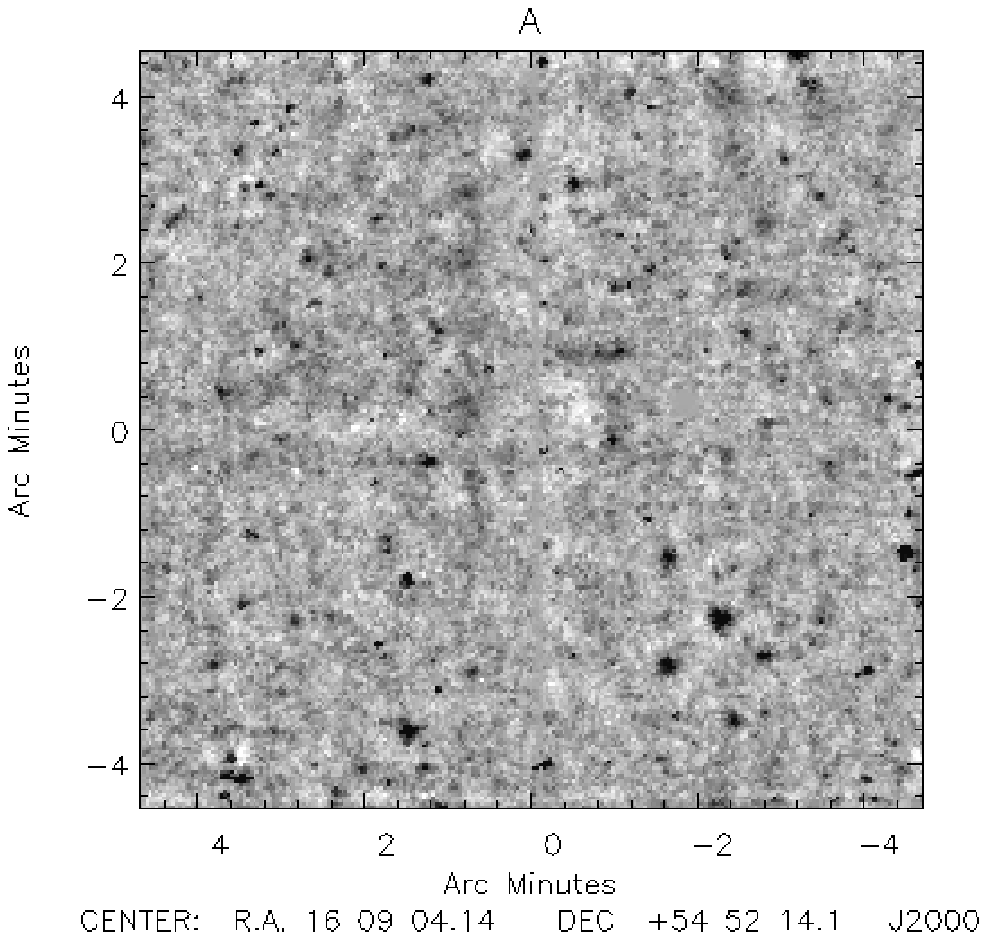}\\
  \includegraphics[width=65mm]{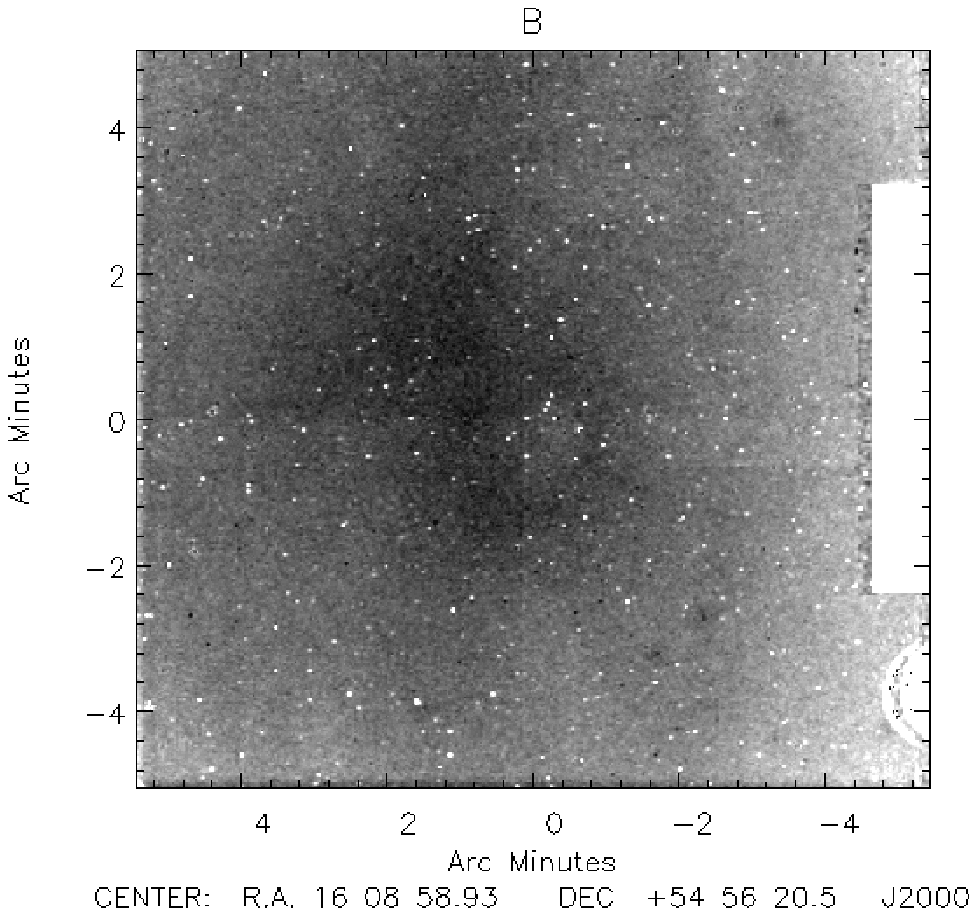}\\
\end{tabular}
\caption{Example of the improvement provided by the optimised toolkit for the ELAIS-N1 L15 image 1320235-003 discussed in the text. (A) shows a single pointing processed by the optimised toolkit. This image has been cropped, so as not to contain the slit area. (B) shows the same pointing processed by the new archival pipeline \citep{Egusa2016}. }
\label{fig:astrometry_problem_comparison} 
\end{figure}

Note that two of the ELAIS-N1 pointings, 1320235-002 and 1320235-003, taking in the IRC L15 band, had very significant offsets of up to 135 arcseconds and 318 arcseconds respectively. It is thought that this was caused by the star tracker and telescope observing different parts of the sky (star tracker problem). 
Neither archival pipelines were able to co-add these two pointings, however both of these large astronomical offsets were successfully corrected by the optimised toolkit (see Figure \ref{fig:astrometry_problem_comparison}). 

\begin{figure*}
\begin{tabular}{cc}
  \includegraphics[width=70mm]{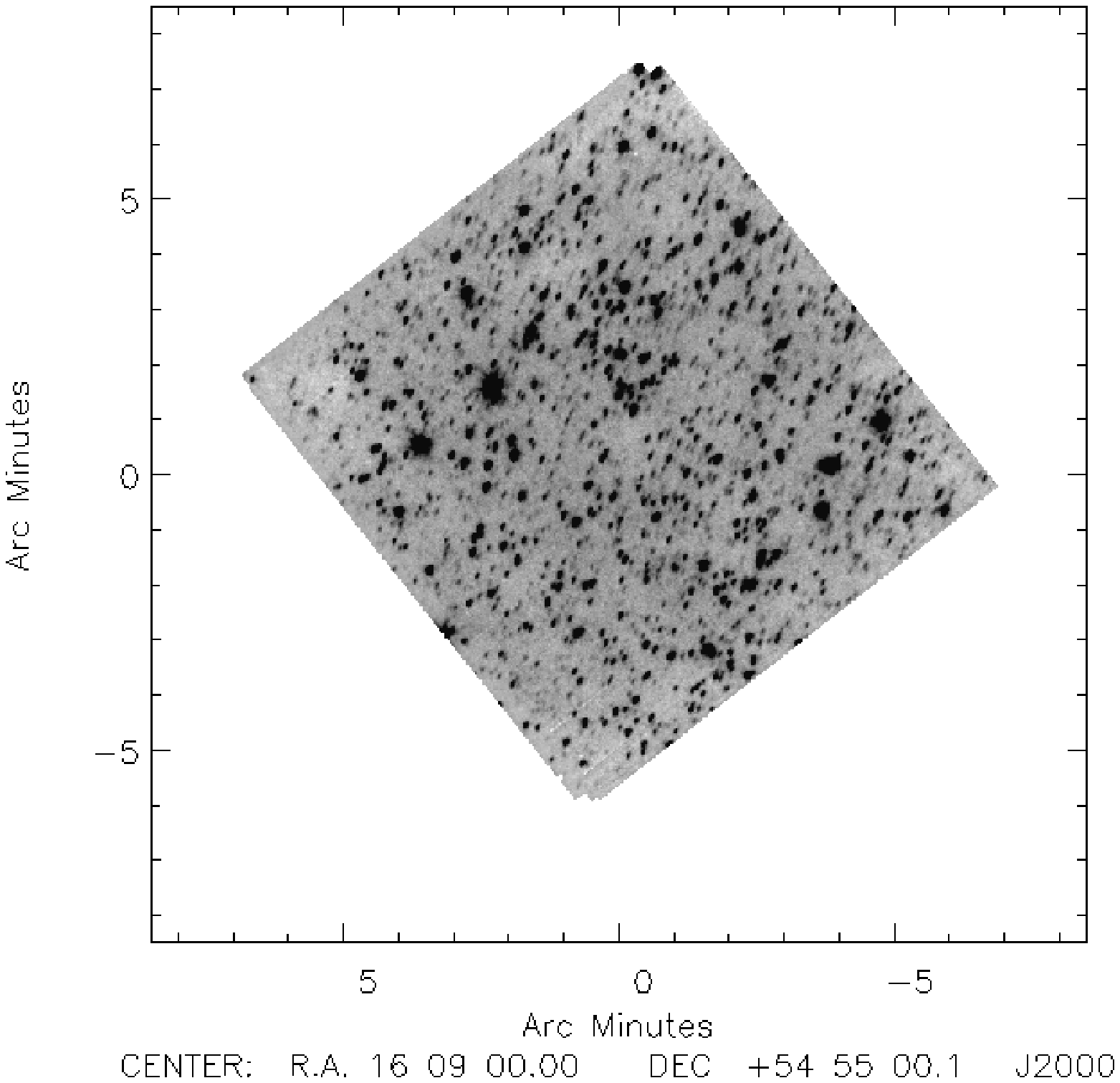} & 
  \includegraphics[width=70mm]{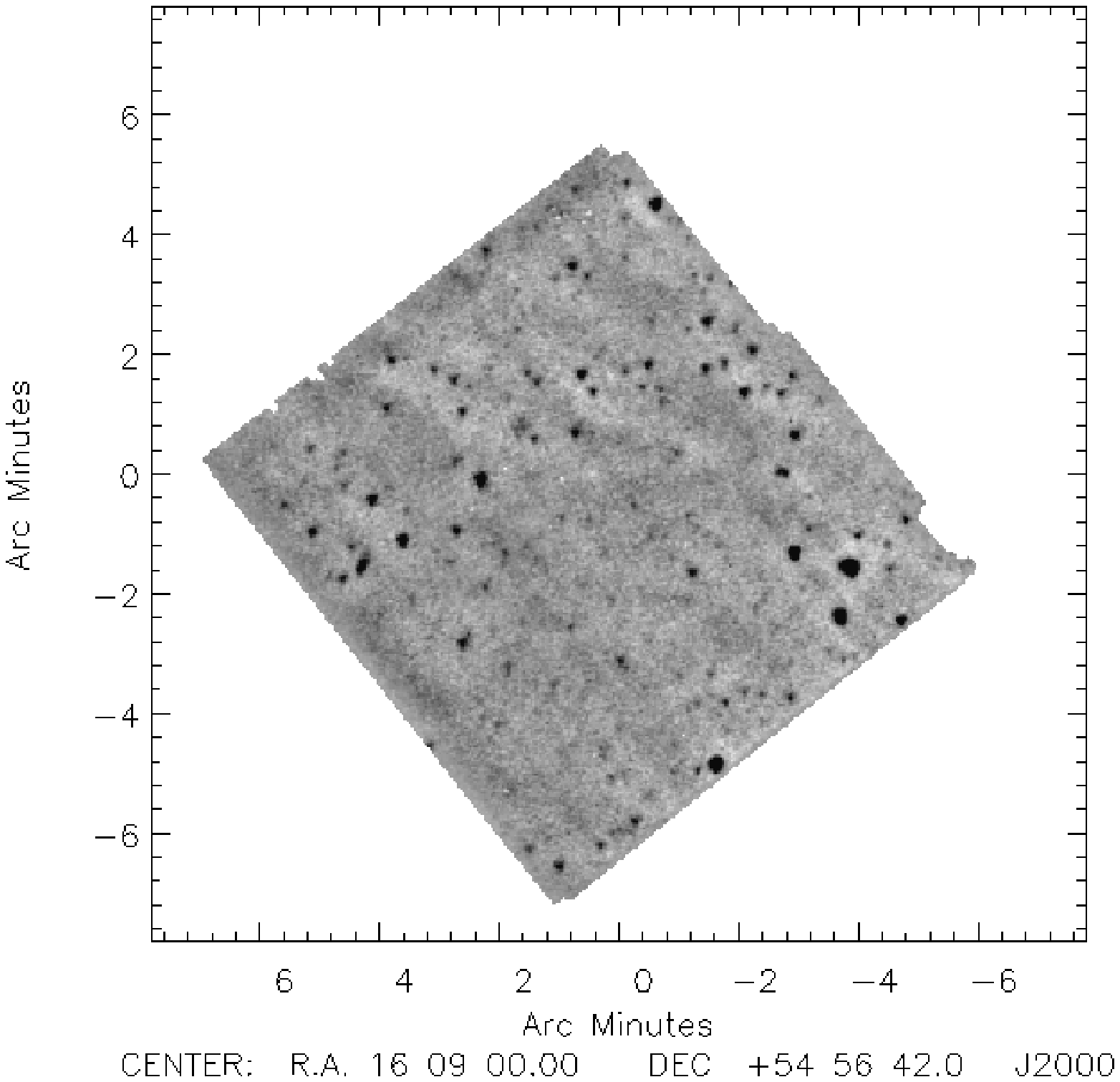} \\
  (a) ELAIS North N4 & (b) ELAIS North S11 \\[6pt]
  \includegraphics[width=70mm]{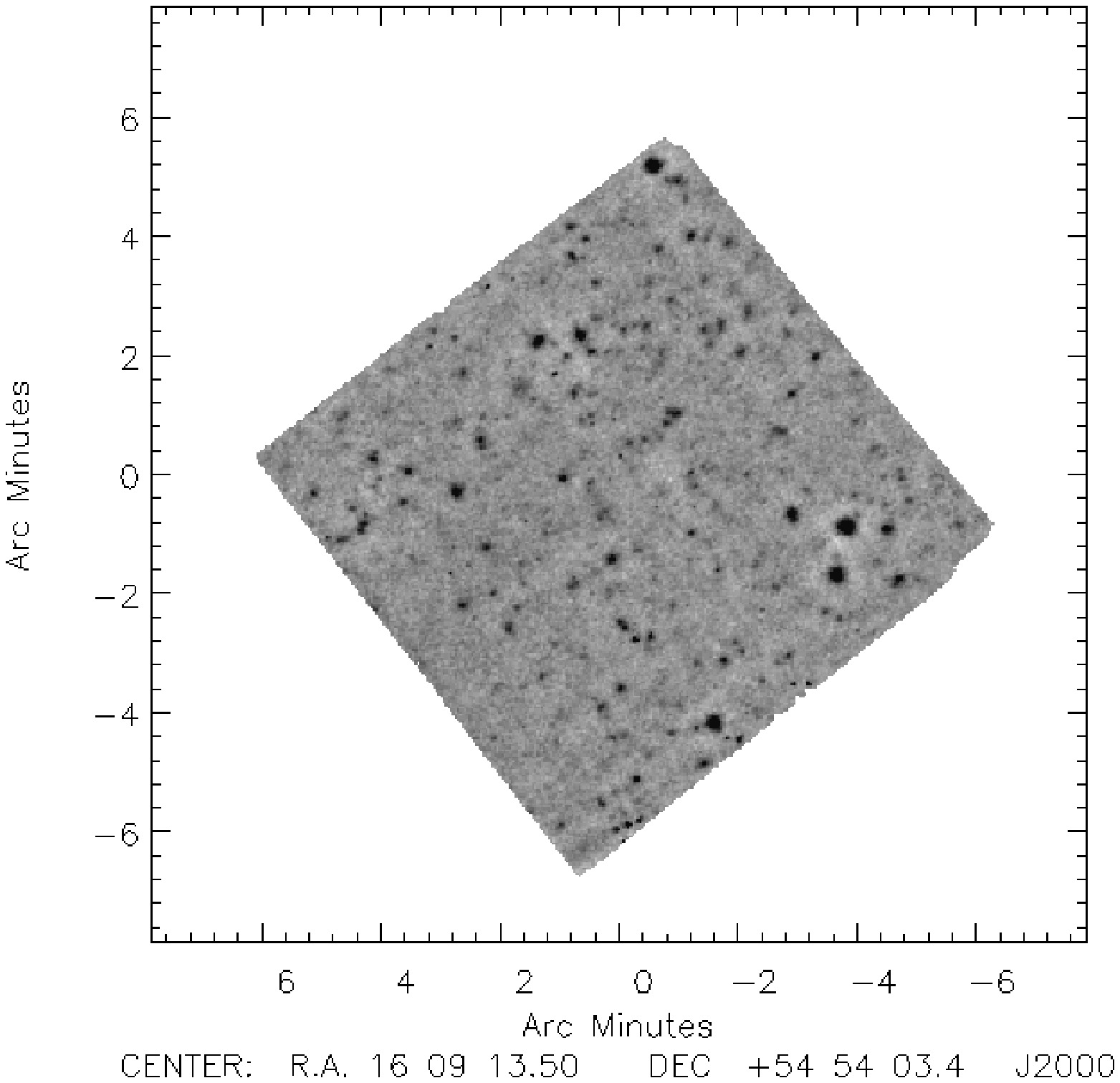} & 
  \includegraphics[width=70mm]{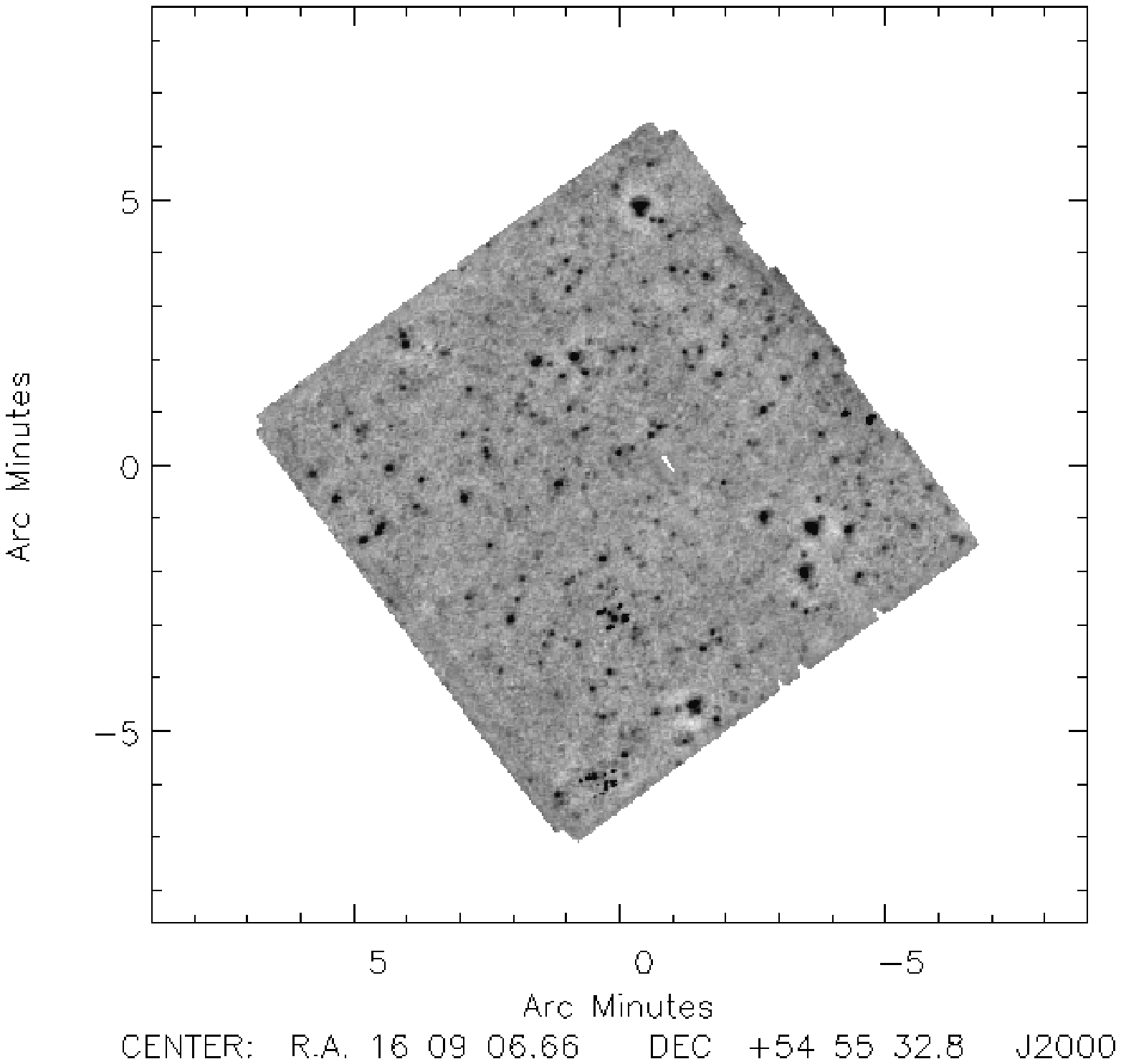} \\
  (c) ELAIS North L15& (d) ELAIS North L18\\[6pt]
\end{tabular}
\caption{The ELAIS-N1 mosaicked images.}
\label{elaismosaickedimages} 
\end{figure*}

The same as for the IRAC Dark Field frames, the images were checked for possible ghosts, but none were found.
Although the optimised toolkit was successful in combining many frames, previously deemed unusable, there were a handful of incidences where frames still had to be discarded. Table \ref{tab:elaisdiscarrdedframes} shows the frames, which were not used to create the new ELAIS-N1 deep field image, with justification. Similar to the IRAC Dark Field frames, those with an artificial stripe pattern, caused by damage from a cosmic ray incident were removed. The single frame with many hot pixels, was probably caused by protons or electrons trapped by the Earth's magnetic field hitting the detector. 
Figure \ref{elaismosaickedimages} shows the ELAIS-N1 final deep field images from the optimised toolkit at 4.1, 11, 15 and 18\,$\mu$m. The survey area of ELAIS-N1 is 0.028 square degrees.

\begin{table}
\scriptsize
  \caption{Observation log for the FU-HYU ELAIS-N1 pointings.}
  \label{tab:ELAISN1}
  \begin{center}
    \begin{tabular}{lllllllll}
      \hline \hline
 Pointing Number&Date&Filter&AOT\\
 \hline
 1320226-001&19/07/2007&N4 \& S11&IRC05\\
 1320226-003&19/07/2007&N4 \& S11&IRC05\\
 1320226-004&19/07/2007&N4 \& S11&IRC05\\
 1320226-005&19/07/2007&N4 \& S11&IRC05\\
 1320226-006&20/07/2007&N4 \& S11&IRC05\\
 1320226-007&20/07/2007&N4 \& S11&IRC05\\
 1320235-001&21/07/2007&L15&IRC05\\
 1320235-002&21/07/2007&L15&IRC05\\
 1320235-003&21/07/2007&L15&IRC05\\
 1320235-005&22/07/2007&L15&IRC05\\
 1320013-001&16/01/2007&L15&IRC05\\
 1320014-001&15/01/2007&L15&IRC05\\
 1320015-001&17/01/2007&L15&IRC05\\
 1320232-001&20/07/2007&L18W&IRC05\\
 1320232-002&20/07/2007&L18W&IRC05\\
 1320232-003&20/07/2007&L18W&IRC05\\
 1320232-004&21/07/2007&L18W&IRC05\\
 1320232-005&22/07/2007&L18W&IRC05\\
 1320232-006&22/07/2007&L18W&IRC05\\
 1320232-007&22/07/2007&L18W&IRC05\\
    \hline
    \end{tabular}
  \end{center}
\end{table}

\begin{table}
\scriptsize
  \caption{ELAIS-N1 discarded frames.}
  \label{tab:elaisdiscarrdedframes}
  \begin{center}
    \begin{tabular}{lllllllll}
      \hline \hline
 Filter&Pointing Number&Frame Number&Reason\\
 \hline
 L18W&1320232\_003&F009136797\_L004&Artificial stripe pattern\\
 L18W&1320232\_003&F009136798\_L002&Artificial stripe pattern\\
 L18W&1320232\_003&F009136798\_L003&Artificial stripe pattern\\
 L18W&1320232\_003&F009136798\_L004&Artificial stripe pattern\\
 L18W&1320232\_004&F009145059\_L004&Many hot pixels\\
    \hline
    \end{tabular}
  \end{center}
\end{table}

\subsection{Case Study Shallow Field: ADF-S}
\label{ADFS}
The {\it AKARI} Deep Field South (hereafter ADF-S) is close to the SEP and was selected as the site for a large area survey (12 square degrees centred on RA = 04h44m00s, Dec = -53$^\circ$20'00" J2000) at far infrared wavelengths \citep{Matsuura2011}. In addition, the central 1 square degree area of the ADF-S was observed in {\it AKARI} Open Time ,with the IRC (PI C. Pearson). The ADF-S has also enjoyed extensive multi-wavelength follow up observations (e.g. \citealt{Valiante2010}, \citealt{Scott2010}, \citealt{Clements2011}, \citealt{Hatsukade2011}, \citealt{Oliver2012}, \citealt{White2012}, \citealt{Barrufetinprep}).

ADF-S was chosen as a test of the optimised toolkit, as the IRC observations of the ADF-S form a shallow survey of overlapping single pointings.  For the toolkit, the main difference with creating a mosaicked shallow field as apposed to a deep field, is that shallow fields need careful removal of cosmic rays and hot pixels in the individual frames, due to the fact that not as many are removed in the coadding stage. Cosmic rays and hot pixels can be mis-identified as point sources. For the case of the ADF-S, it is only 3 to 4 frames deep in the NIR images, compared to 30 frames deep in the case of ELAIS-N1. Unfortunately, this also means that many of the NIR frames required individual masking to remove an artefact only present in that frame, which is also visible in the coadded image. This also occurred in several of the MIR-S and L frames to a lesser extent. The NIR frames also require extra masking due to the {\it muxbleed} discussed in Section~\ref{wraparound_correction}. Time dependent flat field images were created from an ensemble of independent {\it AKARI} observations over the same period as the ADF-S IRC observations (01-02-2007 to 09-02-2007).
The ghosts and artefacts discussed in Section \ref{masking_before_coadding}, are visible in one pointing, see Figure \ref{fig:murata_ghost}, and have been masked. 

Table \ref{tab:ADFS} shows the observation log for the ADF-S pointings. All pointings were mid-Phase 2 and used astronomical template IRC02. Table \ref{tab:adfsdiscarrdedframes} shows the list of frames not used. In the table, `Flux Error' refers to an effect only found on a final frame in a pointing and is caused by the  shutter closing during integration. In addition, frames which have high signal due to cosmic ray impact have also been removed. The effect of the cosmic ray lasts for a few minutes and can be seen in subsequent frames. 

\begin{figure*}
\begin{tabular}{cc}
  \includegraphics[width=70mm]{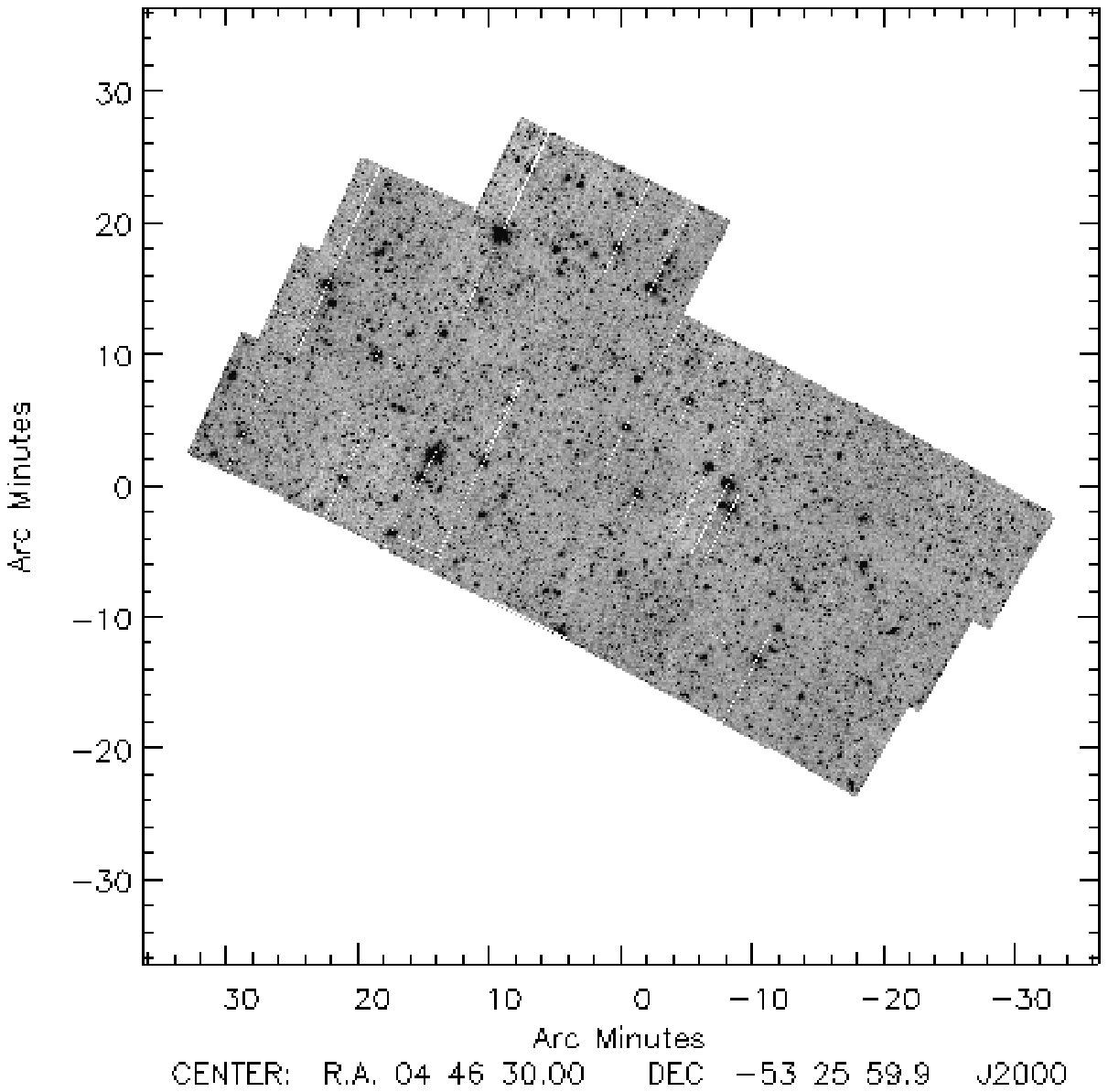} & 
  \includegraphics[width=70mm]{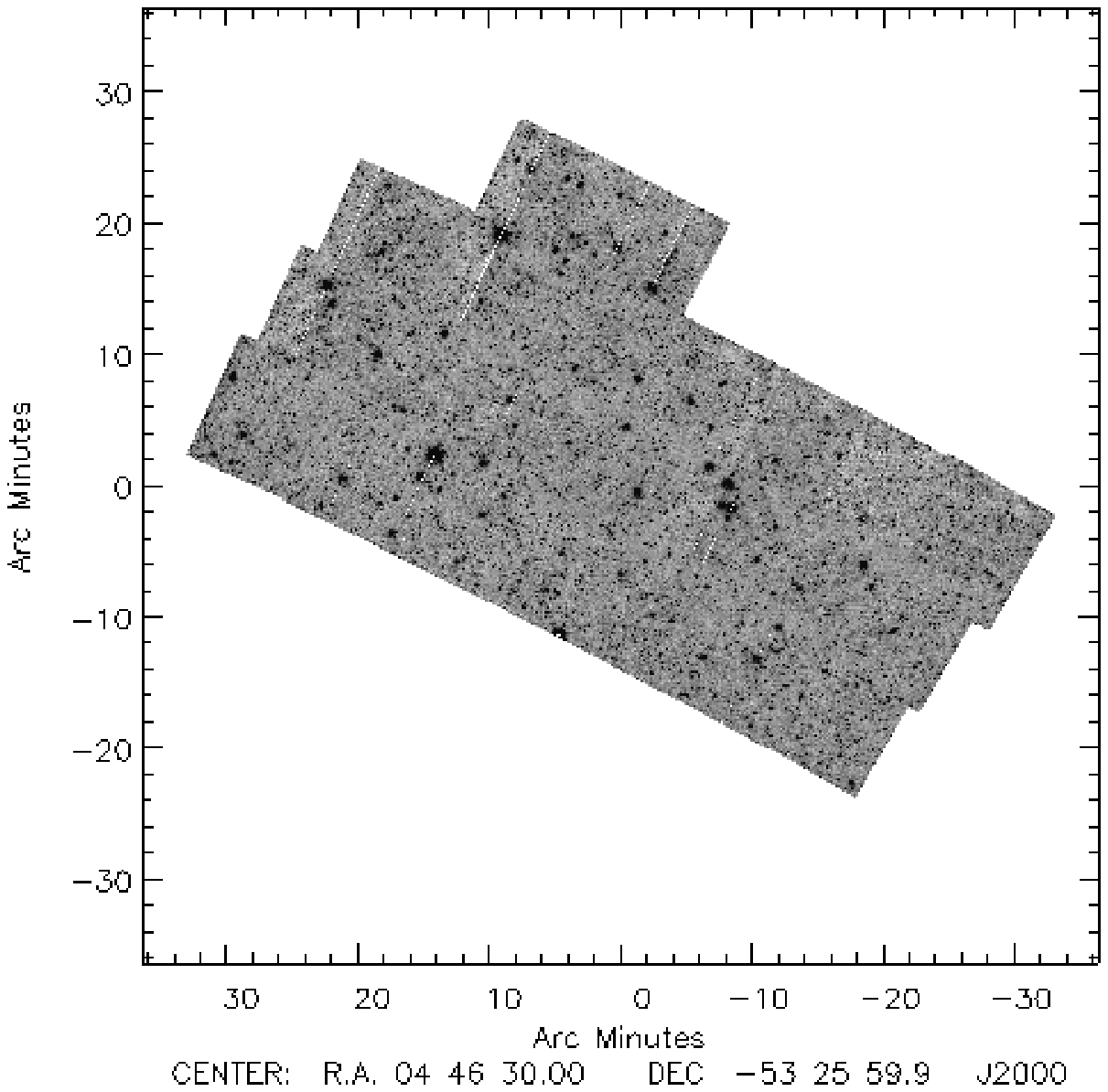} \\
  (a) ADF-S N3 & (b) ADF-S N4 \\[6pt]
  \includegraphics[width=70mm]{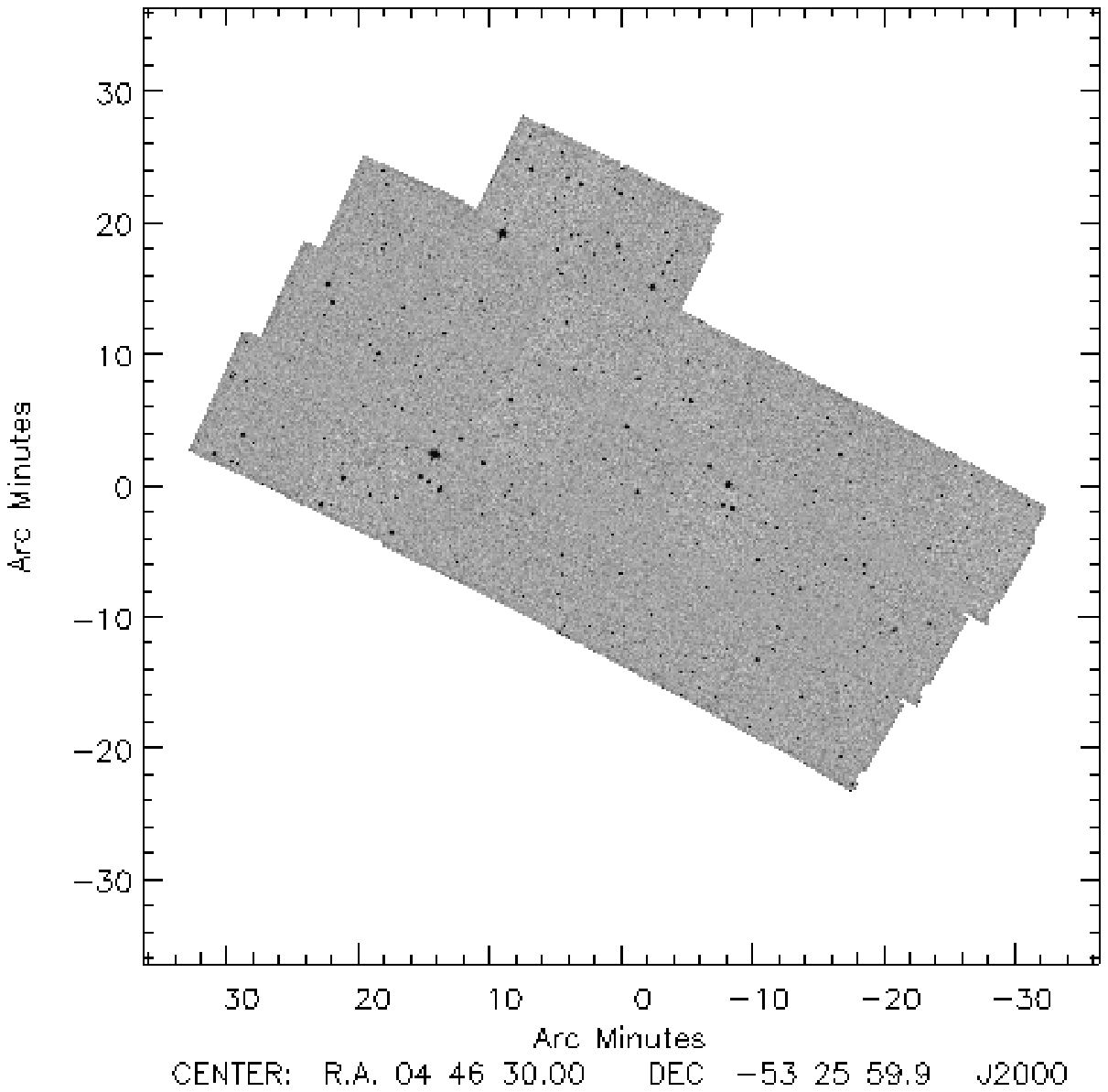} & 
  \includegraphics[width=70mm]{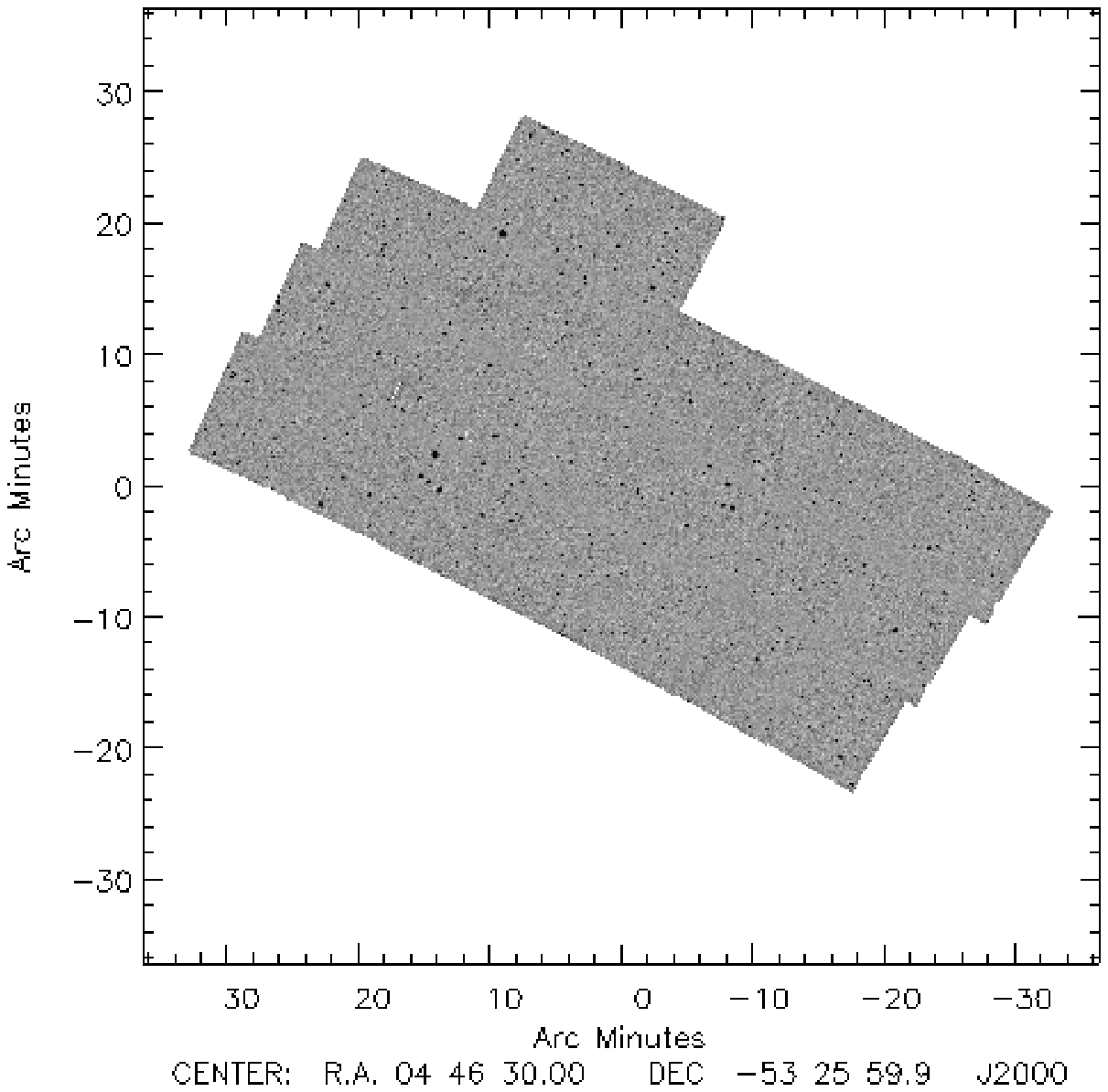} \\
  (c) ADF-S S7 & (d) ADF-S S11 \\[6pt]
    \includegraphics[width=70mm]{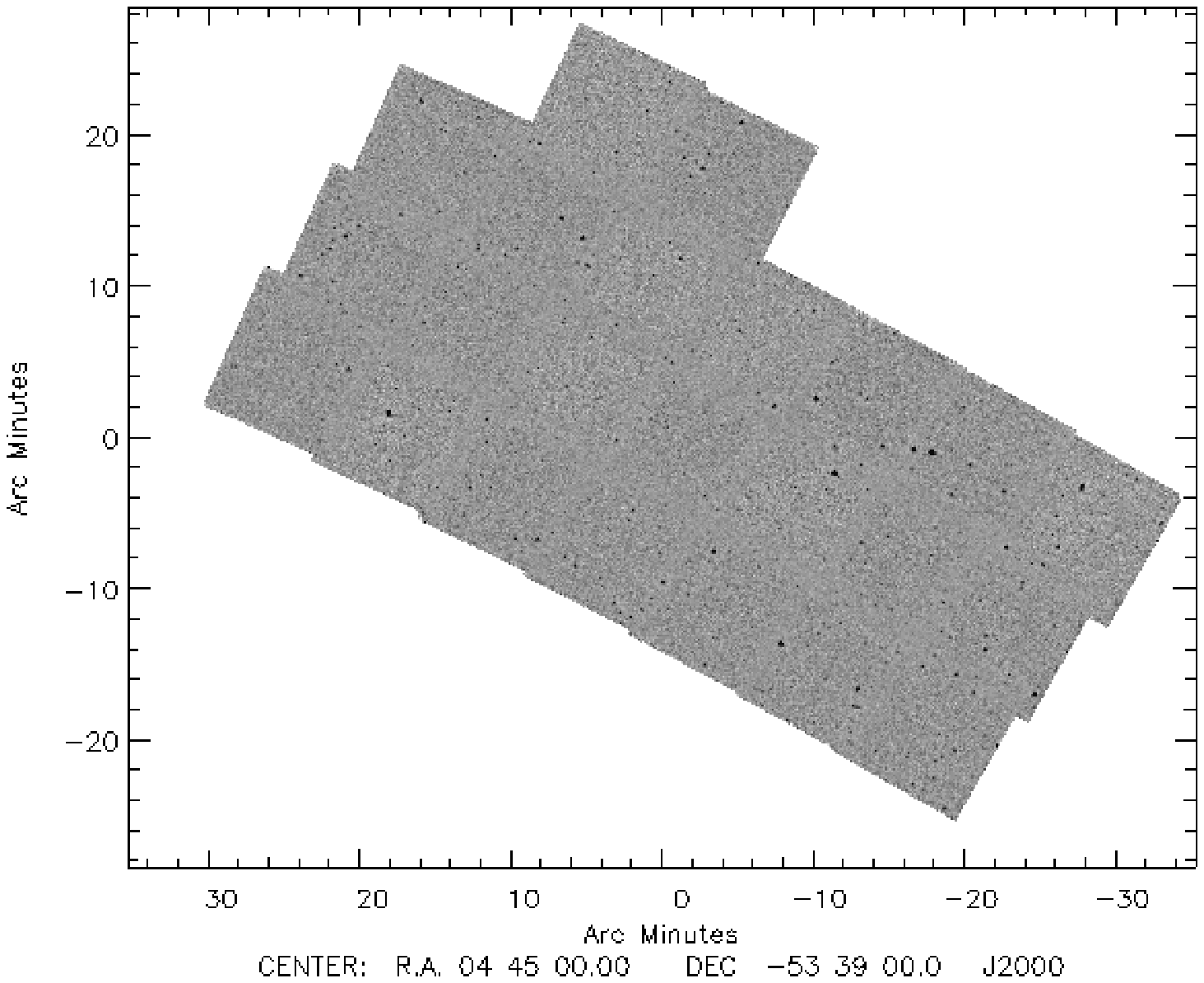} & 
  \includegraphics[width=70mm]{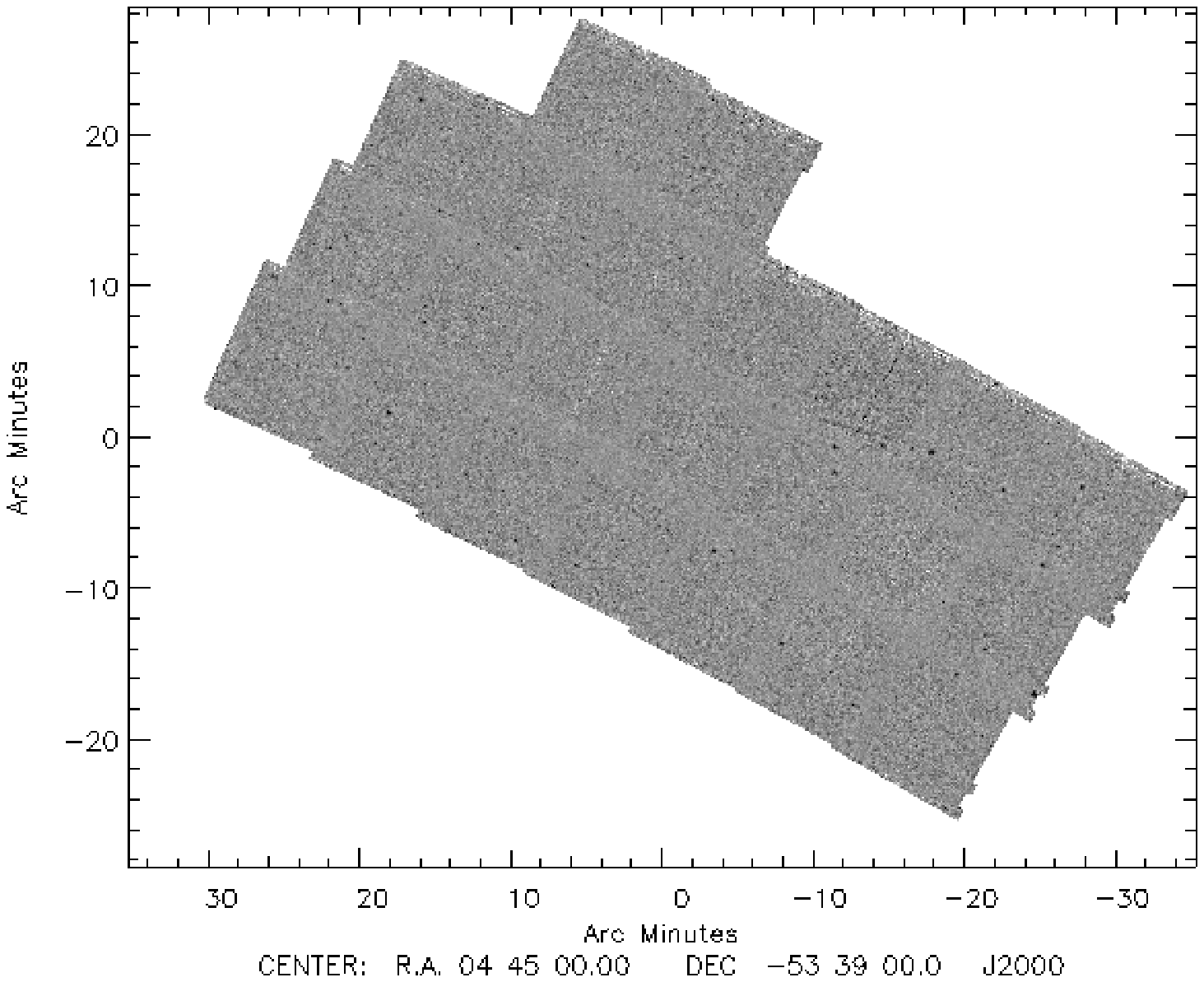} \\
  (e) ADF-S L15 & (f) ADF-S L24 \\[6pt]
\end{tabular}
\caption{The ADF-S mosaicked images.}
\label{adfsmosaickedimages} 
\end{figure*}

Figure \ref{adfsmosaickedimages} shows the mosaicked ADF-S images in the six {\it AKARI}/IRC filters at 3.2, 4.1, 7, 11, 15 and 24$\mu$m. As the L24 filter was the least sensitive of the AKARI/IRC filters and the ADF-S is a shallow field, SExtractor was unable to detect a statistically meaningful number of sources i.e. only fourteen sources. It was decided that further analysis of the ADF-S L24 image with a different source extraction method would not be beneficial, as there already exists a large amount of deep {\it Spitzer}/MIPS 24\,$\mu$m data over the same field area from \cite{Clements2011}. The survey area of the ADF-S is 0.46 square degrees.

\begin{table}
\scriptsize
  \caption{Observation log for ADFS.}
  \label{tab:ADFS}
  \begin{center}
    \begin{tabular}{lllllllll}
      \hline \hline
 Pointing Number&Date&Filter&AOT\\
 \hline
320001-001&03/02/2007&N3 N4 S7 S11 L15 L24&IRC02\\
320002-001&04/02/2007&N3 N4 S7 S11 L15 L24&IRC02\\
320003-001&04/02/2007&N3 N4 S7 S11 L15 L24&IRC02\\
320004-001&05/02/2007&N3 N4 S7 S11 L15 L24&IRC02\\
320005-001&03/02/2007&N3 N4 S7 S11 L15 L24&IRC02\\
320006-001&04/02/2007&N3 N4 S7 S11 L15 L24&IRC02\\
320007-001&05/02/2007&N3 N4 S7 S11 L15 L24&IRC02\\
320008-001&06/02/2007&N3 N4 S7 S11 L15 L24&IRC02\\
320009-001&03/02/2007&N3 N4 S7 S11 L15 L24&IRC02\\
320010-001&05/02/2007&N3 N4 S7 S11 L15 L24&IRC02\\
320011-001&05/02/2007&N3 N4 S7 S11 L15 L24&IRC02\\
320012-001&06/02/2007&N3 N4 S7 S11 L15 L24&IRC02\\
320013-001&09/02/2007&N3 N4 S7 S11 L15 L24&IRC02\\
320014-001&08/02/2007&N3 N4 S7 S11 L15 L24&IRC02\\
320015-001&07/02/2007&N3 N4 S7 S11 L15 L24&IRC02\\
320016-001&07/02/2007&N3 N4 S7 S11 L15 L24&IRC02\\
320017-001&09/02/2007&N3 N4 S7 S11 L15 L24&IRC02\\
320018-001&08/02/2007&N3 N4 S7 S11 L15 L24&IRC02\\
320019-001&07/02/2007&N3 N4 S7 S11 L15 L24&IRC02\\
320021-001&09/02/2007&N3 N4 S7 S11 L15 L24&IRC02\\
320042-001&07/02/2007&N3 N4 S7 S11 L15 L24&IRC02\\
320046-001&08/02/2007&N3 N4 S7 S11 L15 L24&IRC02\\
320050-001&05/02/2007&N3 N4 S7 S11 L15 L24&IRC02\\
    \hline
    \end{tabular}
  \end{center}
\end{table}

\begin{table}
\scriptsize
  \caption{ADF-S discarded frames.}
  \label{tab:adfsdiscarrdedframes}
  \begin{center}
    \begin{tabular}{lllllllll}
      \hline \hline
 Filter&Pointing Number&Frame Number&Reason\\
 \hline
 N4&3200002\_001&F006033902\_N002&Many hot pixels\\
 S7&3200004\_001&F006038908\_S004&Flux error\\
 L15&3200003\_001&F006036697\_L003&Cosmic ray\\
 L15&3200003\_001&F006036697\_L004&Cosmic ray\\
 L15&3200004\_001&F006038908\_L004&Flux error\\
 L24&3200003\_001&F006036693\_L004&Many hot pixels\\
 L24&3200003\_001&F006036699\_L002&Cosmic ray\\
 L24&3200003\_001&F006036699\_L003&Cosmic ray\\
 L24&3200003\_001&F006036699\_L004&Cosmic ray\\
 L24&3200006\_001&F006034162\_L002&Artificial stripe pattern\\
 L24&3200006\_001&F006034162\_L003&Artificial stripe pattern\\
 L24&3200006\_001&F006034162\_L004&Artificial stripe pattern\\
    \hline
    \end{tabular}
  \end{center}
\end{table}

\section{Results}
\label{results2}
\subsection{Source extraction and photometry}
\label{source_extraction}
Source extraction was made using SExtractor \citep{BertinandArnouts1996} on the mosaicked images, to create the source catalogues for the IRAC Dark Field, ELAIS-N1 and ADF-S fields.
The parameters for SExtractor were based on those used in \cite{Murata2013} for the {\it AKARI} North Ecliptic Pole (NEP) survey. The main differences compared to those used on the NEP field were for the NIR and MIR-L channel images where the SExtractor parameters  DETECT\_MINAREA and DETECT\_THRESH were set to 5 pixels and 3$\sigma$ respectively, and for the MIR-S channel image where they were set to 5 pixels and 3.5$\sigma$ respectively. These settings were optimised to have the lowest number of false source detections (see Section~\ref{reliability}) while maintaining the highest number of true source detections. 
SExtractor aperture photometry was used to measure the extracted source flux, using the flux conversion from analogue to digital units (ADU) to Janskys from \cite{Tanabe2008} and aperture corrections from \cite{Arimatsu2011}, both listed in Table \ref{tab:fluxconandapaconver}.

\begin{table}
  \caption{Table showing the flux conversion and aperture corrections for the filters used in the IRAC Dark Field, ELAIS-N1 and ADF-S fields.}
  \label{tab:fluxconandapaconver}
  \begin{center}
    \begin{tabular}{lllllllll}
      \hline \hline
 Filter&Flux Conversion&Aperture Correction\\
    &N$_{ADU}$ to Jy&\\
 N3&0.4394 $\times$ 10$^{-6}$&0.873\\
 N4 IRC03 &0.2584 $\times$ 10$^{-6}$&0.871\\
 N4 IRC05&0.1753 $\times$ 10$^{-6}$&0.871\\
 S7&1.0220 $\times$ 10$^{-6}$&0.918\\
 S11&0.7732 $\times$ 10$^{-6}$& 0.902\\
 L15&1.6910 $\times$ 10$^{-6}$&0.852\\
 L18W&1.1460 $\times$ 10$^{-6}$&0.793\\
 L24&4.8920 $\times$ 10$^{-6}$&0.685\\
 \hline
    \hline
    \end{tabular}
  \end{center}
\end{table}

\subsection{Reliability}
\label{reliability}
To test the reliability of the source extraction, a negative image was created for each mosaicked image during the co-adding stage. SExtractor was then applied to this negative image using the same settings. Table \ref{tab:reliability} shows the percentage of sources detected in the negative image to the source detections in the original image in order to provide a measure of the possible number of false detections. Reliability corrections were performed as a function of source flux, where the number of extracted negative sources was greater than 1\%. The large number of spurious sources in all of the IRAC Dark Field S11 and ELAIS-N1 MIR images is attributed to Earthshine light not having been fully removed.

\begin{table}
  \caption{Table showing the percentage of sources extracted from the negative image compared to the total number of sources in the original image, as a measure of the reliability of the source extraction.}
  \label{tab:reliability}
  \begin{center}
    \begin{tabular}{lllllllll}
      \hline \hline
 Field&Filter&Negative sources (\%)\\
 \hline
 IRAC Dark Field&N4&0.21\\
 	&S11&5.05\\
	&L15&2.67\\
	&L18W&0\\ 
ELAIS-N1&N4&0\\
      &S11&9.48\\
      &L15&2.42\\
      &L18W&6.15\\
ADF-S&N3&0.03\\
      &N4&0.06\\
      &S7&0\\
      &S11&3.57\\
      &L15&0\\
      &L24&0\\
    \hline
    \end{tabular}
  \end{center}
\end{table}

\subsection{Completeness}
\begin{figure}
\centering
\includegraphics[width=80mm]{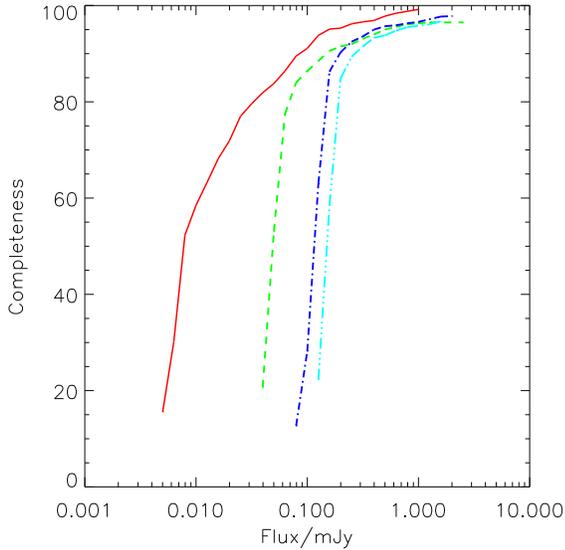}
\caption{The completeness curves for IRAC Dark Field. The completeness for N4 is the red solid line, the completeness for S11 is the green dashed line, the completeness for L15 is the blue dot-dashed line, and the completeness for L18W is the cyan dot-dot-dot-dashed line.}
\label{fig:irac_completeness}
\end{figure}

\begin{figure}
 \centering
 \includegraphics[width=85mm]{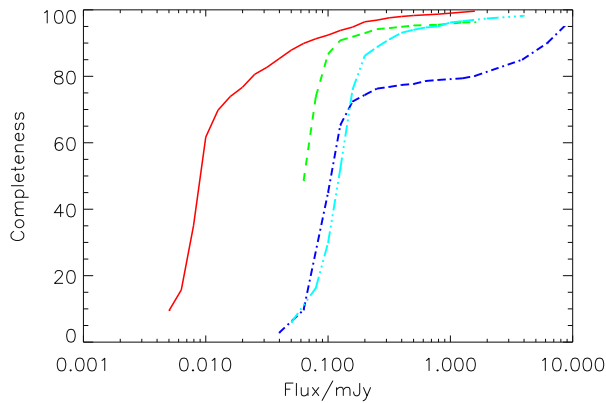}
 \caption{The completeness correction curve for ELAIS-N1. The completeness for N4 is the red solid line, the completeness for S11 is the green dashed line, the completeness for L15 is the blue dot-dashed line, and the completeness for L18W is the cyan dot-dot-dot-dashed line.}
 \label{fig:elais_completeness_correction}
\end{figure}

\begin{figure}
 \centering
 \includegraphics[width=85mm]{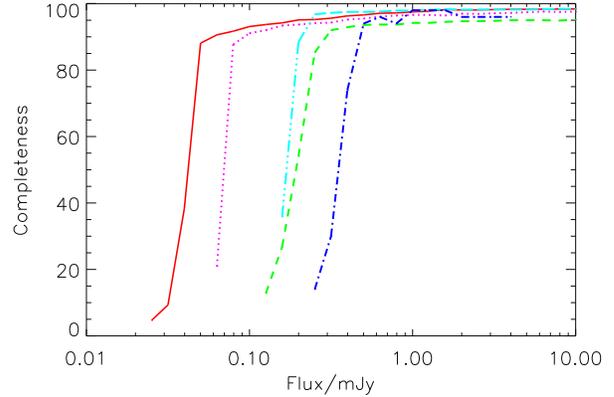}
 \caption{The completeness correction curve for ADF-S. The completeness for N3 is the magenta dotted line, the completeness for N4 is the red solid line, the completeness for S7 is the cyan dot-dot-dot-dashed line, the completeness for S11 is the green dashed line and the completeness for L15 is the blue dot-dashed line.}
 \label{fig:adfs_completeness_correction}
\end{figure}

To correct for sources missed by the source extraction process (completeness), Monte Carlo simulations were used to inject artificial sources of known location and flux into each mosaicked image. After a single artificial source had been injected into an image, the image was run through SExtractor. This was repeated 1000 times for each flux bin for the deep IRAC Dark Field and ELAIS-N1 field in each channel. After performing an initial test of completeness, it was found that the NIR band ADF-S images required much less of a completeness correction that the other bands and fields. It was decided 1000 simulations were not required for the ADF-S N3 and N4 images, whereas it was necessary for the other images. Therefore the ADF-S NIR band images each had 100 simulations and the ADF-S MIR-S and L band images had 1000 simulations. Figures \ref{fig:irac_completeness}, \ref{fig:elais_completeness_correction} and \ref{fig:adfs_completeness_correction} show the completeness curves (the ratio of successfully extracted sources to total number of sources) as a function of input source flux for the IRAC Dark Field, ELAIS-N1 and ADF-S source counts.

The completeness for the deep IRAC Dark Field field drops off steeply with the 50$\%$ completeness levels in the N4, S11, L15 and L18W bands being 0.008, 0.04, 0.105 and 0.107 mJy respectively. Whereas the completeness for the deep ELAIS-N1 field drops off gradually with the 50$\%$ completeness levels in the N4, S11, L15 and L18W bands being 0.009, 0.06, 0.12 and 0.10 mJy respectively. There is not obvious reason for the broad shoulder at around 80$\%$ in the completeness curve for the L15 band in ELAIS-N1 field, and there are likely to be several factors causing this. One possible reason could be that the final image has higher background noise levels than the other images, which could be due to a combination of the filter and detector deterioration (late-Phase 2), or caused by too much smoothing to remove the Earthshine light artefact. Another reason could be related to the small scale noise caused by the confusion in the field, or there may be some structure along the line of sight by change.
The completeness correction for the shallower ADF-S field shows a very steep turnover in all bands at around the 90$\%$ level, with corresponding 50$\%$ completeness levels in the N3, N4, S7, S11 and L15 bands of 0.07, 0.04, 0.18, 0.20 and 0.35 mJy respectively.

\subsection{Stellar Subtraction}
\label{Stellarsubtraction}
Especially in the shorter wavebands of the IRC, the contribution to the source counts from stars can be significant and must be removed. Stellar subtraction was performed using catalogues at optical wavelengths to identify stars in the {\it AKARI} fields. Stars in the IRAC Dark Field and ELAIS-N1 images were identified using the Automatic Plate Measuring (APM) machine data from the Palomar Observatory Sky Survey 1 \citep{Maddox1990}. {\it AKARI} sources was crossed matched with the APM star positions, with a search radius of 4 arcseconds. This search radius was chosen as it was representative of the sources' PSF. Matched sources were removed from the source catalogues. On the rare occasion that there were two or more extracted sources within the radius, the extracted source closest to the APM location was removed. Tables \ref{tab:iracstellarfraction} and \ref{tab:elaisstellarfraction} show the IRAC Dark Field and ELAIS-N1 stellar fraction respectively as a function of flux density. For the ADF-S, the star positions were found using the R-band MOSAIC (CITO) data  \citep{Barrufetinprep} where coverage was available. The stellar fraction for each filter was then calculated assuming the same fraction over the entire ADF-S area. Table \ref{tab:adfsstellarfraction} shows the ADF-S stellar fraction as a function of flux density.

\begin{table}
  \caption{Stellar contribution to the source counts as a function of flux density for the IRAC Dark Field.}
  \label{tab:iracstellarfraction}
  \begin{center}
    \begin{tabular}{lllllllll}
      \hline \hline
 Flux / mJy&N4 \%&S11 \%&L15 \%&L18W \%\\
 \hline
0.008&0.0&-&-&-\\
0.010&1.35&-&-&-\\
0.013&0.0&-&-&-\\
0.016&2.86&-&-&-\\
0.020&1.28&-&-&-\\
0.025&1.30&0.00&-&-\\
0.032&1.72&0.00&-&-\\
0.040&3.28&3.33&-&-\\
0.050&4.17&17.39&-&-\\
0.063&3.03&12.90&-&0.00\\
0.079&15.63&9.09&0.0&0.00\\
0.100&7.14&25.00&0.0&0.00\\
0.126&27.78&0.0&3.03&0.00\\
0.158&30.77&18.75&0.0&4.00\\
0.200&16.67&25.00&6.67&0.00\\
0.251&50.00&8.33&3.03&3.13\\
0.316&55.56&14.29&5.56&0.00\\
0.398&80.00&20.00&0.00&0.00\\
0.501&66.67&20.00&0.00&16.67\\
0.631&66.67&25.00&0.00&0.00\\
0.794&100.00&0.00&0.0&0.00\\
1.000&50.00&0.00&0.00&0.00\\
1.259&-&0.00&-&-&\\ 
1.585&100.00&-&0.00&0.00\\
     \hline
    \end{tabular}
  \end{center}
\end{table}

\begin{table}
  \caption{Stellar contribution to the source counts as a function of flux density for the ELAIS-N1 field.}
  \label{tab:elaisstellarfraction}
  \begin{center}
    \begin{tabular}{lllllllll}
      \hline \hline
 Flux&N4&S11&L15&L18W\\
 / mJy&\%&\%&\%&\%\\
 \hline
0.008&0.0&-&-&-\\
0.010&0.0&-&-&-\\
0.013&0.0&-&-&-\\
0.016&1.32&-&-&-\\
0.020&0.0&-&-&-\\
0.025&2.60&-&-&-\\
0.032&3.45&-&-&-\\
0.040&3.33&-&0.0&-\\
0.050&8.00&-&-&0.0\\
0.063&7.69&00&0.0&-\\
0.079&6.25&11.37&-&0.0\\
0.100&14.29&24.36&0.0&0.0\\
0.126&13.64&4.89&0.0&0.0\\
0.158&11.11&7.99&0.0&0.0\\
0.200&46.15&0.0&4.93&0.0\\
0.251&42.86&6.64&0.0&12.50\\
0.316&71.43&19.57&0.0&0.0\\
0.398&75.00&0.0&0.0&0.0\\
0.501&80.00&0.0&33.333\\
0.631&83.33&31.72&0.0\\
0.794&0.0&0.0&0.0&0.0\\
1.000&0.0&-&-&0.0\\
1.259&-&0.0&0.0&0.0\\
1.585&-&0.0&0.0&-\\
1.995&0.0&-&0.0\\
3.981&100.00&-&-&0.0\\
     \hline
    \end{tabular}
  \end{center}
\end{table}

\begin{table}
  \caption{Stellar contribution to the source counts as a function of flux density for the ADF-S field.}
  \label{tab:adfsstellarfraction}
  \begin{center}
    \begin{tabular}{lllllllll}
      \hline \hline
 Flux&N3&N4&S7&S11&L15\\
  / mJy&\%&\%&\%&\%&\%\\
 \hline
 0.025&-& 0.0&-&-&-\\
 0.032&-&0.0&-&-&-\\
 0.040&-&0.0&-&-&-\\
 0.050&-& 0.0&-&-&-\\
 0.063&0.0& 0.0&-&-&-\\
 0.080&0.0&0.25&-&-&-\\
 0.100&0.0&3.10&-&-&-\\
 0.126&0.0&5.30&-&0.0&-\\
 0.158&0.33&7.81& 0.0&0.0&-\\
 0.200&1.30&22.82& 29.91&8.44&-\\
 0.251&4.02&18.76& 21.19&8.44&0.0\\
 0.316&13.07&38.99&18.31&10.39&0.0\\
 0.398&19.50&41.42&31.22&11.33&0.0\\
 0.501&18.80&71.20&41.74&11.00&3.85\\
 0.631&37.21&68.42&53.63&11.52&12.16\\
 0.794&49.37&84.21&65.28&14.44&18.37\\
 1.000&59.48&84.21&61.74&15.31&12.90\\
 1.259&76.91&90.95&81.35&22.03&18.18\\
 1.585&80.07&90.23&74.51&34.23& 53.85\\
 1.995&86.16&63.16&48.42&50.66& 50.00\\
 2.512&88.77&92.63&86.44&37.25& 50.00\\
 3.162&96.09&89.16&90.79&29.80&50.00\\
 3.981&81.46&-& 81.72&75.99& 33.33\\
 5.012&75.20&84.21&98.87& 70.37&100.00\\
 6.310&100.00&78.95&80.89&46.06&100.00\\
 7.943&95.84&100.00&84.74&50.66&100.00\\
 10.000&83.55&100.00&95.34&54.28&100.00\\
 12.589&100.00&63.16&76.27&100.00&100.00\\
 15.849&96.41&63.16&42.37&100.00&100.00\\
 19.953&83.55&-&95.33&0.0&-\\
 25.119&62.66&100.00&84.74&100.00&100.00\\
 31.623&62.66&-&100.00&100.00&-\\
 39.811&100.00&-&-&-&-\\
 63.096&-&-&100.00&-&-\\
 158.489&-&-&100.00&-&-\\
     \hline
    \end{tabular}
  \end{center}
\end{table}

\subsection{Galaxy Source Counts and Catalogues}
\label{final_galaxy_counts}
\begin{figure*}
\begin{tabular}{cc}
  \includegraphics[width=70mm]{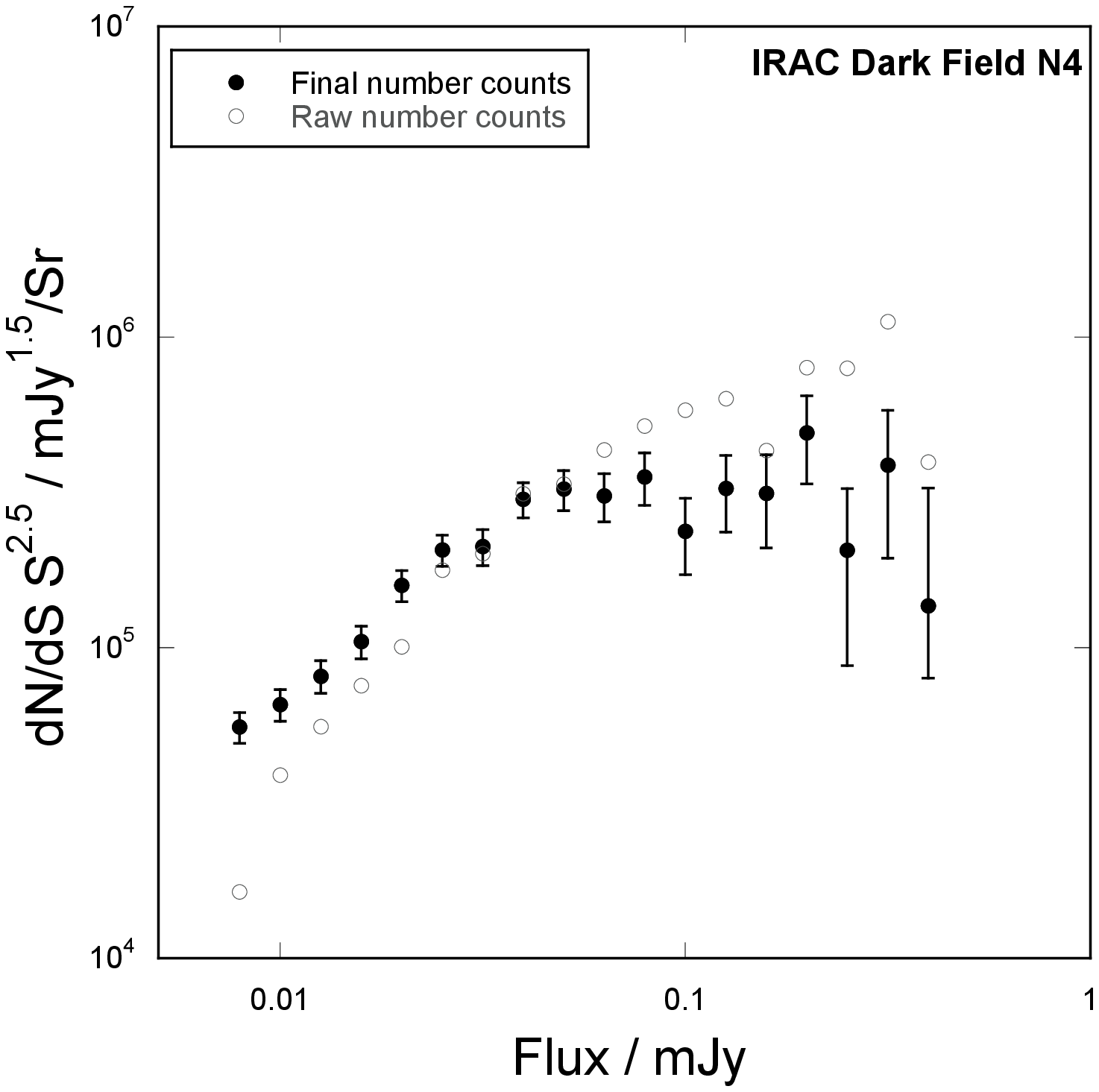} & 
  \includegraphics[width=70mm]{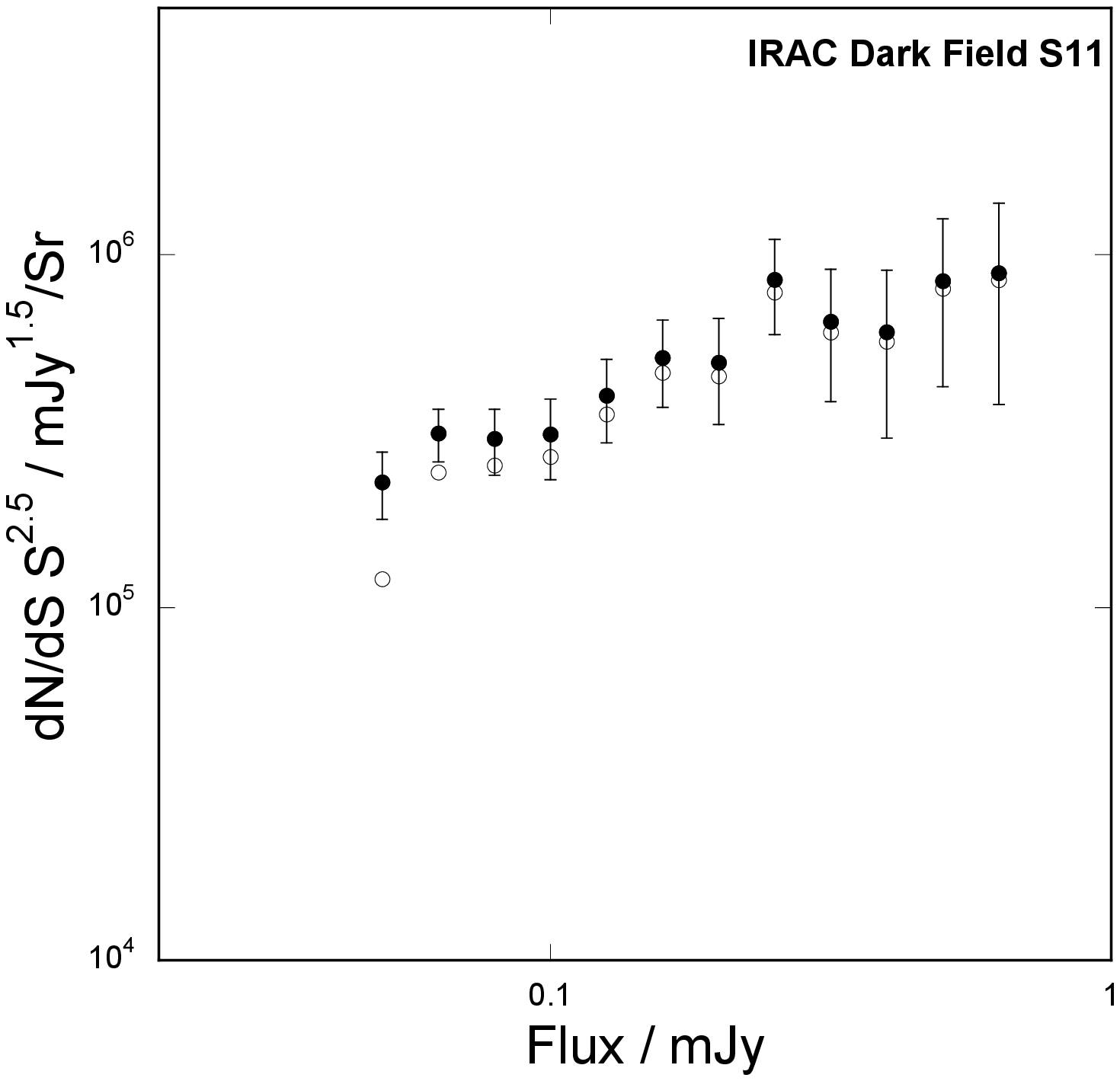} \\
  (a) N4 Filter & (b) S11 Filter \\[6pt]
  \includegraphics[width=70mm]{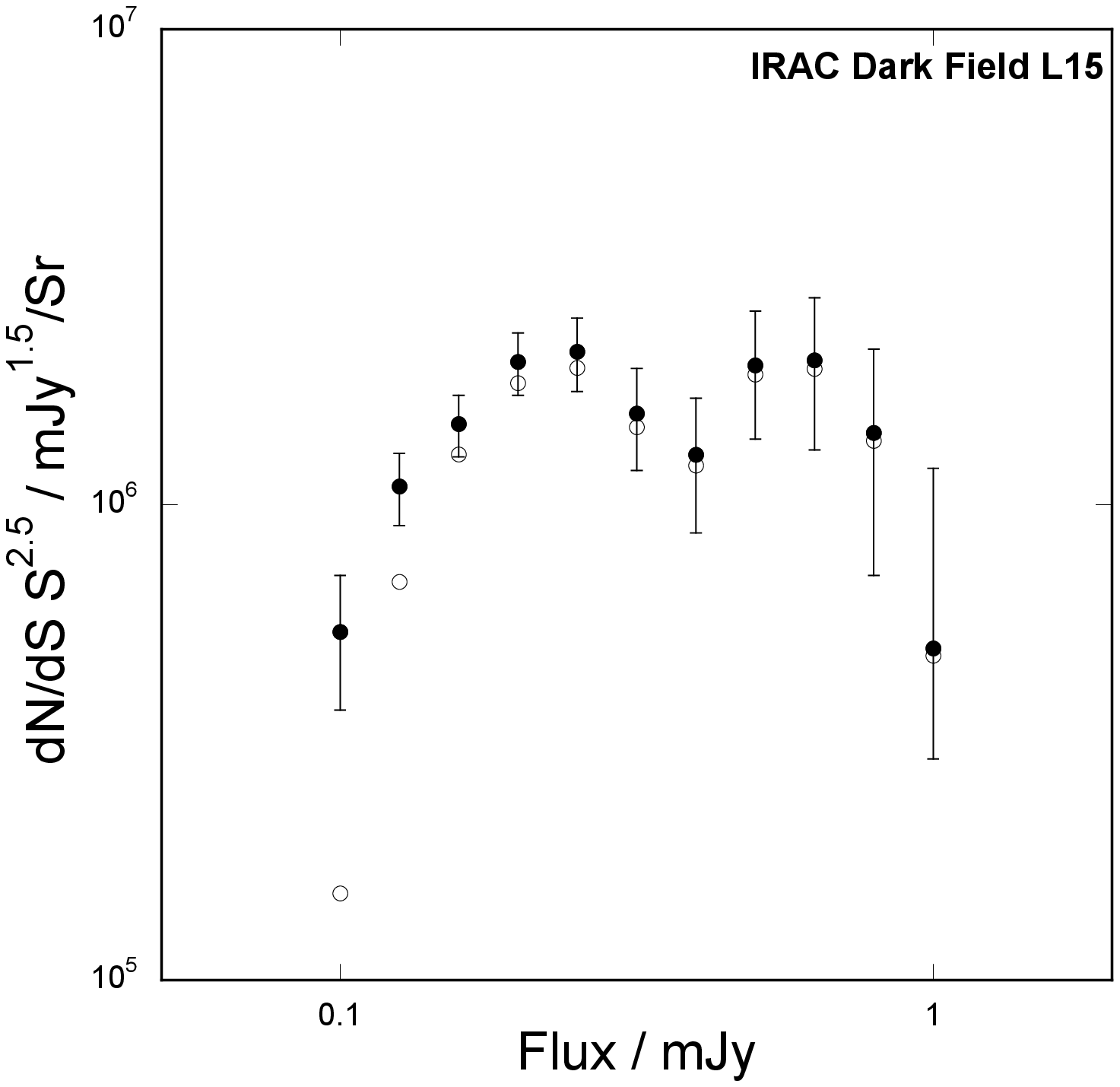} & 
  \includegraphics[width=70mm]{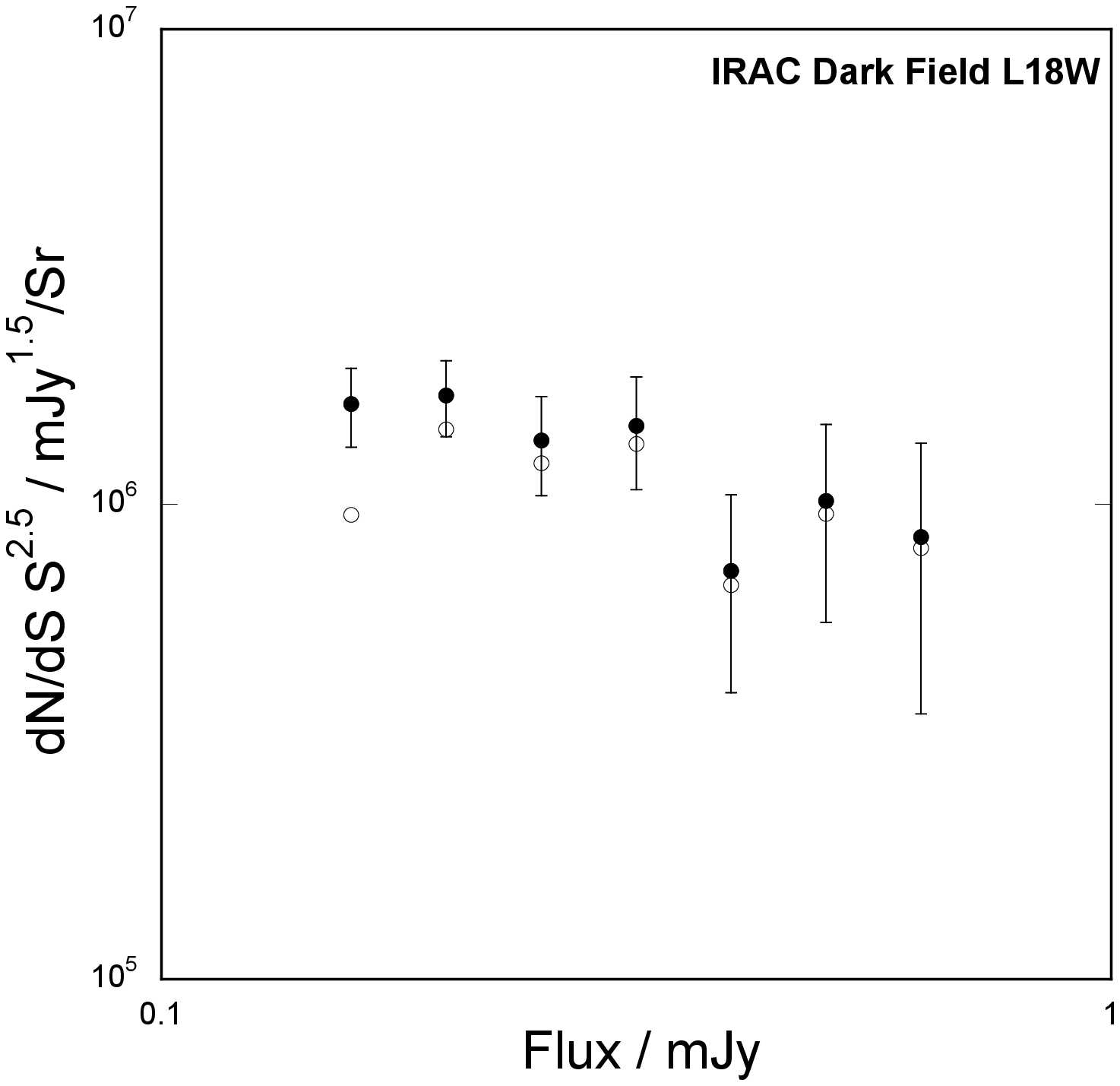} \\
  (c) L15 Filter & (d) L18 Filter\\[6pt]
\end{tabular}
\caption{IRAC Dark Field Euclidean source counts per unit area. The open circles are the raw source counts and the filled circles give the completeness and reliability corrected, stellar subtracted galaxy source counts. Note for image clarity, only the final source counts have their associated errors in the graphs.}
\label{iracnumbercounts} 
\end{figure*}

\begin{figure*}
\begin{tabular}{cc}
  \includegraphics[width=65mm]{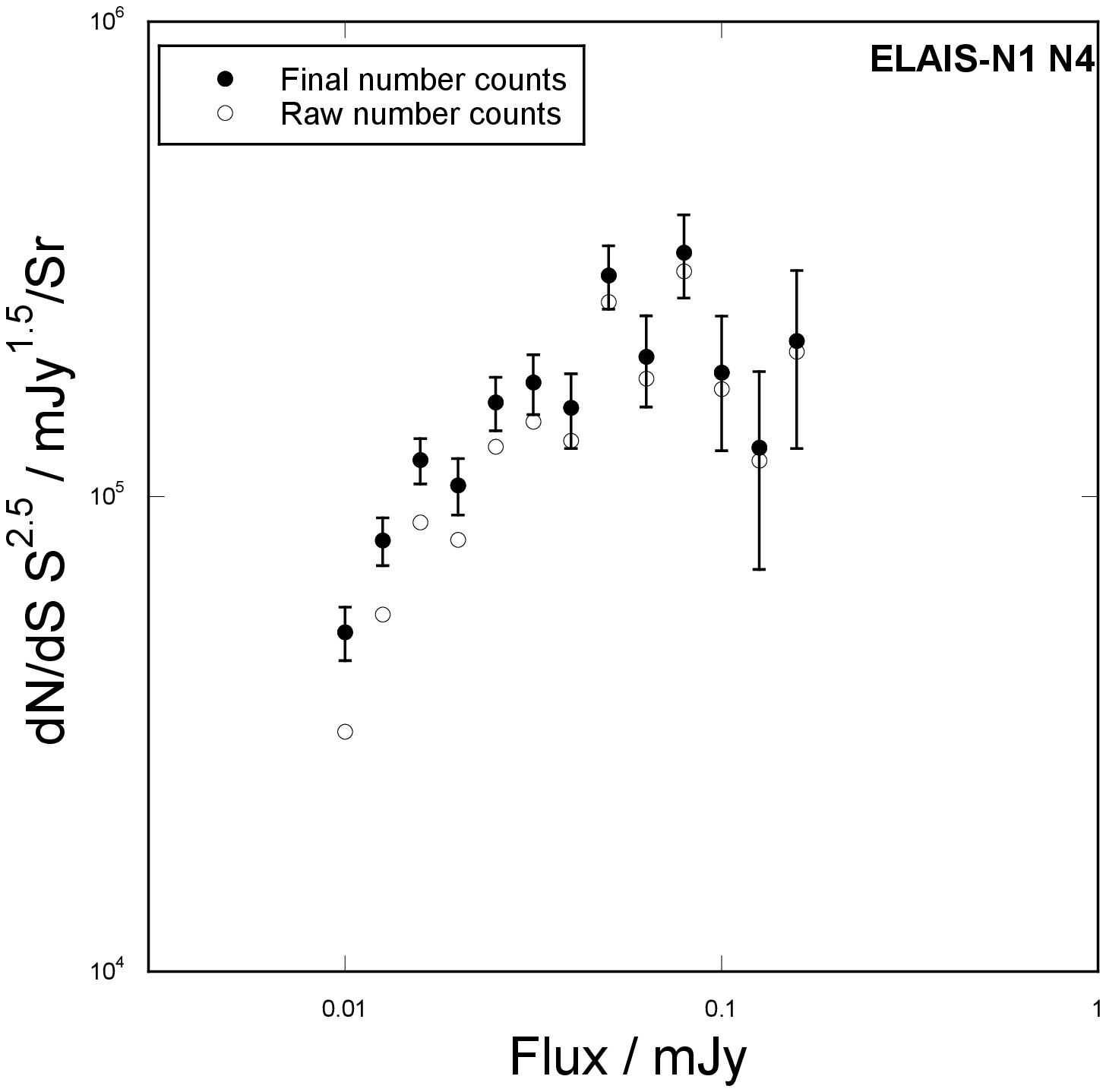} & 
  \includegraphics[width=65mm]{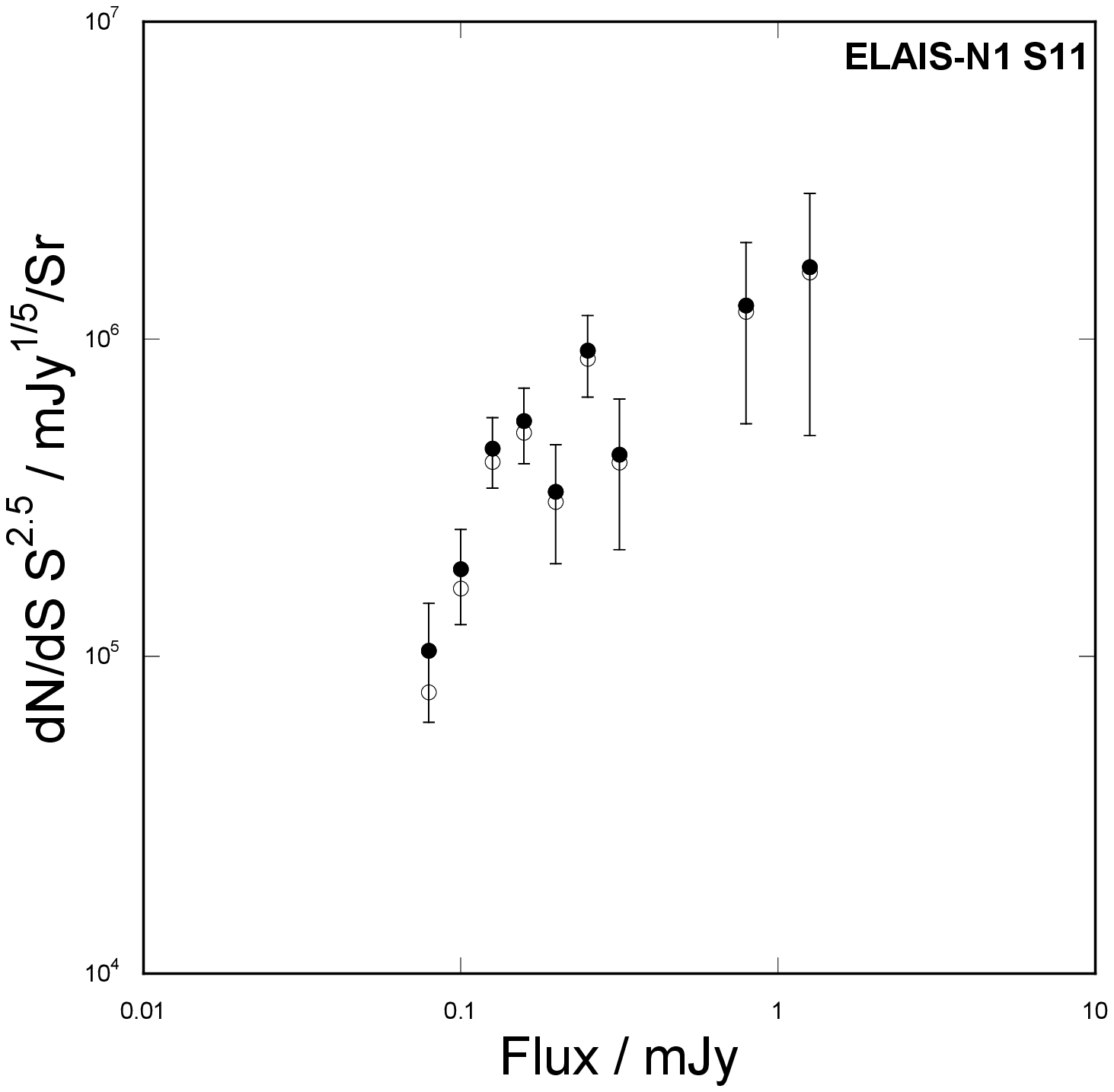} \\
  (a) ELAIS North N4 source counts & (b) ELAIS North S11 source counts \\[6pt]
  \includegraphics[width=65mm]{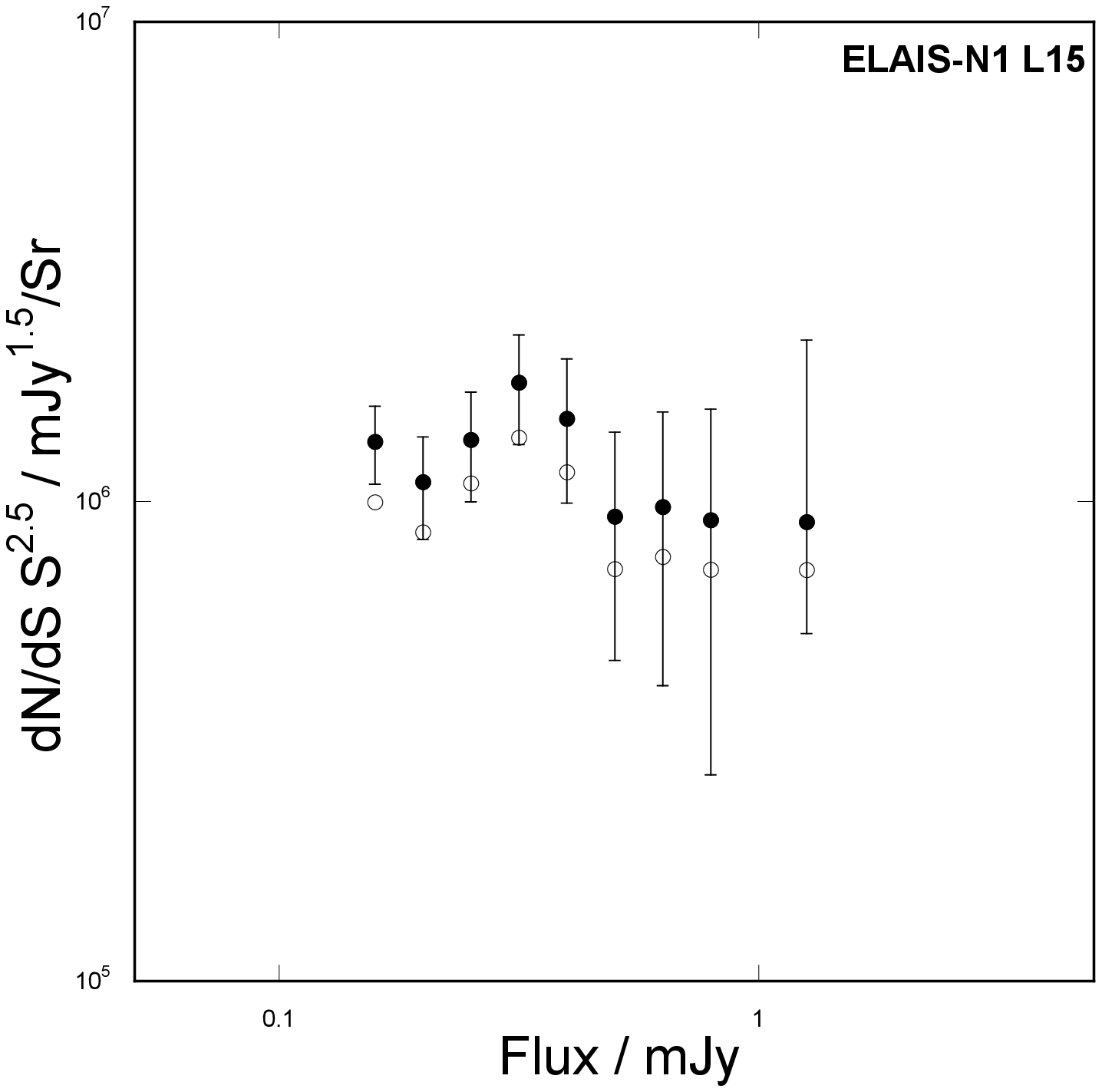} & 
  \includegraphics[width=65mm]{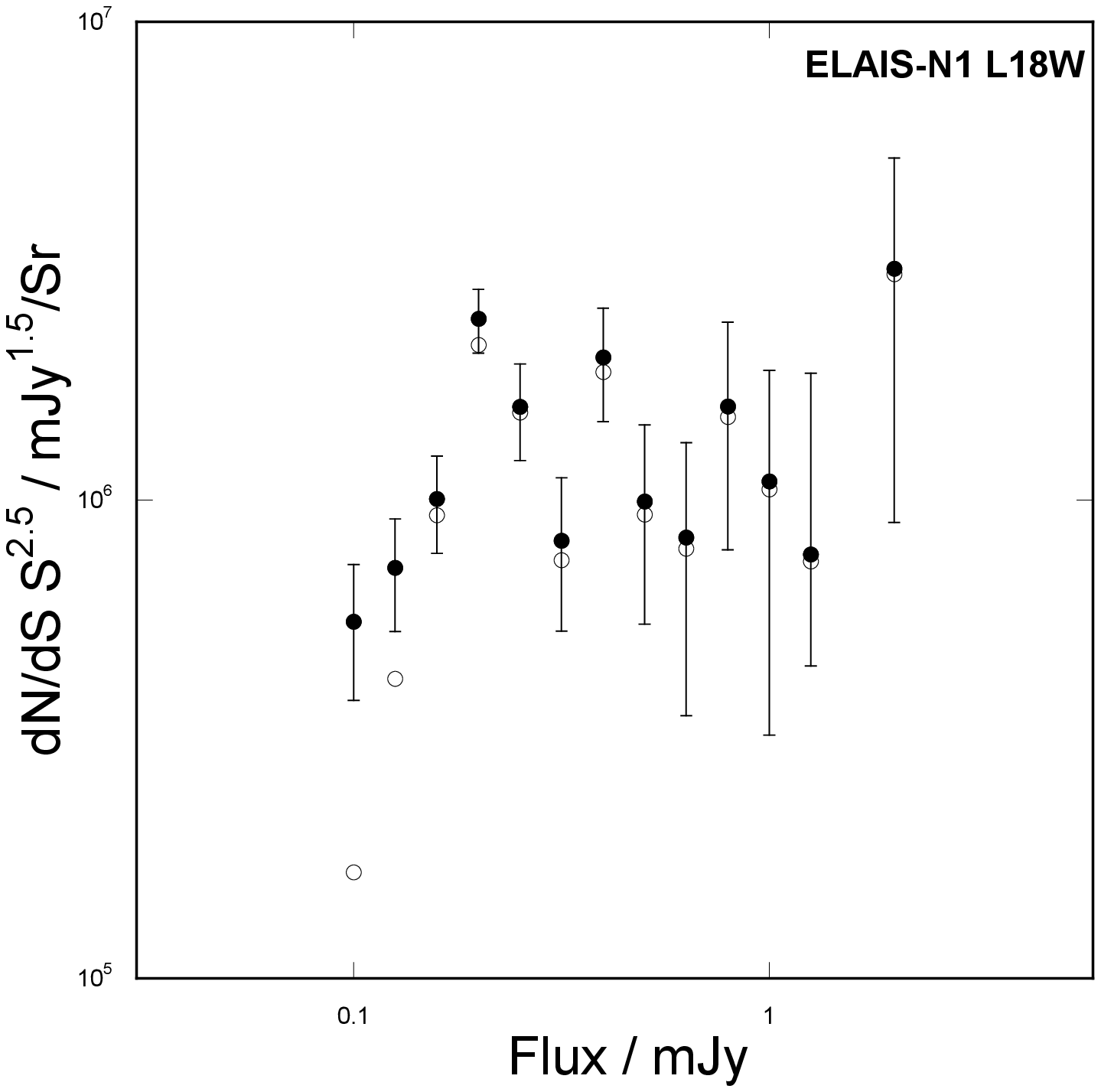} \\
  (c) ELAIS North L15 source counts & (d) ELAIS North L18 source counts \\[6pt]
\end{tabular}
\caption{The ELAIS-N1 source counts. The open circles give the raw counts, and the filled circles give the completeness and reliability corrected, stellar subtracted galaxy source counts. Note for image clarity, only the final source counts have their associated errors in the graphs.}
\label{elaisnumbercounts} 
\end{figure*}

\begin{figure*}
\begin{tabular}{cc}
  \includegraphics[width=65mm]{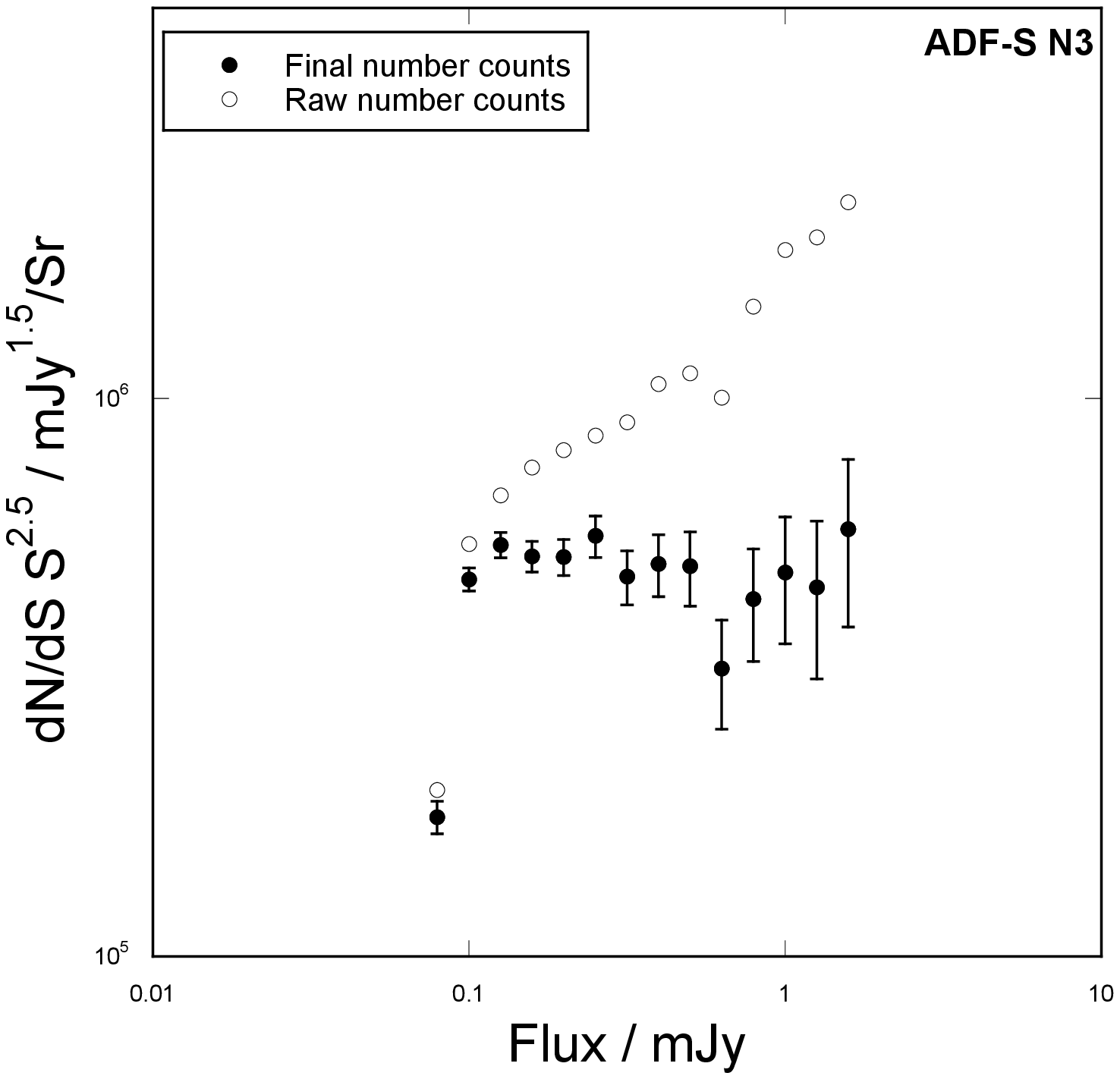}  
  \includegraphics[width=65mm]{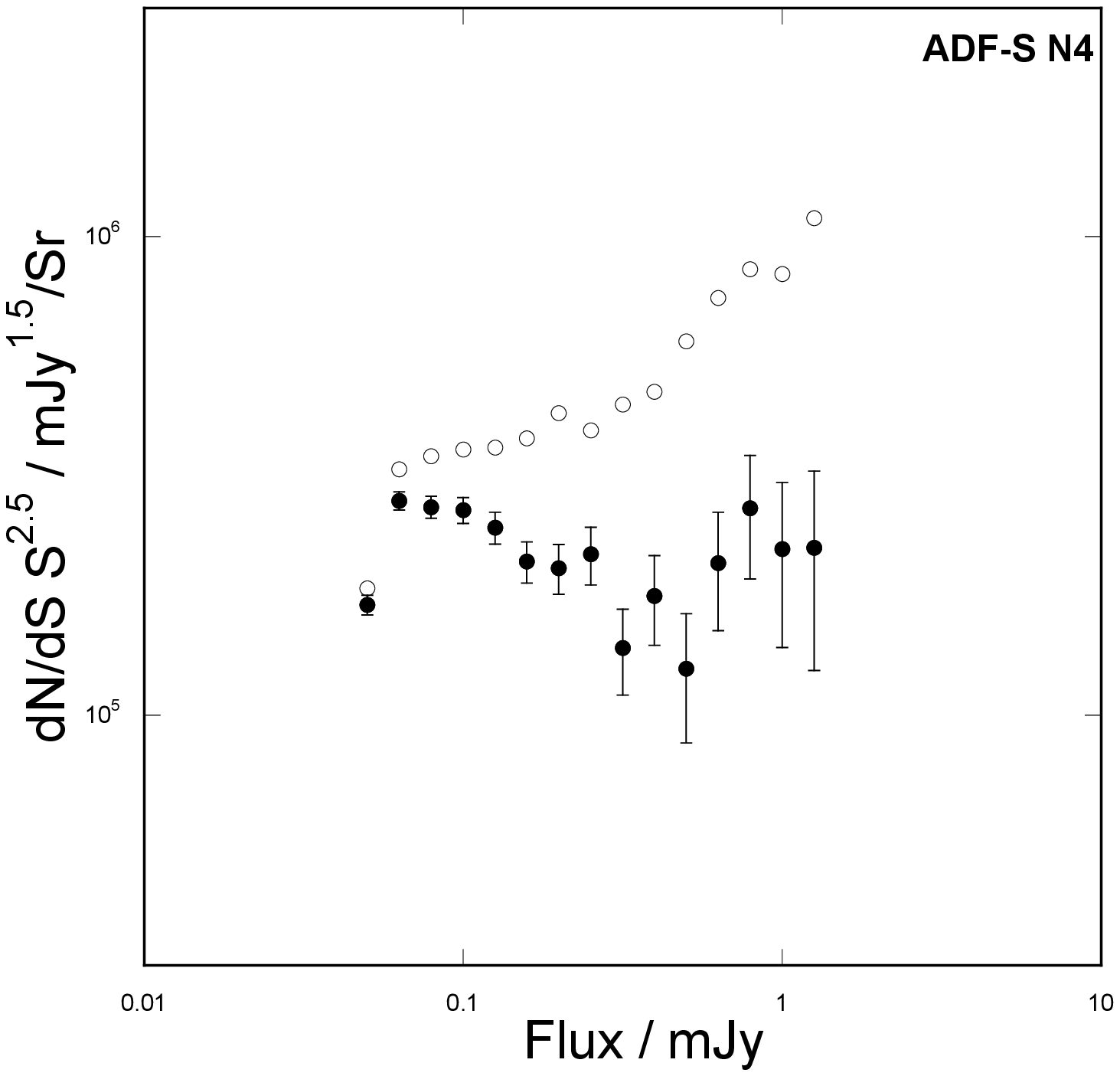} \\
  (a) ADF-S N3 number counts       (b) ADF-S N4 number counts \\[6pt]
  \includegraphics[width=65mm]{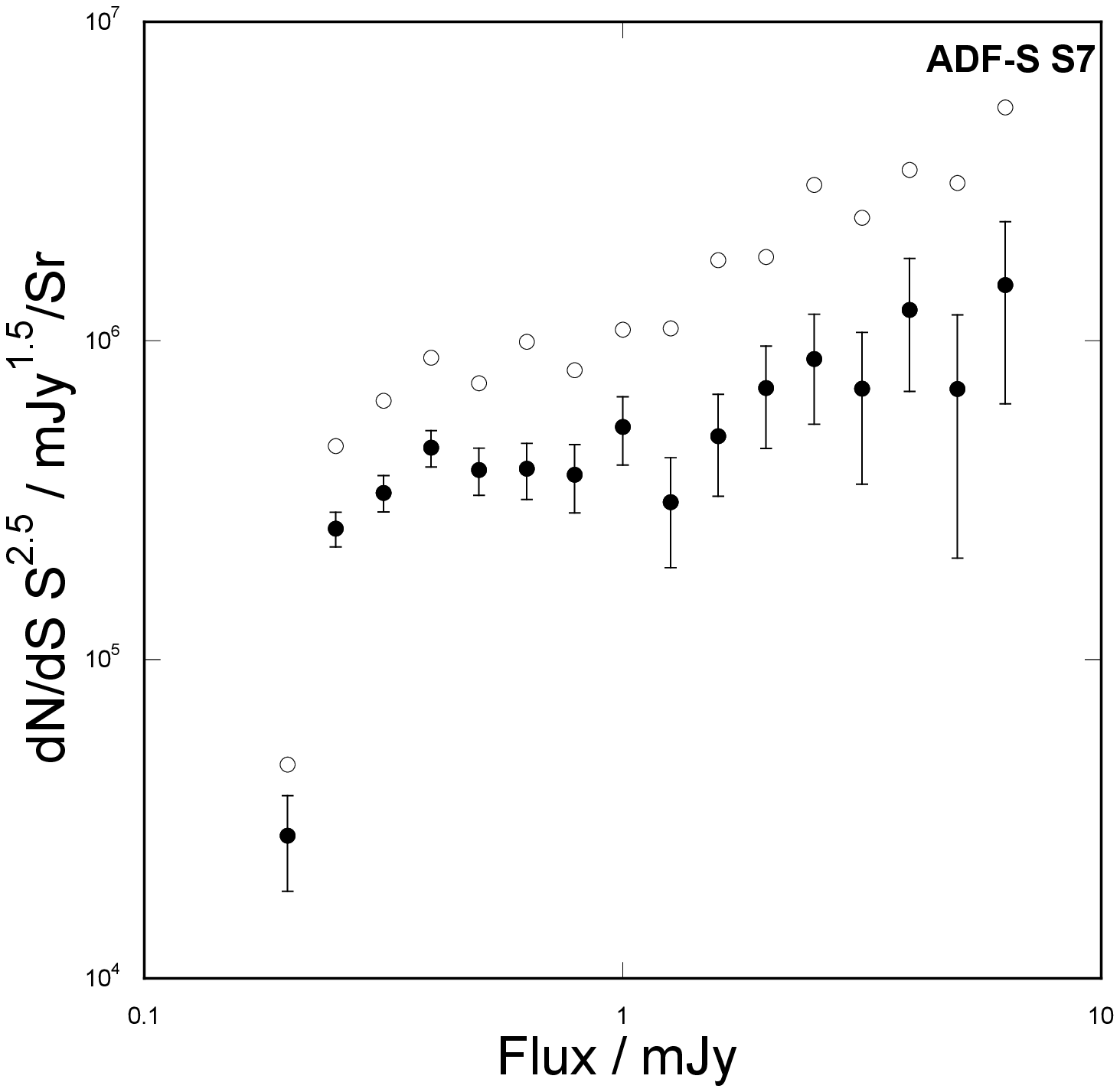}  
  \includegraphics[width=65mm]{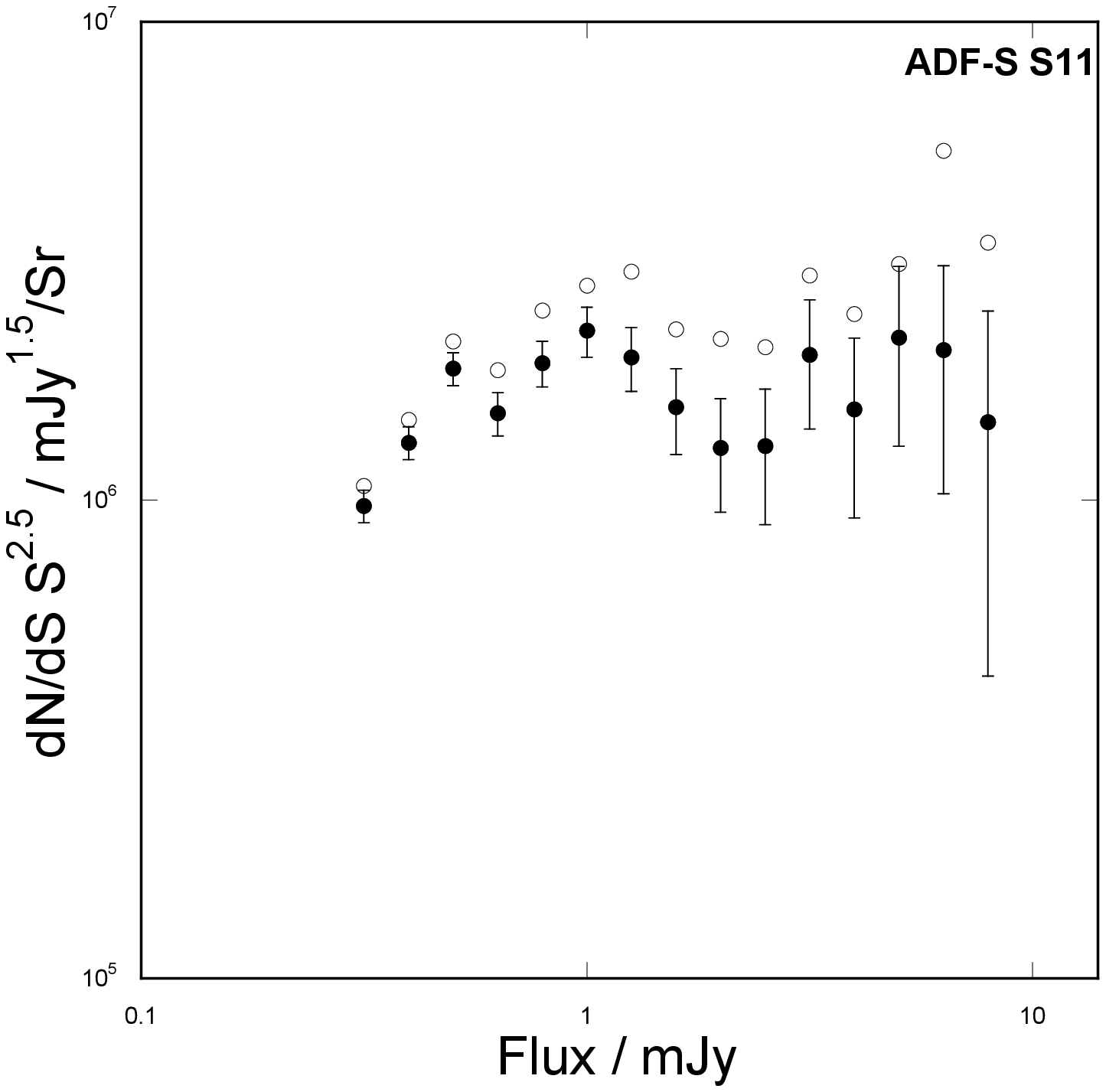} \\
  (c) ADF-S S7 number counts       (d) ADF-S S11 number counts \\[6pt]
    \includegraphics[width=65mm]{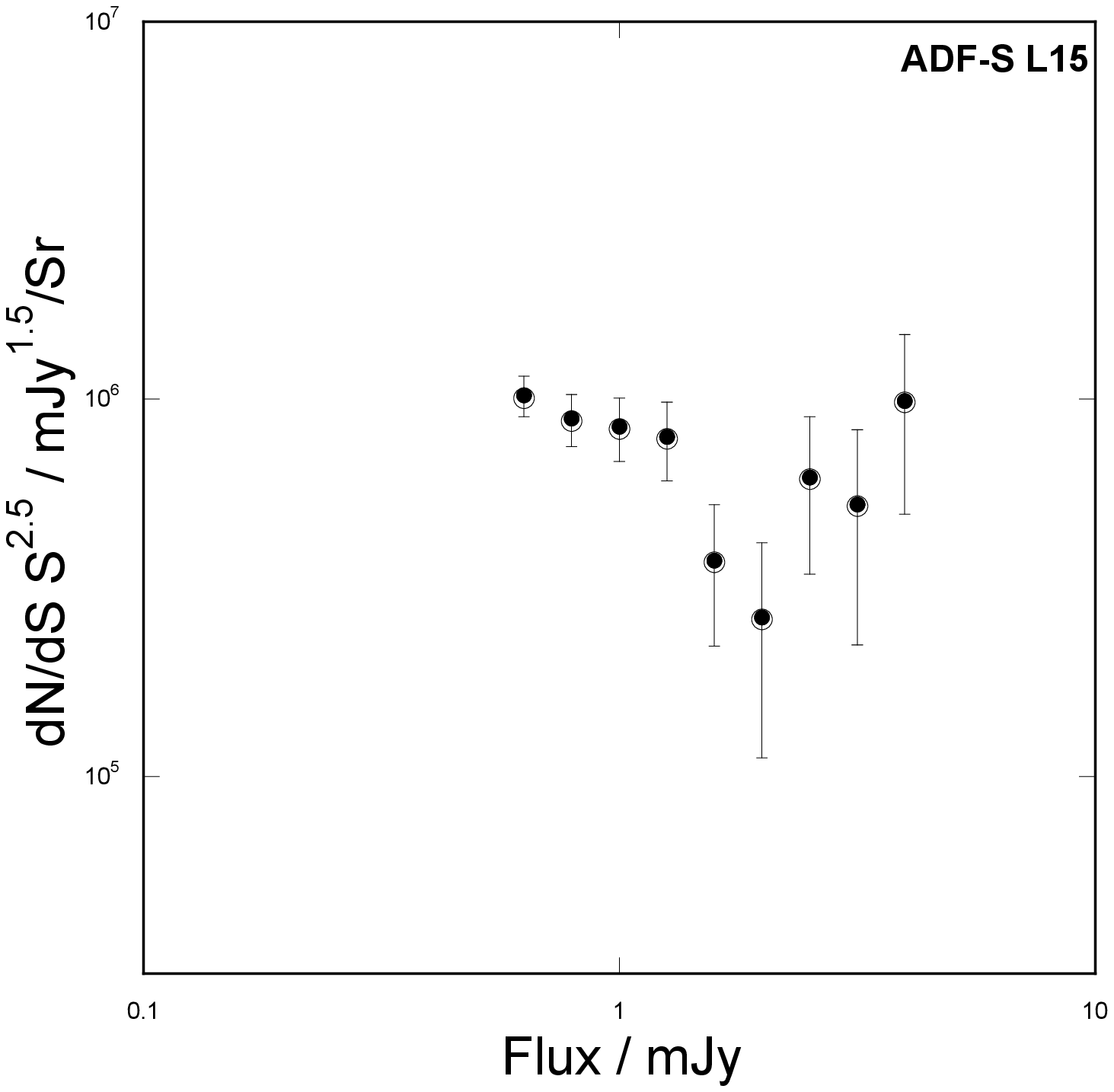}\\
  (e) ADF-S L15 number counts\\[6pt]
\end{tabular}
\caption{The ADF-S number counts. The open circles give the raw counts, and the filled circles give the completeness and reliability corrected, stellar subtracted galaxy source counts. Note the L15 raw and final source counts are very similar. Also note for image clarity, only the final source counts have their associated errors in the graphs.}
\label{adfsnumbercounts} 
\end{figure*}

The final normalised differential source counts per steradian as a function of flux density ($S$) as $(dN/dS) S^{2.5}$ are plotted in Figure \ref{iracnumbercounts}, Figure~\ref{elaisnumbercounts} and Figure~\ref{adfsnumbercounts} for the IRAC Dark Field, ELAIS-N1 and ADF-S fields respectively. The counts have been corrected for completeness, reliability and stellar contributions. The source counts for each band are also tabulated in Tables \ref{tab:iracnumbercountsN4} to \ref{tab:iracnumbercountsL18W} for IRAC Dark Field, Tables \ref{tab:elaisnumbercountsN4} to \ref{tab:elaisnumbercountsL18} for ELAIS-N1, and Tables \ref{tab:adfsnumbercountsN3} to \ref{tab:adfsnumbercountsL15} for the ADF-S field. 

The uncertainty in the source counts, ${\sigma}(N)$, was calculated using Equation~\ref{uncertaintyofF};
\tiny
\begin{equation}
{\sigma}(N)=\sqrt{N_{obs}}\Bigg[\bigg(\frac{1}{f_{com}}\bigg)^2+N_{obs}\Bigg(\bigg({\Delta}f_{com}\frac{1}{f_{com}^2}\bigg)^2\Bigg)\Bigg]^{\frac{1}{2}}\times\frac{S^{2.5}}{{\Delta}S\times A}
\label{uncertaintyofF}
\end{equation}
\normalsize
where $N_{obs}$ is the number of galaxies in a given flux bin centred on flux $S$, $f_{com}$ is the completeness fraction, and $A$ is the area of the survey. For the work of the paper ${\Delta}f_{com}$ (the standard deviation of $f_{com}$) is given by $\frac{\sqrt{f_{com}(1-f_{com})}}{T}$, where $T$ is the total number of injected sources for the flux bin in question.The reliability is assumed to be negligible, and thus an uncertainty term for reliability is not included in the equation. When calculating the uncertainty for the case of small number statistics of a single galaxy in a flux bin, the Poisson term is replaced by $1_{-0.827}^{+0.95}$ \citep{Gehrels1986}.

The {\it AKARI}/IRC galaxy catalogues of the IRAC Dark Field, ELAIS-N1 and ADF-S are available for public download \footnote{ftp://cdsarc.u-strasbg.fr/pub/cats/J/MNRAS/472/4259/}. Table \ref{table:galaxycatexample} gives an example of the IRAC Dark Field galaxy catalogue.

\begin{table}
\footnotesize
  \caption{Final normalised differential source counts, $dN/dS  S^{2.5}$ presented logarithmically as $\rm{mJy^{1.5}/sr}$ for the N4 band in the IRAC Dark Field, with associated errors also presented logarithmically, reliability and completeness.}
  \label{tab:iracnumbercountsN4}
    \begin{tabular}{lllllllll}
      \hline \hline
Log$_{10}$ Flux&Flux&Log$_{10}$ Final&Log$_{10}$&Rel&Com\\
/mJy&/mJy&Counts& Error&\%&\%\\
 \hline
-2.10&0.008&4.74&0.05&100.0&52.3\\
-2.00&0.010&4.82&0.05&100.0&58.5\\
-1.90&0.013&4.91&0.05&100.0&63.3\\
-1.80&0.016&5.02&0.05&100.0&68.2\\
-1.70&0.020&5.20&0.05&100.0&71.9\\
-1.60&0.025&5.32&0.05&100.0&77.0\\
-1.50&0.032&5.37&0.06&100.0&79.6\\
-1.40&0.040&5.48&0.06&100.0&81.9\\
-1.30&0.050&5.51&0.06&100.0&83.8\\
-1.20&0.063&5.49&0.08&100.0&86.4\\
-1.10&0.079&5.55&0.08&100.0&89.5\\
-1.00&0.100&5.38&0.12&100.0&91.1\\
-0.90&0.126&5.51&0.12&100.0&93.8\\
-0.80&0.158&5.50&0.14&100.0&95.1\\
-0.70&0.200&5.69&0.14&100.0&95.3\\
-0.60&0.251&5.32&0.25&100.0&96.2\\
-0.50&0.316&5.59&0.22&100.0&96.6\\
-0.40&0.398&5.13&0.18&100.0&96.9\\
-0.30&0.501&5.28&0.18&100.0&97.8\\
-0.20&0.631&5.43&0.18&100.0&98.4\\
    \hline
    \end{tabular}
\end{table}

\begin{table}
\footnotesize
  \caption{Final normalised differential source counts, $dN/dS  S^{2.5}$ presented logarithmically as $\rm{mJy^{1.5}/sr}$  for the S11 band in the IRAC Dark Field, with associated errors also presented logarithmically, reliability and completeness.}
  \label{tab:iracnumbercountsSS4}
    \begin{tabular}{lllllllll}
      \hline \hline
Log$_{10}$ Flux&Flux&Log$_{10}$ Final&Log$_{10}$&Rel&Com\\
/mJy&/mJy&Counts& Error&\%&\%\\
 \hline
-1.30&0.050&5.36&0.09&89.46&53.1\\
-1.20&0.063&5.49&0.07&77.62&77.5\\
-1.10&0.079&5.48&0.09&89.06&84.0\\
-1.00&0.100&5.49&0.11&100.0&86.4\\
-0.90&0.126&5.60&0.12&100.0&88.5\\
-0.80&0.158&5.71&0.12&100.0&90.6\\
-0.70&0.200&5.69&0.14&100.0&91.5\\
-0.60&0.251&5.93&0.13&100.0&92.0\\
0.50&0.316&5.81&0.18&100.0&93.1\\
-0.40&0.398&5.78&0.22&100.0&94.0\\
-0.30&0.501&5.93&0.22&100.0&95.0\\
-02.0&0.631&5.95&0.25&100.0&95.6\\
-0.10&0.79&5.62&0.18&100.0&96.1\\
0.00&1.000&5.77&0.18&100.0&96.3\\
0.01&1.259&6.22&0.31&100.0&96.5\\
    \hline
    \end{tabular}
\end{table}

\begin{table}
\footnotesize
  \caption{Final normalised differential source counts, $dN/dS  S^{2.5}$ presented logarithmically as $\rm{mJy^{1.5}/sr}$  for the L15 band in the IRAC Dark Field, with associated errors also presented logarithmically, reliability and completeness.}
  \label{tab:iracnumbercountsL15}
    \begin{tabular}{lllllllll}
      \hline \hline
Log$_{10}$ Flux&Flux&Log$_{10}$ Final&Log$_{10}$&Rel&Com\\
/mJy&/mJy&Counts& Error&\%&\%\\
 \hline
-0.90&0.126&6.04&0.08&95.06&63.0\\
-0.80&0.158&6.17&0.06&89.51&86.3\\
-0.70&0.200&6.30&0.07&93.47&90.3\\
-0.60&0.251&6.32&0.08&100.0&92.5\\
-0.50&0.316&6.19&0.11&100.0&93.6\\
-0.40&0.398&6.11&0.14&100.0&95.0\\
-0.30&0.501&6.29&0.13&100.0&95.7\\
-0.20&0.631&6.30&0.15&100.0&95.9\\
-0.10&0.794&6.15&0.22&100.0&96.3\\
0.00&1.000&5.70&0.18&100.0&96.6\\
0.20&1.585&5.99&0.18&100.0&97.7\\
    \hline
    \end{tabular}
\end{table}

\begin{table}
\footnotesize
  \caption{Final normalised differential source counts, $dN/dS  S^{2.5}$ presented logarithmically as $\rm{mJy^{1.5}/sr}$  for the L18W band in the IRAC Dark Field, with associated errors also presented logarithmically, reliability and completeness.}
  \label{tab:iracnumbercountsL18W}
    \begin{tabular}{lllllllll}
      \hline \hline
Log$_{10}$ Flux&Flux&Log$_{10}$ Final&Log$_{10}$&Rel&Com\\
/mJy&/mJy&Counts& Error&\%&\%\\
 \hline
-0.80&0.158&6.21&0.08&100.0&58.5\\
-0.70&0.200&6.23&0.08&100.0&84.9\\
-0.60&0.251&6.14&0.10&100.0&89.4\\
-0.50&0.316&6.17&0.12&100.0&91.5\\
-0.40&0.398&5.86&0.19&100.0&93.3\\
-0.30&0.501&6.01&0.19&100.0&93.8\\
-0.20&0.631&5.93&0.25&100.0&94.8\\
-0.10&0.794&5.60&0.18&100.0&95.5\\
0.10&1.259&5.90&0.18&100.0&96.1\\
    \hline
    \end{tabular}
\end{table}

\begin{table}
\footnotesize
  \caption{Final normalised differential source counts, $dN/dS  S^{2.5}$ presented logarithmically as $\rm{mJy^{1.5}/sr}$  for the N4 band in the ELAIS-N1 field, with associated errors also presented logarithmically, reliability and completeness.}
  \label{tab:elaisnumbercountsN4}
    \begin{tabular}{lllllllll}
      \hline \hline
Log$_{10}$ Flux&Flux&Log$_{10}$ Final&Log$_{10}$&Rel&Com\\
/mJy&/mJy&Counts& Error&\%&\%\\
 \hline
-2.00&0.010&4.71& 0.06&100.0&61.7\\
-1.90&0.013&4.91& 0.05&100.0&69.8\\
-1.80&0.016&5.08& 0.05&100.0&73.9\\
-1.70&0.020&5.02& 0.06&100.0&76.7\\
-1.60&0.025&5.20& 0.06&100.0&80.6\\
-1.50&0.032&5.24& 0.06&100.0&82.7\\
-1.40&0.040&5.19& 0.08&100.0&85.3\\
-1.30&0.050&5.47& 0.07&100.0&87.9\\
-1.20&0.063&5.30& 0.09&100.0&89.9\\
-1.10&0.079&5.51& 0.09&100.0&91.3\\
-1.00&0.100&5.26& 0.14&100.0&92.4\\
-0.90&0.126&5.10& 0.19&100.0&93.8\\
-0.80&0.158&5.33& 0.18&100.0&94.8\\
-0.70&0.200&4.99& 0.31&100.0&96.4\\
-0.60&0.251&4.84& 0.18&100.0&96.9\\
-0.50&0.316&4.99& 0.18&100.0&97.6\\
-0.40&0.398&5.14& 0.18&100.0&98.0\\
-0.30&0.501&5.28& 0.18&100.0&98.3\\
-0.10&0.794&5.58& 0.18&100.0&98.7\\
    \hline
    \end{tabular}
\end{table}

\begin{table}
\footnotesize
  \caption{Final normalised differential source counts, $dN/dS  S^{2.5}$ presented logarithmically as $\rm{mJy^{1.5}/sr}$  for the S11 band in the ELAIS-N1 field, with associated errors also presented logarithmically, reliability and completeness.}
  \label{tab:elaisnumbercountsS11}
    \begin{tabular}{lllllllll}
      \hline \hline
Log$_{10}$ Flux&Flux&Log$_{10}$ Final&Log$_{10}$&Rel&Com\\
/mJy&/mJy&Counts& Error&\%&\%\\
 \hline
-1.10&0.079&5.02& 0.18&100.0&73.9\\
-1.00&0.100&5.28& 0.14&100.0&86.7\\
-0.90&0.126&5.66& 0.11&62.2&90.8\\
-0.80&0.158&5.74& 0.12&78.8&91.9\\
-0.70&0.200&5.52& 0.18&100.0&93.0\\
-0.60&0.251&5.96& 0.13&100.0&94.2\\
-0.50&0.316&5.64& 0.22&100.0&94.5\\
-0.40&0.398&5.18& 0.18&100.0&94.9\\
-0.30&0.501&5.33& 0.18&100.0&95.3\\
-0.10&0.794&6.11& 0.25&100.0&95.5\\
 0.10&1.259&6.23& 0.31&100.0&96.2\\
 0.20&1.585&6.08& 0.18&100.0&96.3\\
    \hline
    \end{tabular}
\end{table}

\begin{table}
\footnotesize
  \caption{Final normalised differential source counts, $dN/dS  S^{2.5}$ presented logarithmically as $\rm{mJy^{1.5}/sr}$  for the L15 band in the ELAIS-N1 field, with associated errors also presented logarithmically, reliability and completeness.}
  \label{tab:elaisnumbercountsL15}
    \begin{tabular}{lllllllll}
      \hline \hline
Log$_{10}$ Flux&Flux&Log$_{10}$ Final&Log$_{10}$&Rel&Com\\
/mJy&/mJy&Counts& Error&\%&\%\\
 \hline
-0.90&0.126&5.39&0.16&100.0&65.3\\
-0.80&0.158&6.12&0.08&96.8&72.4\\
-0.70&0.200&6.04&0.10&94.8&74.4\\
-0.60&0.251&6.13&0.11&94.1&76.3\\
-0.50&0.316&6.25&0.11&100.0&76.8\\
-0.40&0.398&6.17&0.14&100.0&77.4\\
-0.30&0.501&5.97&0.22&100.0&77.7\\
-0.20&0.631&5.99&0.25&100.0&78.6\\
-0.10&0.794&5.96&0.31&100.0&78.9\\
 0.10&1.259&5.96&0.18&100.0&79.4\\
 0.20&1.585&6.58&0.25&100.0&80.1\\
    \hline
    \end{tabular}
\end{table}

\begin{table}
\footnotesize
  \caption{Final normalised differential source counts, $dN/dS  S^{2.5}$ presented logarithmically as $\rm{mJy^{1.5}/sr}$  for the L18W band in the ELAIS-N1 field, with associated errors also presented logarithmically, reliability and completeness.}
  \label{tab:elaisnumbercountsL18}
    \begin{tabular}{lllllllll}
      \hline \hline
Log$_{10}$ Flux&Flux&Log$_{10}$ Final&Log$_{10}$&Rel&Com\\
/mJy&/mJy&Counts& Error&\%&\%\\
 \hline
-0.90&0.126&5.86&0.11&89.0&52.1\\
-0.80&0.158&6.00&0.10&82.2&76.0\\
-0.70&0.200&6.38&0.07&97.8&86.2\\
-0.60&0.251&6.20&0.10&91.2&88.8\\
-0.50&0.316&5.92&0.15&100.0&91.1\\
-0.40&0.398&6.30&0.12&100.0&93.1\\
-0.30&0.501&6.00&0.19&100.0&94.0\\
-0.20&0.631&5.92&0.25&100.0&94.7\\
-0.10&0.794&6.20&0.22&100.0&95.0\\
0.00&1.000&6.04&0.31&100.0&96.2\\
0.10&1.259&5.89&0.18&100.0&96.6\\
0.30&1.996&6.48&0.31&100.0&97.4\\
0.60&3.981&6.63&0.18&100.0&98.2\\
    \hline
    \end{tabular}
\end{table}

\begin{table}
\footnotesize
  \caption{Final normalised differential source counts, $dN/dS  S^{2.5}$ presented logarithmically as $\rm{mJy^{1.5}/sr}$  for the N3 band in the ADF-S field, with associated errors also presented logarithmically, reliability and completeness.}
  \label{tab:adfsnumbercountsN3}
    \begin{tabular}{lllllllll}
      \hline \hline
Log$_{10}$ Flux&Flux&Log$_{10}$ Final&Log$_{10}$&Rel&Com\\
/mJy&/mJy&Counts& Error&\%&\%\\
 \hline
-1.10&0.079&5.25& 0.03&100.0&87.7\\
-1.00&0.100&5.67& 0.02&100.0&91.1\\
-0.90&0.126&5.73& 0.02&100.0&92.0\\
-0.80&0.158&5.71& 0.03&100.0&93.4\\
-0.70&0.200&5.71& 0.03&99.9&93.7\\
-0.60&0.251&5.75& 0.04&100.0&94.1\\
-0.50&0.316&5.68& 0.05&100.0&94.3\\
-0.40&0.398&5.72& 0.06&100.0&95.2\\
-0.30&0.501&5.70& 0.07&100.0&95.5\\
-0.20&0.631&5.51& 0.10&100.0&96.1\\
-0.10&0.794&5.64& 0.10&100.0&96.4\\
0.00&1.000&5.69& 0.11&100.0&96.6\\
0.10&1.259&5.66& 0.14&100.0&96.6\\
0.20&1.585&5.77& 0.14&100.0&96.4\\
0.30&1.995&5.96& 0.14&100.0&97.0\\
0.40&2.512&5.89& 0.18&100.0&96.9\\
0.50&3.162&6.10& 0.16&100.0&97.1\\
0.60&3.981&6.19& 0.18&100.0&97.2\\
0.70&5.012&5.86& 0.31&100.0&97.4\\
0.80&6.310&6.01& 0.31&100.0&97.6\\
0.90&7.943&6.33& 0.25&100.0&97.5\\
1.00&10.000&6.01& 0.18&100.0&97.6\\
1.10&12.589&6.16& 0.18&100.0&97.7\\
1.20&15.848&6.31& 0.18&100.0&97.7\\
    \hline
    \end{tabular}
\end{table}

\begin{table}
\footnotesize
  \caption{Final normalised differential source counts, $dN/dS  S^{2.5}$ presented logarithmically as $\rm{mJy^{1.5}/sr}$  for the N4 band in the ADF-S field, with associated errors also presented logarithmically, reliability and completeness.}
  \label{tab:adfsnumbercountsN4}
    \begin{tabular}{lllllllll}
      \hline \hline
Log$_{10}$ Flux&Flux&Log$_{10}$ Final&Log$_{10}$&Rel&Com\\
/mJy&/mJy&Counts& Error&\%&\%\\
 \hline
-1.30&0.050&5.23&0.02&100.0&88.1\\
-1.20&0.063&5.45&0.02&100.0&90.6\\
-1.10&0.079&5.43&0.0&100.0&91.7\\
-1.00&0.100&5.43&0.03&100.0&93.1\\
-0.90&0.126&5.39&0.03&100.0&93.7\\
-0.80&0.158&5.32&0.04&100.0&94.2\\
-0.70&0.200&5.31&0.05&100.0&95.1\\
-0.60&0.251&5.34&0.06&100.0&95.2\\
-0.50&0.316&5.14&0.09&100.0&95.6\\
-0.40&0.398&5.25&0.09&100.0&96.3\\
-0.30&0.501&5.10&0.13&100.0&96.6\\
-0.20&0.631&5.32&0.12&100.0&97.1\\
-0.10&0.794&5.43&0.13&100.0&97.2\\
0.00&1.000&5.35&0.16&100.0&97.5\\
0.10&1.259&5.35&0.19&100.0&97.7\\
0.20&1.585&5.80&0.14&100.0&98.2\\
0.30&1.995&5.43&0.25&100.0&98.1\\
0.40&2.512&5.70&0.22&100.0&98.1\\
0.50&3.162&5.25&0.18&100.0&98.2\\
0.60&3.981&5.70&0.31&100.0&98.3\\
0.70&5.012&5.85&0.31&100.0&98.3\\
1.00&10.000&6.30&0.31&100.0&98.4\\
1.40&25.119&6.60&0.18&100.0&98.4\\
    \hline
    \end{tabular}
\end{table}

\begin{table}
\footnotesize
  \caption{Final normalised differential source counts, $dN/dS  S^{2.5}$ presented logarithmically as $\rm{mJy^{1.5}/sr}$  for the S7 band in the ADF-S field, with associated errors also presented logarithmically, reliability and completeness.}
  \label{tab:adfsnumbercountsS7}
    \begin{tabular}{lllllllll}
      \hline \hline
Log$_{10}$ Flux&Flux&Log$_{10}$ Final&Log$_{10}$&Rel&Com\\
/mJy&/mJy&Counts& Error&\%&\%\\
 \hline
-0.70&0.200&4.45&0.14&100.0&88.4\\
-0.60&0.251&5.41&0.05&100.0&96.8\\
-0.50&0.316&5.52&0.06&100.0&97.2\\
-0.40&0.398&5.67&0.06&100.0&97.5\\
-0.30&0.501&5.60&0.07&100.0&97.5\\
-0.20&0.631&5.60&0.09&100.0&97.6\\
-0.10&0.794&5.58&0.11&100.0&97.9\\
0.00&1.000&5.73&0.11&100.0&97.9\\
0.10&1.259&5.49&0.16&100.0&98.1\\
0.20&1.585&5.70&0.15&100.0&98.3\\
0.30&1.995&5.85&0.15&100.0&98.2\\
0.40&2.512&5.94&0.16&100.0&98.2\\
0.50&3.162&5.85&0.22&100.0&98.2\\
0.60&3.981&6.10&0.19&100.0&98.2\\
0.70&5.012&5.85&0.31&100.0&98.2\\
0.80&6.310&6.18&0.25&100.0&98.3\\
0.90&7.943&5.85&0.18&100.0&98.4\\
1.00&10.000&6.47&0.25&100.0&98.4\\
1.30&19.953&6.45&0.18&100.0&98.5\\
1.70&50.119&7.05&0.18&100.0&98.5\\
    \hline
    \end{tabular}
\end{table}

\begin{table}
\footnotesize
  \caption{Final normalised differential source counts, $dN/dS  S^{2.5}$ presented logarithmically as $\rm{mJy^{1.5}/sr}$  for the S11 band in the ADF-S field, with associated errors also presented logarithmically, reliability and completeness.}
  \label{tab:adfsnumbercountsS11}
    \begin{tabular}{lllllllll}
      \hline \hline
Log$_{10}$ Flux&Flux&Log$_{10}$ Final&Log$_{10}$&Rel&Com\\
/mJy&/mJy&Counts& Error&\%&\%\\
 \hline
-0.80&0.158&4.98&0.12&94.51&54.0\\
-0.70&0.200&5.30&0.07&95.52&85.3\\
-0.60&0.251&5.38&0.06&98.34&92.0\\
-0.50&0.316&5.99&0.03&100.0&92.9\\
-0.40&0.398&6.12&0.03&100.0&93.6\\
-0.30&0.501&6.28&0.03&99.49&93.7\\
-0.20&0.631&6.18&0.05&100.0&93.7\\
-0.10&0.794&6.29&0.05&100.0&94.2\\
0.00&1.000&6.35&0.05&100.0&94.2\\
0.10&1.259&6.30&0.07&100.0&94.5\\
0.20&1.585&6.19&0.09&100.0&94.7\\
0.30&1.995&6.11&0.12&100.0&94.7\\
0.40&2.512&6.11&0.14&100.0&94.8\\
0.50&3.162&6.30&0.13&100.0&95.0\\
0.60&3.981&6.19&0.18&100.0&95.0\\
0.70&5.012&6.34&0.18&100.0&95.0\\
0.80&6.310&6.31&0.22&100.0&94.9\\
0.90&7.943&6.16&0.31&100.0&95.2\\
1.00&10.000&6.61&0.22&100.0&95.1\\
    \hline
    \end{tabular}
\end{table}

\begin{table}
\footnotesize
  \caption{Final normalised differential source counts, $dN/dS  S^{2.5}$ presented logarithmically as $\rm{mJy^{1.5}/sr}$  for the L15 band in the ADF-S field, with associated errors also presented logarithmically, reliability and completeness.}
  \label{tab:adfsnumbercountsL15}
    \begin{tabular}{lllllllll}
      \hline \hline
Log$_{10}$ Flux&Flux&Log$_{10}$ Final&Log$_{10}$&Rel&Com\\
/mJy&/mJy&Counts& Error&\%&\%\\
 \hline
-0.40&0.398&4.72& 0.22&100.0&58.6\\
-0.30&0.501&5.45& 0.09&100.0&97.1\\
-0.20&0.631&6.01& 0.05&100.0&98.3\\
-0.10&0.794&5.95& 0.07&100.0&98.4\\
0.00&1.000&5.93& 0.08&100.0&98.6\\
0.10&1.259&5.90& 0.10&100.0&98.9\\
0.20&1.585&5.57& 0.18&100.0&99.0\\
0.30&1.995&5.42& 0.25&100.0&98.8\\
0.40&2.512&5.79& 0.19&100.0&99.1\\
    \hline
    \end{tabular}
\end{table}

\begin{table*}
  \caption{The first ten lines of the {\it AKARI}/IRC IRAC Dark Field galaxy catalogue. The complete catalogue is available online}
  \label{table:galaxycatexample}
  \begin{center}
    \begin{tabular}{llllllllll}
      \hline \hline
RA&DEC&N4 flux/&N4 error/&S11 flux/&S11 error/&L15 flux/&L15 error/&L18W flux/&L18W error/\\
&&mJy&mJy&mJy&mJy&mJy&mJy&mJy&mJy\\
 \hline
264.96&69.08&0.027&0.0014&0.16&0.0069&0.11&0.016&0.16&0.018\\
264.99&69.08&0.036&0.0016&0.16&0.0070&0.15&0.014&0.20&0.019\\
265.07&69.08&0.030&0.0015&0.048&0.0046&0.16&0.012&0.15&0.018\\
265.08&69.03&0.069&0.0022&0.23&0.0077&0.21&0.015&0.25&0.020\\
265.08&69.04&0.030&0.0015&0.081&0.0054&0.20&0.015&0.14&0.017\\
265.02&69.07&0.045&0.0019&0.058&0.0050&0.18&0.014&0.26&0.020\\
265.05&69.06&0.038&0.0017&0.10&0.0059&0.21&0.015&0.26&0.020\\
265.03&69.06&0.054&0.0020&0.052&0.0048&0.14&0.014&0.27&0.020\\
265.02&69.07&0.098&0.0027&0.058&0.0049&0.18&0.015&0.13&0.017\\
264.88&69.05&0.040&0.0017&0.030&0.0044&0.087&0.012&0.14&0.017\\
    \hline
    \end{tabular}
  \end{center}
\end{table*}

\section{Comparison with Other Surveys and Models}
\label{discussion}
\begin{figure*}
\begin{tabular}{cc}
  \includegraphics[width=65mm]{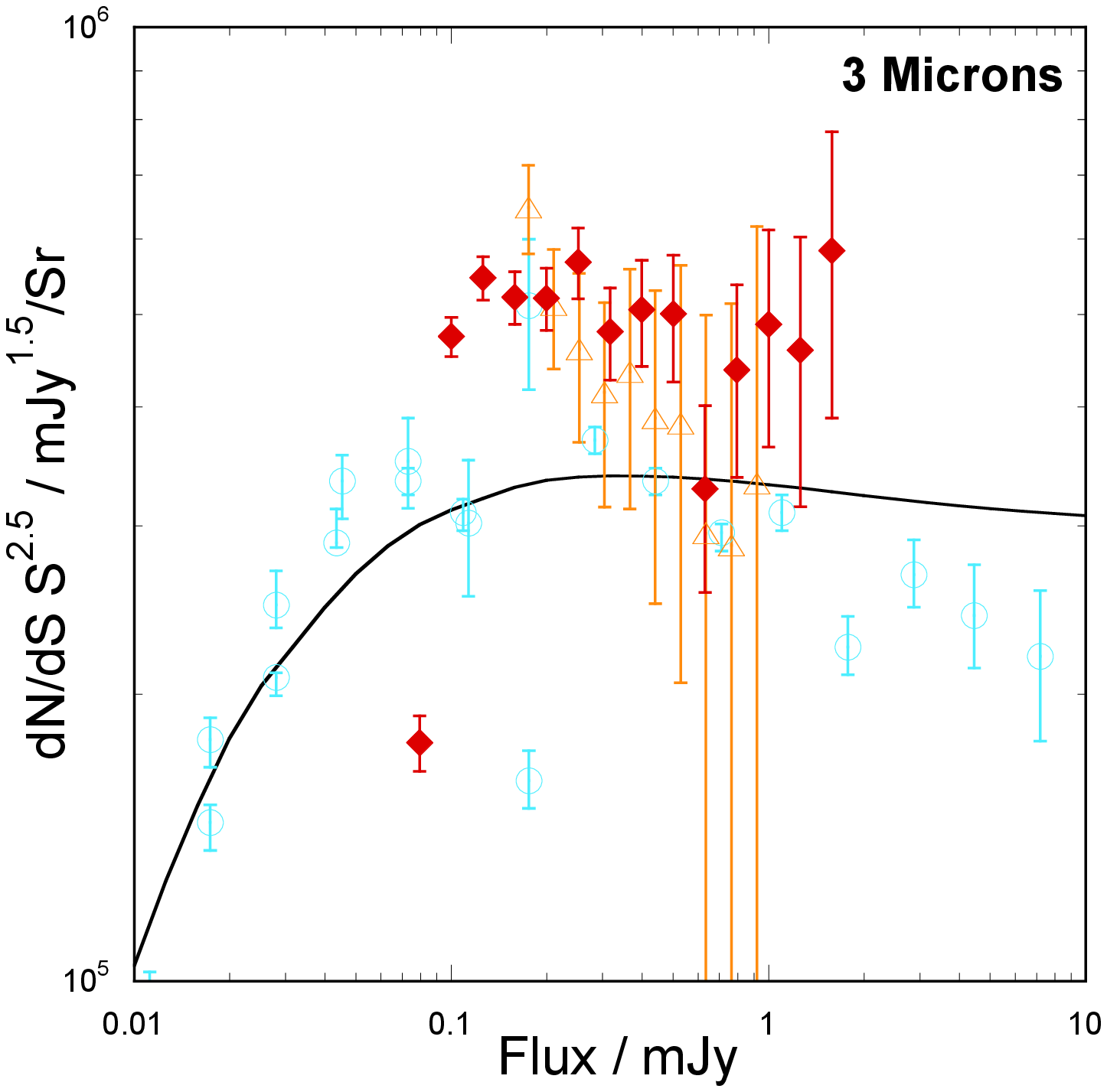} & 
  \includegraphics[width=65mm]{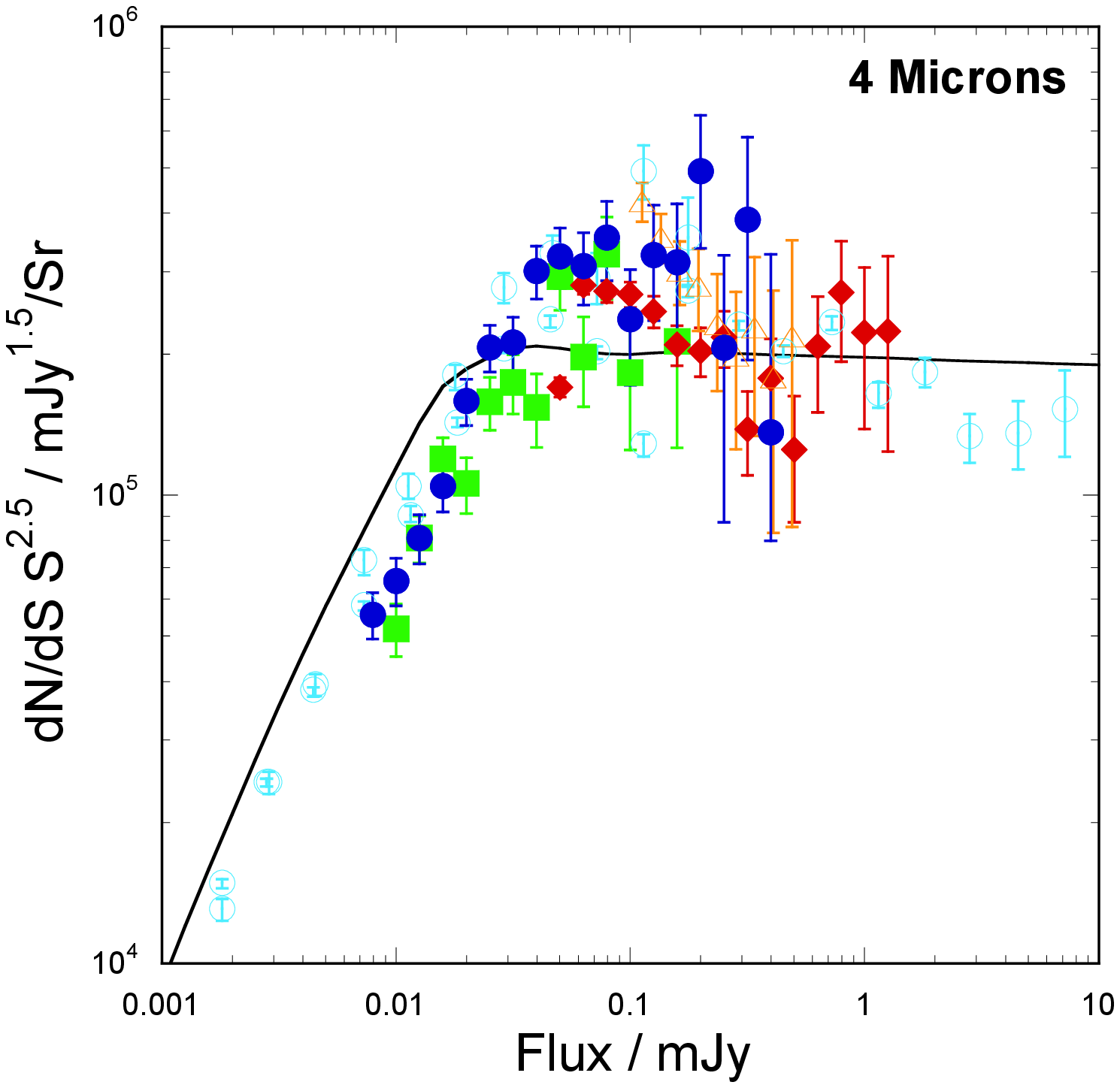} \\
  (a) 3 micron  & (b) 4 micron  \\[6pt]
  \includegraphics[width=65mm]{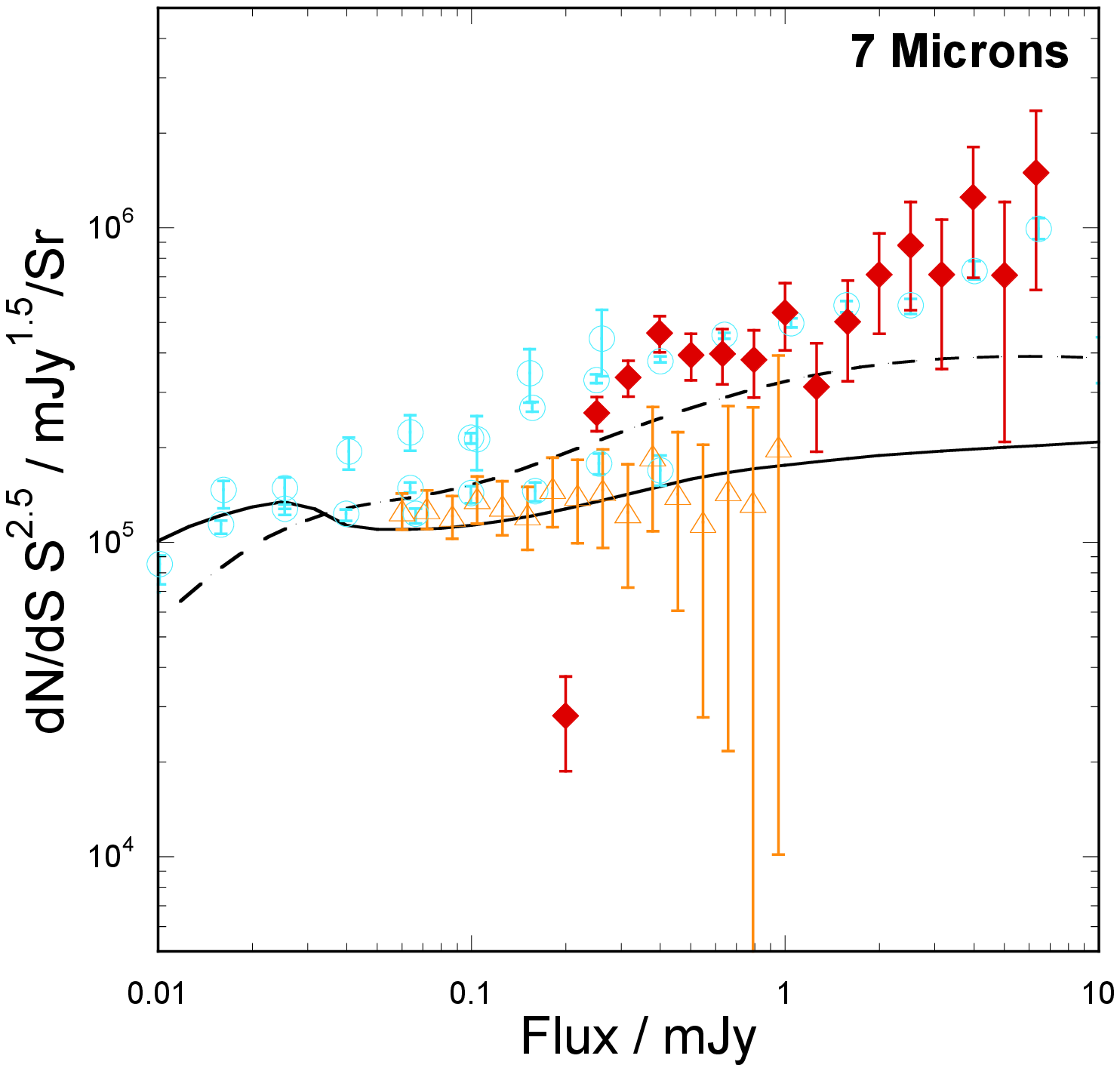} & 
  \includegraphics[width=65mm]{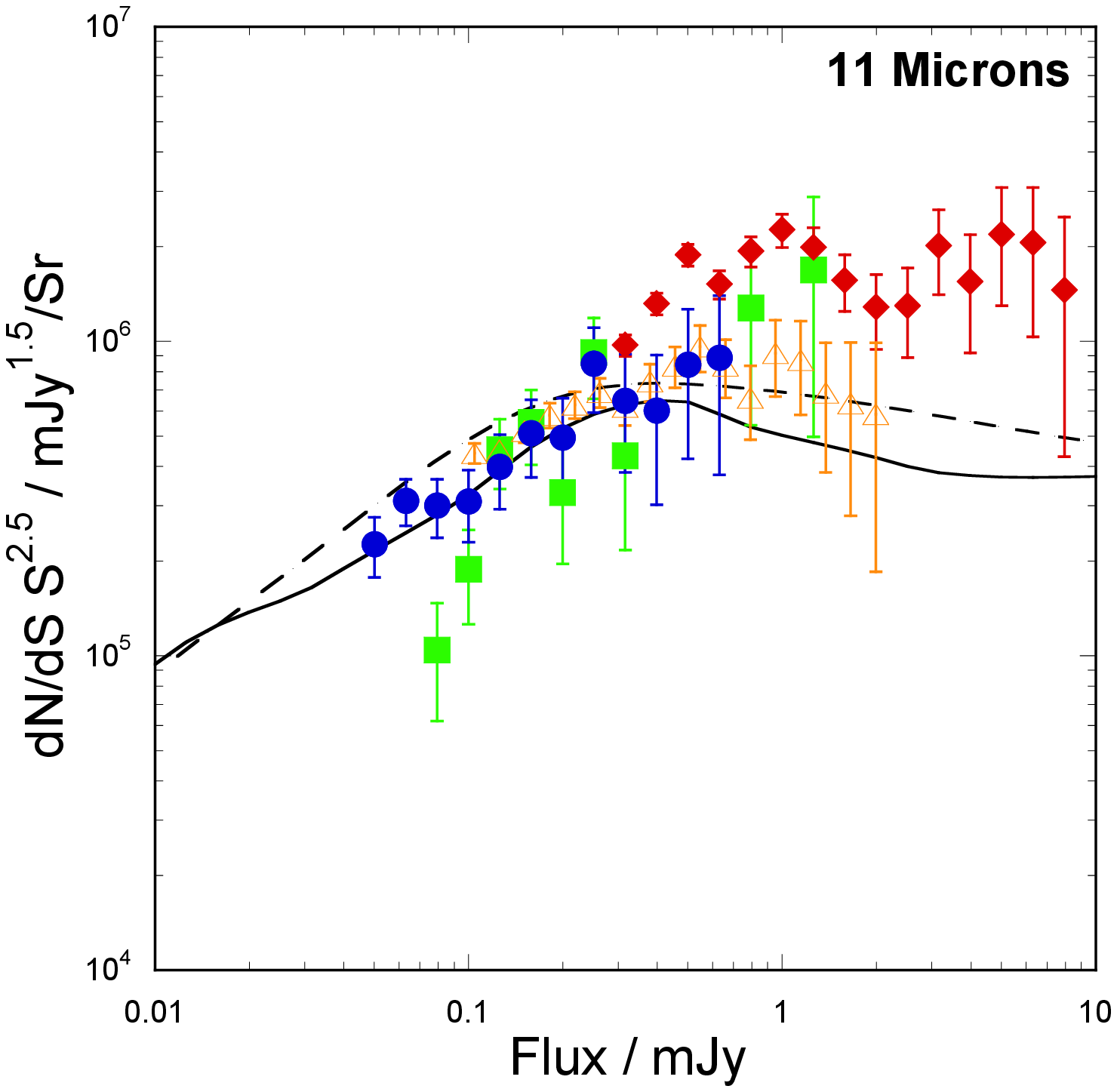} \\
  (c) 7 micron  & (d) 11 micron  \\[6pt]
    \includegraphics[width=65mm]{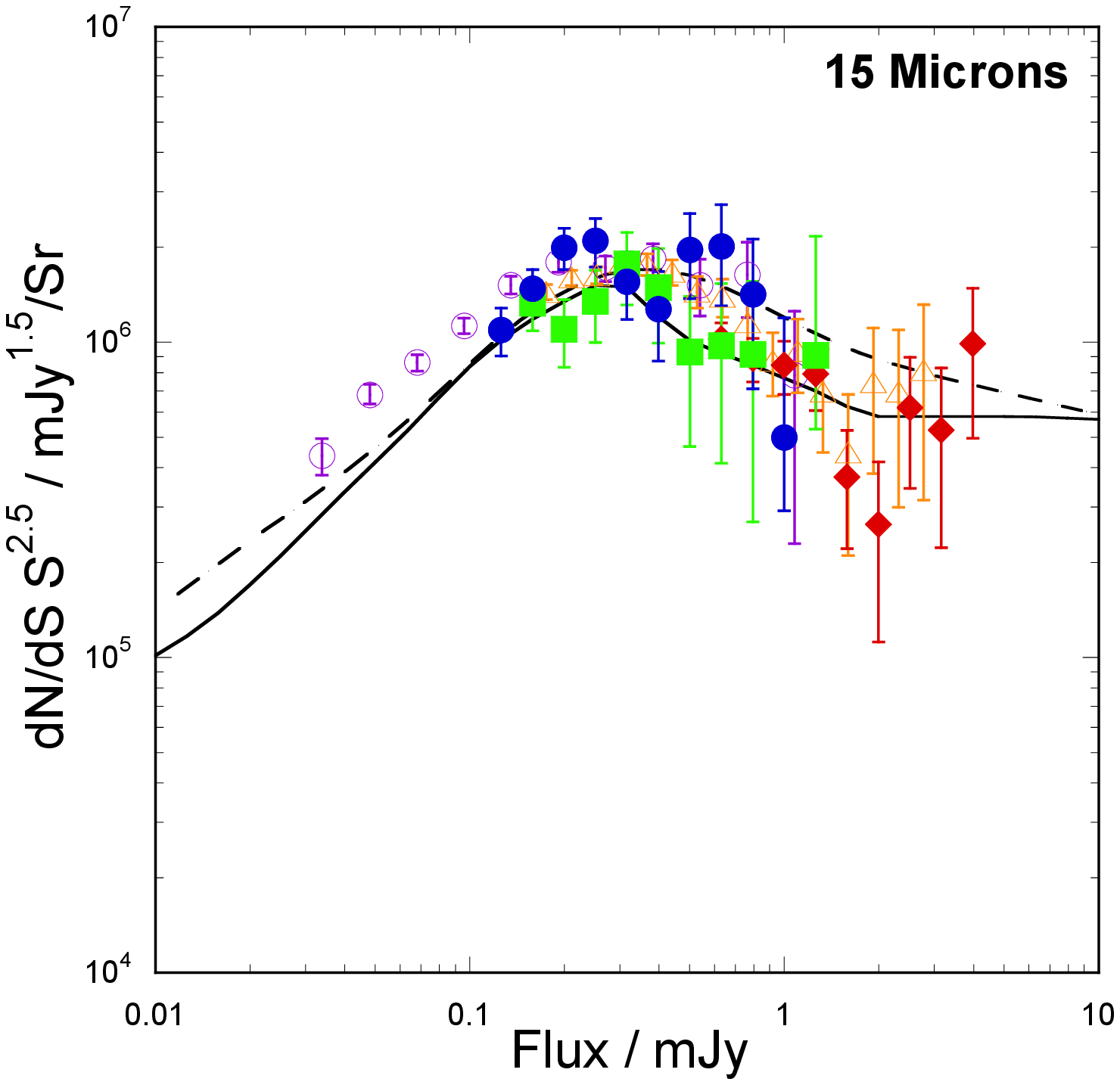} & 
  \includegraphics[width=65mm]{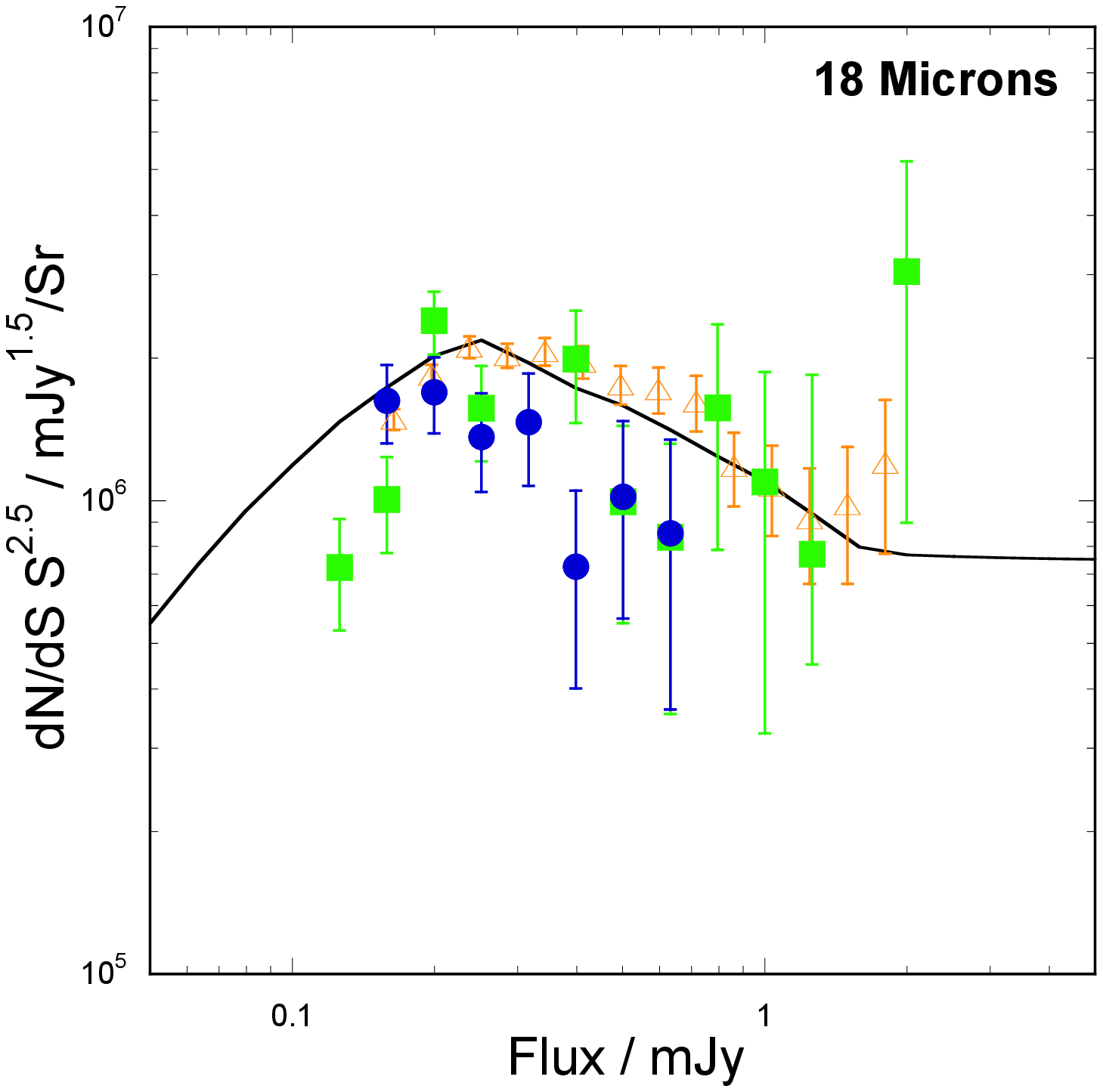} \\
  (e) 15 micron  & (f) 18 micron  \\[6pt]
\end{tabular}
\caption{Comparison of the {\it AKARI} IRAC Dark Field, ELAIS-N1 and ADF-S  source counts from this work, with the source counts from the {\it AKARI} NEP survey and {\it Spitzer}/IRAC and {\it Spitzer}/IRS surveys. The IRAC Dark Field counts are shown as dark blue circles, ELAIS-N1 number counts as green squares, and the ADF-S counts  as red diamonds. The {\it AKARI} NEP number counts are shown as open orange triangles \citep{Murata2014}, the {\it Spitzer}/IRAC counts as open cyan circles \citep{Fazio2004} and the {\it Spitzer}/IRS as open purple circles \citep{Teplitz2011}. The SISSA model is shown as a dashed line. The Pearson model is shown as a solid line, the grey line is the contribution of normal type galaxies, the magenta line is the contribution of elliptical type galaxies, the red line is the contribution of star forming galaxies, the cyan line is the contribution of LIRGs, the blue line is the contribution of ULIRGs and the green line is the contribution of AGN.}
\label{finalnumbercounts}
\end{figure*}

The source counts derived from the optimised toolkit are compared in Figure~\ref{finalnumbercounts} with previously published work from \cite{Murata2014} who presented {\it AKARI}  source counts from the North Ecliptic Pole (NEP) field, a $\sim$0.5 deg$^2$  area, using all  9 photometric bands of the IRC, located at $\rm (RA=17^h56^m,DEC=66^\circ 37')$ just offset from the ecliptic pole. Also shown are the {\it Spitzer} galaxy counts where available, from \cite{Fazio2004}, and the 16\,$\mu$m  counts from \cite{Teplitz2011} from the {\it Spitzer} InfraRed Spectrograph (IRS) peak-up camera.

To interpret the results, the source counts are compared to two different galaxy evolution models.
The Pearson phenomenological backwards evolution model (\citealt{Pearson2005}, \citealt{Pearson2009}) has provided a good fit to source counts from the near-infrared to millimetre wavelengths. The Pearson model is made up of six different population types; normal quiescent, elliptical, star-forming, luminous infrared galaxies (LIRGs), ultra luminous infrared galaxies (ULIRGs) and AGN. The model evolves the separate galaxy populations in luminosity (F($z$)) and density (G($z$)), both as a function of redshift. The parameterisation for both the evolution in luminosity and density are described as a double power law up to $z\sim2$. The power laws are dependent on galaxy type. At $z>2$ both the luminosity and density evolution decline to higher redshifts. The Pearson galaxy evolution model has strong links with AKARI, as it was influential in deciding the wavelength of the near and mid-infrared filters.

The galaxy evolution model of \cite{Cai2013} (also known as and hereafter referred to as the SISSA model) is a hybrid model using a combination of physical and phenomenological models. The galaxy populations have been divided into high-$z$ ($z\geqslant1.5$ ), proto-spheriods and AGN, and low-$z$ ($z<1.5$), late type `cold' galaxies and `warm' starburst galaxies. To model the high-$z$ population, \cite{Cai2013} use a bolometric luminosity function for spheroidal galaxies and SEDs for the high-$z$ AGN; these are used in a physical forward evolution model, based on \cite{Granato2004}. To model the low-$z$ populations \cite{Cai2013} use a parametric phenomenological backward evolution model. The low-$z$ population is subdivided into `warm' starburst galaxies, `cold' late type galaxies, type 1 AGN and type 2 AGN; where the AGN have been `reactivated'. Using their galaxy evolution models, \cite{Cai2013} have created luminosity functions, SEDs and source count.

Both the Pearson and the SISSA models are plotted with the observed source counts in Figure \ref{finalnumbercounts}.

The source counts at 3.2\,$\mu$m are shown in Figure \ref{finalnumbercounts}.a. The source counts from this work lie slightly above the {\it AKARI} NEP counts of \cite{Murata2014} but are consistent within the errors. However, there appears to be a significant offset when compared with the IRAC counts of \citep{Fazio2004}, although the scatter is large. Some of this difference may be be due to stellar subtraction. Note that the optical data used for stellar subtraction did not provide full coverage of the area, therefore the fractional stellar contribution was derived statistically. As there are very few stars in the observed area, the scaling is likely to be affected by small number statistics. In this band, the Pearson model  is in better agreement with the IRAC 3.6\,$\mu$m counts. The counts of \cite{Murata2014}  exhibit a faint upturn at the faintest fluxes not seen in the ELAIS-N1 number counts or predicted by the model.

The source counts at 4.1\,$\mu$m are shown in Figure \ref{finalnumbercounts}.b for both the IRAC Dark Field, ELAIS-N1 and ADF-S fields. The counts are compared with the {\it AKARI} NEP, the IRAC 4.5\,$\mu$m counts and the Pearson evolutionary model. 
The observed counts span a large range in flux density from $\sim$2\,mJy to  0.01\,mJy with the wider ADF-S field constraining the bright end of the source counts (S$>$2\,mJy) and the fainter counts covered by the deeper IRAC Dark Field and ELAIS-N1 field. The source counts derived from the optimised toolkit are in good agreement with the IRAC 4.5\,$\mu$m counts at all flux densities.
The faint end slope of the source counts appears steeper for the counts presented here compared with the IRAC counts, and the Pearson model predicts an even shallower slope to the faint end of the source counts. At this faint level, the counts are probing the region between the dust emission and stellar contribution within galaxies where the assumed spectral energy distribution for the models are not well constrained.

Figure \ref{finalnumbercounts}.c shows the {\it AKARI} 7\,$\mu$m counts in the ADF-S (neither the IRAC Dark Field nor ELAIS-N1 were observed in this band) plotted together with the {\it AKARI} NEP and the IRAC 8\,$\mu$m counts. The ADF-S counts are in good agreement with the IRAC 8\,$\mu$m over the 0.2$<$S$<$10\,mJy flux range. The SISSA models appear to fit the ADF-S counts better than the Pearson model. However, the Pearson model and {\it AKARI} NEP counts are in agreement, although the errors on the NEP counts are very large. It should be noted that even at this longer wavelength, the reliability of the stellar subtraction method can affect the source counts at the brightest flux levels.

The 11\,$\mu$m  source counts are shown in Figure \ref{finalnumbercounts}.d for both the IRAC Dark Field, ELAIS-N1 and ADF-S fields compared with the {\it AKARI} NEP counts and the Pearson and SISSA models. The 11\,$\mu$m band was unique to {\it AKARI} and there are no {\it Spitzer} counts in this waveband. The IRAC Dark Field and the NEP counts appear to be in agreement. There is a significant scatter in the ELAIS-N1 counts possibly due to remnants of Earthshine, since out of the four filters observed in the ELAIS-N1 field, the S11 band had the worst Earthshine effect. In this band the observations cannot distinguish between the models (which are themselves quite similar), but the Pearson model appears to fit the IRAC Dark Field better at fainter fluxes, whereas the SISSA model appears to fit the brighter counts better. Figure \ref{finalnumbercounts}.d shows that the S11 number counts are deeper than those of \cite{Murata2014}, the previous deepest 11\,$\mu$m counts. Thus the AKARI IRAC Dark Field 11\,$\mu$m image created in the work of this paper, shown in Figure \ref{iracdarkfieldmosaickedimages}.b, is currently the deepest 11\,$\mu$m image of the sky.

Figure \ref{finalnumbercounts}.e shows the source counts at 15\,$\mu$m for the IRAC Dark Field, ELAIS-N1 and the ADF-S fields. Also plotted are the {\it AKARI} NEP counts, the {\it Spitzer} IRS Peak-Up counts \citep{Teplitz2011}, and the Pearson and SISSA models. All the source counts are in relatively good agreement and consistent with each other. Although the Pearson and SISSA evolutionary predictions peak at slightly different flux levels, both models are consistent with the observed counts (except at the faint end where they under predict compared to the IRS counts). However, it should be noted that the IRS counts were made over a very small area of 36 square arc minutes and only 153 sources.

Figure \ref{finalnumbercounts}.f shows both the IRAC Dark Field and ELAIS-N1 18\,$\mu$m counts with the {\it AKARI} NEP counts and the Pearson model. The counts presented here are in broad agreement with the NEP counts given that  the ELAIS-N1 L18W data were taken towards the end of Phase 2 and therefore were severely affected by Earthshine, which is likely to be responsible for the large scatter. There is some hint that the counts fall off steeper than the evolutionary model predicts but given the large uncertainties this is not well constrained.

A small disparity in source counts from two different deep fields, is believed be due to cosmic variance \citep{Somerville2004}, the uncertainty in the number of galaxies in a given volume density, caused by large-scale density fluctuations. The percentage of cosmic variance is greater for deep, narrow fields, e.g. the IRAC Dark Field, ELAIS-N1 and NEP. A discrepancy in the SISSA model and the shorter source counts could be due to the fact that elliptical galaxies are not fully modelled. At high-$z$ the SISSA model includes a term for the evolution of proto-spheroid galaxies, the progenitors to elliptical galaxies, but they do not fully model the evolution of proto-spheroids to ellipticals. Also at low-$z$ there is no specific term for elliptical galaxies, and there are no discussion about the formation of ellipticals through major merger of spiral galaxies. Thus including the evolution of elliptical galaxies in the SISSA model should increase the predicted source counts. The biggest increase would be expected at the shorter wavelengths (where the contribution from the older stellar population is significant. This could explain the discrepancy between the SISSA model and the source counts at 7\,$\mu$m.

Overall, a combination of the deep IRAC Dark Field and ELAIS-N1, and the shallower ADF-S fields covers a large range in flux density from 0.1$<$S$<$10\,mJy. Over this range, both evolutionary models fit the majority of the source counts well. There is evidence that in the mid-short infrared bands (7, 11\,$\mu$m) that some modification is needed in the models. At these wavelengths the SISSA model performs slightly better than the Pearson model. The inconsistencies may be due to incorrect stellar subtraction on the counts, or may suggest incorrect modelling of stellar populations, e.g. thermally pulsating asymptotic giant branch (TP-AGB) stars.

\section{Investigation of AKARI colours}
\label{akari_colours}

\begin{figure*}
\begin{tabular}{cc}
  \includegraphics[width=70mm]{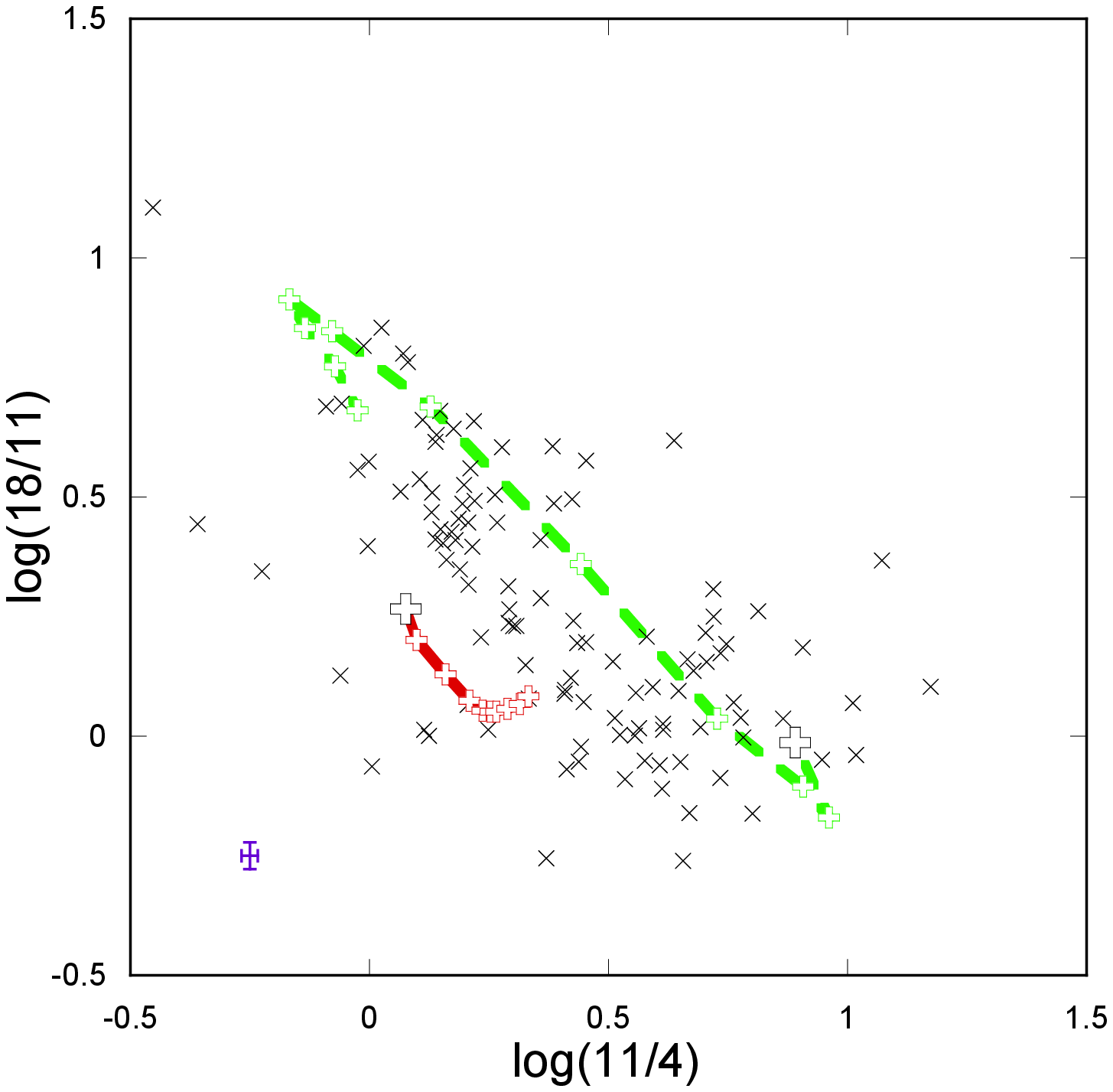} & 
  \includegraphics[width=70mm]{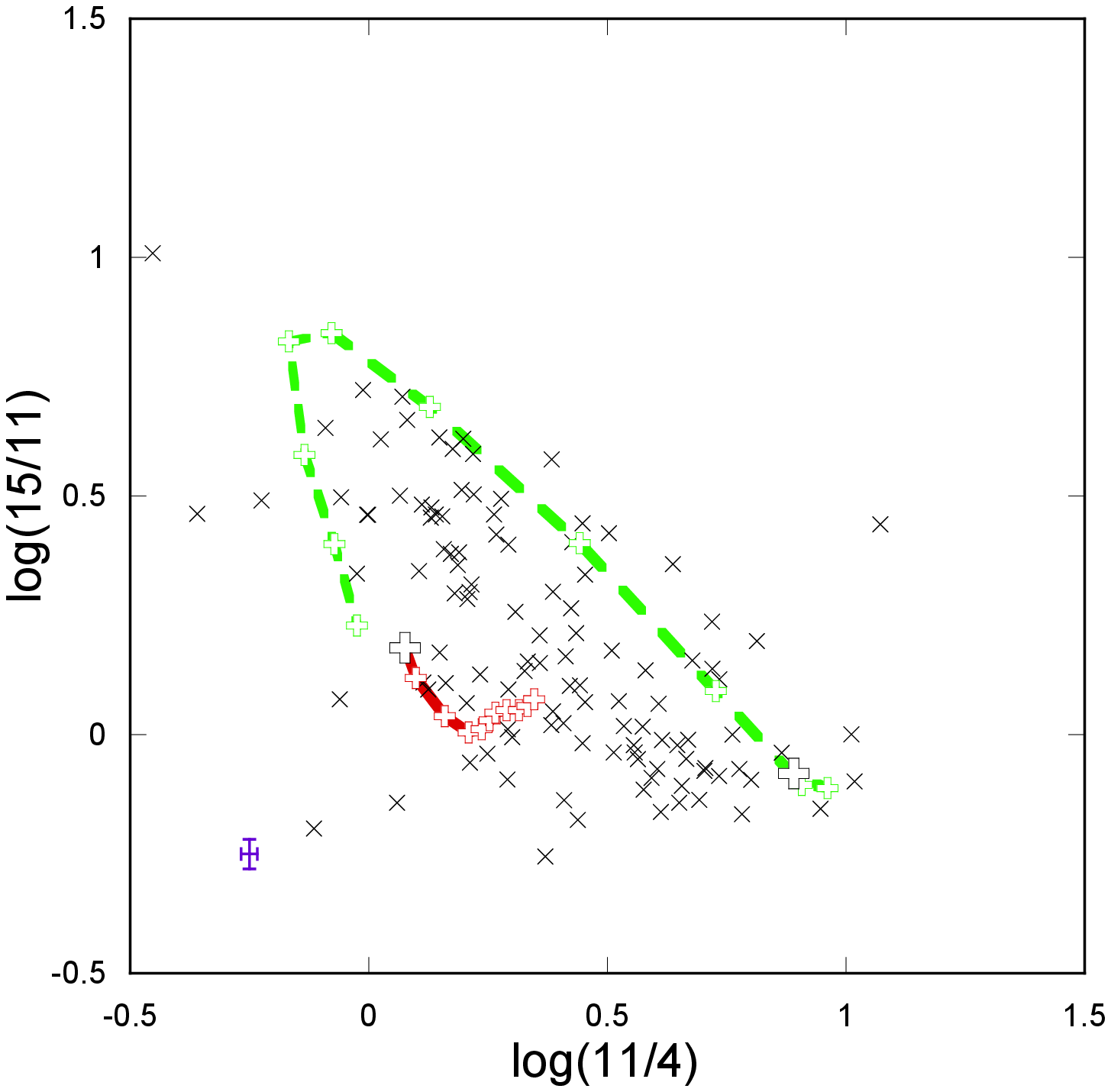} \\
  (a) 11/4 against 18/11 & (b) 11/4 against 15/11 Filter \\[6pt]
\end{tabular}
\caption{Colour-colour diagrams of the AKARI N4, S11, L15 and L18W filters. The black crosses are the ELAIS-N1 and IRAC Dark Field galaxy colours, the starburst track is in the dashed green line and the AGN track is in solid red line. The tracks in figures a and b have been cut at $z=2$. $z=0$ is marked by a larger cross. An example error for the galaxy flux is shown as the purple error bars in the lower left corner.}
\label{fig:agnseparation} 
\end{figure*}

In this section the colour-colour diagrams in the mid-infrared, specifically AKARI N4, S11, L15 and L18W are investigated. Most of the work using mid-infrared colour-colour diagrams to differentiate between AGN and star forming galaxies has been performed using the {\it Spitzer}/IRAC and MIPS filters (\citealt{Lacy2004}, \citealt{Stern2005} and \citealt{Donley2012}). There is some published work on using the AKARI/IRC mid-infrared filters to separate AGN dominated galaxies and star forming galaxies \citep{Hanami2012}. Investigation of the AKARI colours was mainly performed to trace the PAH emission bands, which was used for galaxy separation and to calculate star formation rate (SFR). The study of PAH in galaxy separation may not always work, as PAH features can be suppressed in star forming galaxies, and in some cases AGN are known to have PAH emission (\citealt{Murata2014b} \& \citealt{Laurent2000}).

Due to the fact that the {\it AKARI} mid-infrared filters were designed to study the PAH features, which are prominent in star forming and AGN composite objects, star forming and spiral galaxies, and to a lesser extent in AGN; the tracks of these types of galaxies cover much of the {\it AKARI} colour-colour space. Thus many different galaxy types lie in the same region of the {\it AKARI} colour-colour plots. This is a feature seen in all {\it AKARI} colour-colour plots, and an indication that the {\it AKARI} colour-colour space is not ideally suited for galaxy separation. 

Figure \ref{fig:agnseparation} shows two {\it AKARI} colour-colour plots, with a starburst (green) and AGN (red) track made from spectral energy distribution (SED) templates of \cite{Berta2013}, and {\it AKARI} IRAC Dark Field and ELAIS-N1 sources. In the mid-infrared , at $z<2$, the starburst track is dominated by PAH features, as the features travel through the different filters. It is the PAHs which are mainly causing the changes in track angle. On the {\it AKARI} colour-colour diagrams the AGN tracks cover a smaller area of the plot than the starburst track. This is due to the fact that PAH emission tends to be suppressed in AGN dominated galaxies, due to the intense radiation from the AGN, thus the AGN tracks do not have as prominent PAH features travelling through the filters, and causing the tracks to move around the diagram to the same extent as for the starburst track. The changes in direction of the starburst track can be explained by the PAH features. Figure \ref{fig:agnseparation}.a shows S11/N4 against L18W/S11. At $z\sim0.2$ the S11/N4 flux ratio decreases, while the L18W/S11 flux ratio increases. This is due to the 7.7\,$\mu$m PAH feature leaving the S11 filter at increasing redshift. Figure \ref{fig:agnseparation}.b shows S11/N4 against L15/S11. The sharp decrease in the flux ratio L15/S11 and only a small increase in the S11/N4 flux ratio at the $z\sim1.4$ is due to the 6.2\,$\mu$m PAH feature leaving the L15 filter.

It was investigated whether AGN dominated galaxies could be separated from starburst galaxies by just using the IRAC Dark Field and ELAIS-N1 AKARI colours. Figures \ref{fig:agnseparation}.a and \ref{fig:agnseparation}.b show colour-colour diagrams of the AKARI filters. The starburst and the AGN track go from $0<z<2$. As the AGN track clearly occupies a different region from the starburst track, for this range of redshifts an AGN selection area on the colour-colour plot has been identified, i.e. the area around the AGN track. For the galaxies in these two plots $\sim75$\% have $z<2$, so this is not an unreasonable redshift cut off. As can be seen in Figure \ref{fig:agnseparation}, the galaxies do not just populate the starburst and the AGN tracks, but are also found in regions between, indicating that many of the sources are composite objects. This indicates that to reliably confirm galaxy type, SED fitting is required. This investigation has only just four of the nine {\it AKARI} filters. An investigation including filters not observed by the IRAC Dark Field and ELAIS-N1 (e.g. the S9W filter) may find further AGN selection criteria. 

\section{Conclusions}
\label{conclusions}
A new toolkit for observations made with the {\it AKARI} IRC instrument has been presented that is specifically optimised for the data analysis of extragalactic fields. The main differences between the optimised toolkit and the archival pipeline are the removal of hot pixels, creation of noise images, distortion correction, Earthshine light correction, astrometry correction and the masking of artefacts. 

This optimised toolkit was applied to three {\it AKARI} example test fields; the narrow-deep early to mid-Phase 2 IRAC Dark Field, the narrow-deep late-Phase 2 ELAIS-N1 field, and the wide-shallow ADF-S field. These fields were selected on the basis of difficulty of reduction using the archival pipeline.
Source catalogues and completeness plus reliability corrected source counts were produced for the {\it AKARI} bands from $3-18\,\mu$m spanning the wavelength desert between the {\it Spitzer} IRAC and MIPS instruments.

The 4, 11, 15 and 18\,$\mu$m {\it AKARI} colours have been investigated, and AGN selection criteria has been found for galaxies $z<2$. When plotting the {\it AKARI} IRAC Dark Field and ELAIS-N1 sources on colour-colour plots, many objects were identified as star forming and AGN composite objects.

The source count results presented in this paper are in broad agreement with the previously published results from {\it AKARI} surveys and {\it Spitzer} surveys with the IRAC instrument and IRS Peak-Up observations. We do however also find discrepancies at brighter fluxes (in the ADF-S) in some instances that we attribute to problems with the stellar subtraction in either the {\it AKARI} or IRAC data.

The observed source counts were also compared with the galaxy evolutionary models of \cite{Pearson2005}, \cite{Pearson2009} and \cite{Cai2013}. We find that no single evolutionary model can fit all the wavebands simultaneously. The models are in good agreement with the counts at wavelengths longer than 11\,$\mu$m but have difficulties at shorter wavelengths. This could be due to incorrect stellar subtraction in the source counts, or incorrect modelling of stellar populations.

Further data reduction of other extragalactic fields in the  {\it AKARI} data archive using the optimised toolkit is expected to constrain the evolutionary models further and will be presented in future work.

\section*{Acknowledgments}

The {\it AKARI} Project is an infrared mission of the Japan Space Exploration Agency (JAXA) Institute of Space and Astro-nautical Science (ISAS), and is carried out with the participation of mainly the following institutes; Nagoya University, The University of Tokyo, National Astronomical Observatory Japan, The European Space Agency (ESA), Imperial College London, University of Sussex, The Open University (UK), University of Groningen / SRON (The Netherlands), Seoul National University (Korea). The far-infrared detectors were developed under collaboration with The National Institute of Information and Communications Technology.

The authors would like to acknowledge the STFC Doctoral training grant ST/K502212/1 and support from the Open University.

Helen Davidge would also like to thank the Japan Society for the Promotion of Science for financial support on the JSPS Summer Programme 2015.

\appendix
\section[Distortion Polynomials]{Distortion Polynomials}\label{appendixA}
The distortion correction polynomials used in the optimised toolkit.

N2 filter\\
\begin{tabular}{c |  c c c}
$x_{ij}$ & 0 & 1 & 2\\
\hline
0 & -5.83 & 0.00190 & 0.0000111\\
1 & 1.02 & 0.0000372 & -0.0000000809\\
2 & -0.0000106 & 0.00000000946 & 0.0000000000218\\
\end{tabular}

\begin{tabular}{c |  c c c}
$y_{ij}$ & 0 & 1 & 2\\
\hline
0 & 5.53 & 0.986 & 0.0000157\\
1 & -0.0140 & -0.0000182 & 0.0000000933\\
2 & 0.0000151 & 0.0000000286 & -0.000000000217\\
\end{tabular}

N3 filter\\
\begin{tabular}{c |  c c c}
$x_{ij}$ & 0 & 1 & 2\\
\hline
0 & -3.35 & 0.00356 & 0.00000682\\
1 & 1.02 & 0.0000157 & -0.00000000996\\
2 & -0.0000131 & 0.0000000653 & -0.000000000155\\
\end{tabular}

\begin{tabular}{c |  c c c}
$y_{ij}$ & 0 & 1 & 2\\
\hline
0 & 5.55 & 0.985 & 0.0000183\\
1 & -0.0117 & -0.0000119 & 0.0000000434\\
2 & 0.0000101 & 0.00000000766 & -0.0000000000696\\
\end{tabular}

N4 filter\\
\begin{tabular}{c |  c c c}
$x_{ij}$ & 0 & 1 & 2\\
\hline
0 & -4.01 & 0.00472 & 0.00000255\\
1 & 1.02 & 0.0000165 & 0.00000000153\\
2 & -0.00000648 & 0.0000000232 & -0.0000000000770\\
\end{tabular}

\begin{tabular}{c |  c c c}
$y_{ij}$ & 0 & 1 & 2\\
\hline
0 & 2.79 & 0.984 & 0.0000210\\
1 & -0.0117 & -0.0000182 & 0.0000000562\\
2 & 0.0000121 & 0.0000000243 & -0.000000000123\\
\end{tabular}

S7 filter\\
\begin{tabular}{c |  c c c}
$x_{ij}$ & 0 & 1 & 2\\
\hline
0 & -1.64 & -0.00206 & 0.0000597\\
1 & 0.987 & 0.000424 & -0.00000127\\
2 & 0.0000754 & -0.00000124 & 0.00000000372\\
\end{tabular}

\begin{tabular}{c |  c c c}
$y_{ij}$ & 0 & 1 & 2\\
\hline
0 & 1.186 & 1.03 & -0.0000342\\
1 & -0.0150 & -0.0000464 & 0.000000140\\
2 & 0.00000631 & -0.000000149 & 0.000000000357\\
\end{tabular}

S9W filter\\
\begin{tabular}{c |  c c c}
$x_{ij}$ & 0 & 1 & 2\\
\hline
0 & -6.03 & 0.0116 & 0.0000231\\
1 & 1.02 & 0.0000693 & -0.000000256\\
2 & -0.0000120 & -0.000000104 & 0.000000000476\\
\end{tabular}

\begin{tabular}{c |  c c c}
$y_{ij}$ & 0 & 1 & 2\\
\hline
0 & 0.930 & 1.03 & -0.0000246\\
1 & -0.0141 & -0.0000357 & 0.0000000382\\
2 & -0.00000399 & -0.0000000922 & 0.000000000486\\
\end{tabular}

S11 filter\\
\begin{tabular}{c |  c c c}
$x_{ij}$ & 0 & 1 & 2\\
\hline
0 & -3.71 & -0.0265 & 0.000167\\
1 & 0.994 & 0.000635 & -0.00000234\\
2 & 0.0000643 & -0.00000219 & 0.00000000804\\
\end{tabular}

\begin{tabular}{c |  c c c}
$y_{ij}$ & 0 & 1 & 2\\
\hline
0 & -2.99 & 1.06 & -0.000186\\
1 & 0.00653 & -0.000683 & 0.00000332\\
2 & -0.0000683 & 0.00000231 & -0.0000000116\\
\end{tabular}

L15 filter\\
\begin{tabular}{c |  c c c}
$x_{ij}$ & 0 & 1 & 2\\
\hline
0 & 8.06 & 0.00799 & -0.0000285\\
1 & 0.949 & -0.000125 & 0.000000418\\
2 & -0.000101 & -0.000000206 & 0.000000000793\\
\end{tabular}

\begin{tabular}{c |  c c c}
$y_{ij}$ & 0 & 1 & 2\\
\hline
0 & -2.09 & 1.01 & 0.00000684\\
1 & -0.0111 & 0.000205 & -0.000000563\\
2 & 0.0000543 & -0.00000124 & 0.00000000396\\
\end{tabular}

L18W filter\\
\begin{tabular}{c |  c c c}
$x_{ij}$ & 0 & 1 & 2\\
\hline
0 & 7.38 & 0.0125 & -0.0000459\\
1 & 0.951 & -0.000221 & 0.000000677\\
2 & -0.000125 & 0.000000662 & -0.00000000172\\
\end{tabular}

\begin{tabular}{c |  c c c}
$y_{ij}$ & 0 & 1 & 2\\
\hline
0 & -2.96 & 1.01 & -0.00000412\\
1 & 0.00101 & -0.0000439 & 0.000000199\\
2 & -0.00000396 & 0.000000152 & -0.000000000453\\
\end{tabular}

L24 filter\\
\begin{tabular}{c |  c c c}
$x_{ij}$ & 0 & 1 & 2\\
\hline
0 & 8.25 & -0.00220 & 0.00000134\\
1 & 0.933 & -0.0000637 & 0.000000380\\
2 & -0.0000709 & 0.000000267 & -0.00000000142\\
\end{tabular}
\smallskip
\begin{tabular}{c |  c c c}
$y_{ij}$ & 0 & 1 & 2\\
\hline
0 & -0.178 & 1.012 & 0.0000161\\
1 & -0.00356 & 0.0000791 & -0.000000312\\
2 & 0.0000135 & -0.000000220 & 0.000000000965\\
\end{tabular}

\bsp

\label{lastpage}

\end{document}